\documentclass[aps,prb,twocolumn,amsmath,amssymb,superscriptaddress,floatfix,eqsecnum]{revtex4-1}

\usepackage{graphicx}
\usepackage{multirow}
\usepackage{color}
\usepackage{bm}
\usepackage{hyperref}
\usepackage[normalem]{ulem}
\renewcommand{\Vec}[1]{\mbox{\boldmath$#1$}}
\def\infinity{\infty}
\def\t#1{\textrm{#1}}

\def\tensor{\otimes}

\newcommand{\blue}[1]{{\textcolor{blue}{#1}}}

\begin{document}
\title{
Breakdown of the topological classification $\mathbb{Z}$ for
gapped phases of noninteracting fermions by quartic interactions
      }

\author{Takahiro Morimoto}
%\affiliation{Condensed Matter Theory Laboratory, RIKEN, Wako, Saitama, 351-0198, Japan}
\affiliation{RIKEN Center for Emergent Matter Science (CEMS), 
Wako, Saitama, 351-0198, Japan}

\author{Akira Furusaki}
\affiliation{Condensed Matter Theory Laboratory, 
RIKEN, Wako, Saitama, 351-0198, Japan}
\affiliation{RIKEN Center for Emergent Matter Science (CEMS), 
Wako, Saitama, 351-0198, Japan}

\author{Christopher Mudry}
\affiliation{Condensed Matter Theory Group,
  Paul Scherrer Institute, CH-5232 Villigen PSI, Switzerland}

\date{\today}

\begin{abstract}
The conditions for both the stability and the breakdown of 
the topological classification 
of gapped ground states of noninteracting fermions,
the tenfold way, 
in the presence of quartic fermion-fermion interactions are given
for any dimension of space. This is achieved by encoding the effects of
interactions on the boundary gapless modes 
in terms of boundary dynamical masses. 
Breakdown of the noninteracting  topological classification occurs
when the quantum nonlinear sigma models 
for the boundary dynamical masses
favor quantum disordered phases.
For the tenfold way, 
we find that 
(i)
the noninteracting topological classification $\mathbb{Z}^{\,}_{2}$ 
is always stable,
(ii) 
the noninteracting topological classification $\mathbb{Z}$ 
in even dimensions is always stable,
(iii) the noninteracting topological classification $\mathbb{Z}$ 
in odd dimensions is unstable and reduces to 
$\mathbb{Z}^{\,}_{N}$ 
that can be identified explicitly for any dimension and any defining
symmetries. 
We also apply our method to the three-dimensional topological
crystalline insulator SnTe from the symmetry class AII$+R$,
for which we establish the reduction 
$\mathbb{Z}\to\mathbb{Z}^{\,}_{8}$
of the noninteracting topological classification. 
\end{abstract}

%72.10.-d 	Theory of electronic transport; scattering mechanisms
%73.20.-r 	Electron states at surfaces and interfaces
%71.27.+a 	Strongly correlated electron systems; heavy fermions
\maketitle

%\tableofcontents

\newpage 

\section{Introduction}

Topological insulators (TIs) and topological superconductors (TSs) 
of noninteracting fermions are characterized by topological numbers 
($\mathbb{Z}$ or $\mathbb{Z}^{\,}_{2}$) that encode the
non-trivial 
topology of the occupied single-particle 
wave functions and are accompanied by gapless excitations that are 
localized along any boundary.%
~\cite{hasan-kane10,qi-zhang-rmp11}
The integer quantum Hall effect (IQHE) is characterized by
the Hall conductivity quantized by the integer $\nu=1,2,\cdots$ in units
of $e^{2}/h$. The topological integer $\nu$ counts the number of extended 
chiral edge modes propagating at the boundary of the sample.
The $\mathbb{Z}^{\,}_{2}$ topological insulator
is characterized by the parity of the number of Kramers' doublets of extended
boundary modes.
Together with polyacetylene and a two-dimensional $p+\mathrm{i}p$
superconductor,%
~\cite{Su79,Read00}
both instances are now understood 
to be non-trivial entries in the periodic table (i.e., the tenfold way) 
for noninteracting topological insulators and superconductors.%
~\cite{schnyder-ryu08,kitaev09,ryu-njp10}

The gapless modes appearing at the boundary in the IQHE 
are robust to both elastic and
inelastic scattering resulting from
one-body impurity potentials and
many-body electron-electron interactions.%
~\cite{Laughlin81,Niu85}
Similarly, the gapless modes in the $\mathbb{Z}_2$ TIs
are immune to \textit{both} backscattering resulting from
one-body impurity potentials and
many-body electron-electron interactions, 
provided time-reversal symmetry (TRS) 
is neither explicitly nor spontaneously broken.%
~\cite{Kane05a,Kane05b,Levin09,Neupert11b}

Given the robustness to many-body fermion-fermion interactions
of the edge states in the IQHE, 
it was a remarkable observation made by Fidkowski and Kitaev in 2010 
that it is is possible to gap out eight Majorana zero modes
localized at the end of a one-dimensional topological superconducting wire
through many-body interactions without closing the spectral gap in the bulk.%
~\cite{Fidkowski10,Fidkowski11}
In the terminology of the tenfold way,%
~\cite{schnyder-ryu08,kitaev09,ryu-njp10} 
it was demonstrated in
Refs.\ \onlinecite{Fidkowski10,Fidkowski11}
that the $\mathbb{Z}$ topological classification for the noninteracting
one-dimensional symmetry class BDI, when interpreted as a superconductor,
is (i) unstable to quartic contact interactions
that neither break explicitly nor spontaneously
the TRS, and (ii) this instability reduces the noninteracting 
topological classification $\mathbb{Z}$ 
to $\mathbb{Z}^{\,}_{8}$. 

Subsequently, noninteracting two-dimensional topological crystalline
superconductors (TCSs) from the symmetry class DIII$+R$
(where ``$+R\,$'' indicates the presence of an additional reflection symmetry)
and three-dimensional topological superconductors
from the symmetry class DIII were
shown in Refs.\ \onlinecite{Yao13,Qi13}
and Refs.\ 
\onlinecite{kitaev-topomat11,Fidkowski13,Metlitski14,Wang14,You14} 
to display the reduction patterns
$\mathbb{Z}\to\mathbb{Z}^{\,}_{8}$
and
$\mathbb{Z}\to\mathbb{Z}^{\,}_{16}$,
respectively, when perturbed by quartic contact interactions
that neither break explicitly nor spontaneously
the defining symmetries.~% 
\footnote{
In particular, the breakdown of the noninteracting 
topological classifications with the group $\mathbb{Z}$
in three-dimensional SPT phases was diagnosed in 
Refs.\ \onlinecite{Metlitski14,Wang14} 
through the proliferation of certain types of vortices 
in order parameters that spontaneously break one of the defining symmetries.
In this approach, a fully gapped surface phase
is realized at certain values of $\nu$ with all protecting symmetries restored 
by the proliferation of vortices. 
This strategy was also applied to four-dimensional SPT phases 
in Ref.\
\onlinecite{You-Bentov-Xu14}.
         }
The reductions
$\mathbb{Z}\to\mathbb{Z}^{\,}_{4}$
and
$\mathbb{Z}\to\mathbb{Z}^{\,}_{8}$
for the three-dimensional symmetry classes CI and AIII were 
obtained in Ref.\ \onlinecite{Wang14}.

We present in Sec.\ \ref{sec: Definition and strategy} 
a method that allows to derive the
reduction pattern of all noninteracting topological
insulators and superconductors without and with 
reflection symmetries for any dimensionality $d$ of space 
in the presence of quartic contact interactions
that neither break explicitly nor spontaneously the 
defining symmetries. This method relies on
the topology of the classifying spaces from K-theory.
It extends the applicability of K-theory
for obtaining the tenfold way of noninteracting fermions,%
~\cite{kitaev09,Morimoto13}
to obtaining the breakdown of the tenfold way induced by interactions.

This method is applied first to the breakdown of the tenfold way
in Sec.\ \ref{sec: Reduction of the periodic table for strong TI and TS}.~%
\footnote{
We shall also call the topological TIs and TSs entering the periodic table
strong TIs and strong TSs.
         }
In doing so, we prove the following properties that we report
in Table
\ref{table: topo classification short-rnaged entangles AZ classes}.
\begin{enumerate}
\item All $\mathbb{Z}^{\,}_{2}$ entries of the periodic table
irrespectively of the dimensionality of space are stable to
quartic contact interactions.
\item All $\mathbb{Z}$ entries of the periodic table when the
dimensionality of space is even are stable to
quartic contact interactions.
\item Only the $\mathbb{Z}$ entries of the periodic table when the
dimensionality of space is odd are unstable to
quartic contact interactions with a reduction pattern that
is computed explicitly and shown to break the Bott periodicity
of two for the complex symmetry classes and of eight for
the real symmetry classes.
\end{enumerate}

This method is then applied to the three-dimensional
topological crystalline insulators (TCIs) from the 
symmetry class AII$+R$,
which are of relevance to SnTe, 
in Sec.\ \ref{eq: Reduction for TCS and TCI}.
We show the reduction $\mathbb{Z}\to\mathbb{Z}^{\,}_{8}$
in the presence of quartic local fermion-fermion interactions.

The strategy that we use to study the robustness
of $\nu$ boundary modes to quartic contact fermion-fermion interactions 
is inspired by the (unpublished) approach pioneered by Kitaev
in Refs.\ 
\onlinecite{kitaev-topomat11}
and
\onlinecite{kitaev-ipam15}, see also Ref.\ \onlinecite{You14}.
It consists of three steps. First, 
a noninteracting topological phase is represented by 
the many-body ground state of a massive Dirac Hamiltonian
with a matrix dimension that depends on $\nu$.
Second, a Hubbard-Stratonovich transformation is used to trade a
generic quartic contact interaction in favor of
dynamical Dirac mass-like bilinears coupled to their conjugate fields
(that will be called Dirac masses).
These dynamical Dirac masses may violate any
symmetry constraint other than the
particle-hole symmetry (PHS).~%
\footnote{
Any Hamiltonian made exclusively
of fermion bilinears can be written in the Nambu representation.
This representation is redundant and as such comes with a 
particle-hole symmetry (PHS).
         }
Third, the $\nu$ boundary modes that are coupled with 
a suitably chosen subset of dynamical masses are integrated over.
The resulting dynamical theory on the $(d-1)$-dimensional 
boundary is a bosonic one, a quantum nonlinear sigma model (QNLSM)
in $[(d-1)+1]$-dimensional space and time
with a target space that depends on $\nu$.
The reduction pattern is then obtained by identifying the smallest
value of $\nu$ for which this QNLSM cannot be augmented by a topological term.
The presence or absence of topological terms in the 
relevant QNLSM is determined by 
the topology of the spaces of boundary dynamical Dirac masses, 
i.e., the topology of classifying spaces. Now, K-theory
provides a systematic way to study the topology of the classifying
space.  Hence, this is why the same approach that was used to obtain 
the tenfold way of noninteracting fermions can be relied on
to deduce a classification of 
topological short-range entangled (SRE) phases 
[also known as symmetry-protected topological (SPT) phases]
for interacting fermions.~%
\footnote{
The question that we address in this paper is
whether or not the topological classification
of noninteracting fermions is reduced by interactions.
A complete classification of fermionic SPT phases
(combined with that for the bosonic SPT phases)
is beyond the scope of this paper.
         }

Other topological phases are also interesting on their own right.
For example, bosonic SPT (SRE) phases show many novel topological
phases driven by strong interactions.
They have been reviewed in Ref.\ \onlinecite{senthil-review14}. 
The classification of bosonic SPT (SRE) states has been obtained by 
diverse approaches that include
group cohomology,~\cite{Chen11,Chen12} 
the K-matrix approach,~\cite{Lu12b}
enumerating surface topological order,%
\cite{vishwanath-senthil13} 
wire constructions,~\cite{Wang13,Jian14}
and so on. Topological order with long-range entanglement (LRE) 
is also a subject of intensive studies, which
have relied on parton constructions,%
~\cite{Maciejko10,Witczak10,Swingle11,Maciejko12}
topological field theories
~\cite{Cho11,Walker12,Kapustin13} 
exactly soluble models,%
~\cite{Levin11,Keyserlingk13,Chen13,Mesaros13,Ye14,Hung14} 
and wire constructions.%
~\cite{kane02,Teo14,Mong14,Neupert14}

\section{Strategy}
\label{sec: Definition and strategy}

In this section, we present our strategy to obtain
a topological classification for interacting fermions
with gapped ground states as an application of K-theory 
to certain \textit{dynamical Dirac masses for boundary fermions}.

Noninteracting fermions always belong to one of 
the ten Altland-Zirnbauer (AZ) symmetry classes defined by
the presence or absence of the following three symmetries,
time-reversal symmetry (TRS), 
particle-hole symmetry (PHS), 
and chiral symmetry (CHS) (see Appendix 
\ref{appsec: Defining symmetries of strong topological insulators}).
Within any one of these ten symmetry classes,
the defining topological attributes of noninteracting 
topological insulators and superconductors 
are shared by equivalence classes of Hamiltonians. 
Any two members within a topological class can be
deformed into each other by a smooth (adiabatic) deformation 
of the matrix elements of these Hamiltonians without closing
the bulk energy gap.
These equivalence classes
are endowed with an Abelian group structure 
$\mathfrak{G}$. 
For any given dimensionality $d$ of space, 
topological invariants $\mathfrak{G}$ are non-trivial for five
out of the ten AZ classes.
Specifically, three of the ten AZ classes support Abelian groups
$
\mathfrak{G}=\mathbb{Z},
$
while two of the ten AZ classes support Abelian groups
$
\mathfrak{G}=\mathbb{Z}^{\,}_{2}
$. 
The TRS, PHS, and CHS can be augmented by crystalline symmetries.
Noninteracting fermions obeying crystalline symmetries can
also be understood as realizing topologically distinct equivalence classes,
i.e., topological crystalline insulators (TCIs).~%
\footnote{
When the crystalline symmetry operator squares to the unity,
the Abelian groups $\mathfrak{G}$ for noninteracting TCIs 
are given by 
$\mathbb{Z}$, $\mathbb{Z}^{\,}_{2}$ or some direct product of them.%
\cite{Chiu13,Morimoto13,Shiozaki14}
         }

The topological classification with the Abelian group $\mathfrak{G}$
for noninteracting TIs, TSs, or TCIs can break down
in the presence of many-body interactions.
Namely, an Abelian group
$\mathfrak{G}^{\,}_{\mathrm{int}}$
that encodes the topological equivalence classes of gapped ground states
for interacting fermions can be smaller than $\mathfrak{G}$ as a group
(some quotient group of $\mathfrak{G}$).

In order to establish the
instability of the noninteracting classification
of TIs, TSs, and TCIs,
we choose a family of massive Dirac Hamiltonians,
\begin{equation}
\mathcal{H}^{(0)}:=
-\mathrm{i}
\sum_{j=1}^{d}
\frac{\partial}{\partial x^{j}}\,
\widetilde \alpha^{\,}_{j}\,\otimes\openone\,
+m(\Vec x) \widetilde \beta \,\otimes\openone\,
\label{eq: def H bulk}
\end{equation}
as representative single-particle Hamiltonians.
Here, Dirac matrices $\widetilde{\bm{\alpha}}$ and $\widetilde \beta$
anticommute with each other and 
have the minimal dimension (rank) $r^{\,}_{\mathrm{min}}$
under the symmetry constraints, i.e., $r^{\,}_{\mathrm{min}}$
is the minimal rank to realize a Dirac Hamiltonian of the form 
(\ref{eq: def H bulk}).
The dimension of the unit matrix $\openone$ is $\nu=1,2,\cdots$.
The integer 
$\nu\in\mathfrak{G}$
is then related to the dimension $r(\nu)=r^{\,}_{\min}\,\nu$
of the Dirac matrices that we choose.
The question that we want to address is that of the 
stability or instability of 
the boundary states of a noninteracting TI, TS, or TCI
in the presence of many-body interactions that do not break the protecting
symmetries of the noninteracting limit.%
~\footnote{
We consider interactions that do not break the protecting
symmetries of the noninteracting limit, that are strong on the boundary, 
yet are not-too-strong as measured by the single-particle gap 
for the bulk states of insulators.
         }
Here,
whenever $\nu\neq0$, the extended single-particle boundary states 
are governed by the massless Dirac Hamiltonian
\begin{equation}
\mathcal{H}^{(0)}_{\mathrm{bd}}:=
-\mathrm{i}
\sum_{j=1}^{d-1}
\frac{\partial}{\partial x^{j}}\,
\alpha^{\,}_{j}\,\otimes\openone\,
\equiv
-\mathrm{i}
\bm{\partial}\,
\cdot
\bm{\alpha}\otimes\openone,
\label{eq: def mathcal{H}^{(0)}{mathrm{bd}}}
\end{equation}
which is obtained by introducing a domain wall
in the mass $m(\bm{x})$ along the
$x^{d}$-direction that enters Hamiltonian (\ref{eq: def H bulk}).  
The Dirac matrices $\bm{\alpha}\otimes\openone$
have a dimension $r(\nu)/2$ that is half that of the
bulk massive Dirac Hamiltonian $\mathcal{H}^{(0)}$.  
The dimension of the matrices $\bm{\alpha}$ is
$r^{\,}_{\mathrm{min}}/2$.

The breakdown (reduction) 
of the topological classification for noninteracting fermions
takes place when the boundary states of the TIs, TSs, or TCIs
can be gapped by many-body interactions that preserve 
their defining symmetries.
By assumption, we consider many-body
interactions that are weak relative to the bulk gap. If so, it is sufficient
to treat the effects of many-body interactions for the massless
Dirac fermions propagating on the $(d-1)$-dimensional boundary.
To establish an instability of the noninteracting topological
classification, we need not consider all possible many-body interactions.
It suffices to establish that at least one family of strong
(on the boundary) interactions
implies the instability of the noninteracting classification $\mathfrak{G}$
by gapping out all boundary Dirac fermions. To this end,
we limit ourselves to contact interactions.
 
Contact interactions are constructed from taking squares
of local bilinears in the Dirac fermions.
We have two options for these bilinears.
The bilinear under consideration either commutes or anticommutes
with the kinetic contribution to the 
Dirac Hamiltonian. We shall call the latter option a Dirac mass.
In this paper, we only consider
the contact interactions obtained from taking squares
of those bilinears built out of Dirac mass matrices, 
for only these can gap the noninteracting massless boundary Dirac 
fermions in a mean-field approximation. 
Because we assume that the protecting symmetries forbid
the presence of Dirac masses on the boundary that
are consistent with the
protecting symmetries, the only possible Dirac masses induced
by a mean-field treatment of a symmetry-preserving quartic interaction
on the boundary must be odd under at least one of the protecting symmetries.
We shall call such a boundary Dirac mass a boundary dynamical mass and 
label it with the Greek letter $\beta$.

We are thus led to consider the 
many-body interacting Dirac boundary Hamiltonian
\begin{subequations}
\label{eq: def interacting boundary massless Dirac fermions}
\begin{equation}
\widehat{H}^{\,}_{\mathrm{bd}}:=
\widehat{H}^{(0)}_{\mathrm{bd}}
+
\widehat{H}^{(\mathrm{int})}_{\mathrm{bd}},
\end{equation}
where (the subscript ``bd'' stands for boundary)
\begin{equation}
\widehat{H}^{(0)}_{\mathrm{bd}}:=
\int\mathrm{d}^{d-1}\bm{x}\,
\hat{\Psi}^{\dag}(t,\bm{x})\,
\mathcal{H}^{(0)}_{\mathrm{bd}}\,
\hat{\Psi}(t,\bm{x})
\end{equation}
and
\begin{equation}
\widehat{H}^{(\mathrm{int})}_{\mathrm{bd}}:=
\lambda
\sum_{\{\beta\}}
\int\mathrm{d}^{d-1}\bm{x}\,
\left[
\hat{\Psi}^{\dag}(t,\bm{x})\,
\beta\,
\hat{\Psi}(t,\bm{x})
\right]^{2}.
\end{equation}
\end{subequations}
We have chosen the real-valued coupling $\lambda$ with the dimension of
$\mathrm{(length)}^{d-2}$ to be independent of $\beta$ for simplicity.
This coupling constant is marginal in $d=2$ and irrelevant when
$d>2$. (Of course, it can very well be that the set $\{\beta\}$ is empty.
If so, we anticipate that 
$\mathfrak{G} = \mathfrak{G}^{\,}_{\mathrm{int}}$
must hold. This is what happens for the strong topological insulators
in the symmetry classes A, D, and C when $d=2$.)
At this stage, it is convenient to treat the many-body Hamiltonian
(\ref{eq: def interacting boundary massless Dirac fermions})
with the help of the path integral
\begin{subequations}
\begin{equation}
Z^{\,}_{\mathrm{bd}}:=
\int\mathcal{D}[\Psi,\Psi^{\dag}]\,
e^{-S^{\,}_{\mathrm{bd}}},
\end{equation}
where the action in Euclidean time $\tau$ is
\begin{equation}
S^{\,}_{\mathrm{bd}}:=
\int\mathrm{d}\tau
\int\mathrm{d}^{d-1}\bm{x}\,
\mathcal{L}^{\,}_{\mathrm{bd}},
\end{equation}
with the Lagrangian density
\begin{equation}
\mathcal{L}^{\,}_{\mathrm{bd}}:=
\Psi^{\dag}
\left(
\partial^{\,}_{\tau}+\mathcal{H}^{(0)}_{\mathrm{bd}}
\right)
\Psi
+
\lambda
\sum_{\{\beta\}}
\left(
\Psi^{\dag}
\beta\,
\Psi
\right)^{2}.
\end{equation}
\end{subequations}
The path integral is over Grassmann-valued Dirac spinors.

We rewrite the quartic interaction terms by performing 
a Hubbard-Stratonovich transformation
with respect to the bosonic fields 
$\phi^{\,}_{\beta}$
conjugate to 
$\Psi^{\dag}\beta\,\Psi$,
\begin{subequations}
\label{eq: partition fct after HS}
\begin{equation}
Z^{\,}_{\mathrm{bd}}\propto
\int\mathcal{D}[\Psi,\Psi^{\dag},\phi^{\,}_{\beta}]\,
e^{-S^{\prime}_{\mathrm{bd}}}.
\label{eq: partition fct after HS a}
\end{equation}
Here, the action in Euclidean time $\tau$ is
\begin{equation}
S^{\prime}_{\mathrm{bd}}:=
\int\mathrm{d}\tau
\int\mathrm{d}^{d-1}\bm{x}\,
\mathcal{L}^{\prime}_{\mathrm{bd}},
\label{eq: partition fct after HS b}
\end{equation}
with the Lagrangian density
\begin{equation}
\mathcal{L}^{\prime}_{\mathrm{bd}}:=
\Psi^{\dag}
\left(
\partial^{\,}_{\tau}
+
\mathcal{H}^{(\mathrm{dyn})}_{\mathrm{bd}}
\right)
\Psi
+
\frac{1}{\lambda}
\sum_{\{\beta\}}
\phi^{2}_{\beta},
\label{eq: partition fct after HS c}
\end{equation}
where we have introduced the dynamical one-body single-particle
Hamiltonian
\begin{equation}
\mathcal{H}^{(\mathrm{dyn})}_{\mathrm{bd}}(\tau,\bm{x}):=
\mathcal{H}^{(0)}_{\mathrm{bd}}(\bm{x})
+
\sum_{\{\beta\}}
2\mathrm{i}\,
\beta\,
\phi^{\,}_{\beta}(\tau,\bm{x}),
\label{eq: partition fct after HS d}
\end{equation}
\end{subequations}
under the assumption that the sign $\lambda>0$ corresponds to a
repulsive interaction. In a saddle-point approximation, the magnitude
of the vector $\bm{\phi}$ with the components $\phi^{\,}_{\beta}$ 
can be frozen both in imaginary time and in $(d-1)$-dimensional space. 
Fluctuations that change this frozen magnitude
are suppressed by the second term on the right-hand side of Eq.\
(\ref{eq: partition fct after HS c}). 
We will restrict the set
$\{\beta\}$ to pairwise anticommuting Dirac mass matrices. If so,
the direction in which the vector $\bm{\phi}$
with the components $\phi^{\,}_{\beta}$ freezes in the saddle-point
approximation is arbitrary.%
~\footnote{
The saddle-point equation for $\bm{\phi}$ is given as follows.
Integrating the fermionic degrees of freedom leads to
the effective Lagrangian,
\begin{equation*}
\begin{split}
\mathcal{S}^{\,}_{\mathrm{eff}}[\bm{\phi}]:=&\,
(-1)\mathrm{Tr}\,
\log
\left[
\partial^{\,}_{\tau}
+ 
\sum_{j=1}^{d-1} 
(-\mathrm{i}\partial^{\,}_{j})\,
\alpha^{\,}_{j} 
+ 
\sum_{ \{\beta\} } 
2\mathrm{i} 
\beta\,\phi^{\,}_{\beta} 
\right] 
\\
&\,
+ 
\frac{1}{\lambda\,r} 
\sum_{ \{\beta\} }
\mathrm{Tr}\,(\phi^{2}_{\beta}).
\end{split}
\end{equation*}
The symbol $\mathrm{Tr}$ 
represents tracing over the single-particle Hilbert space of the
Dirac Hamiltonian with the Dirac matrices $\bm{\alpha}$ and $\beta$ of 
dimension $r$.
The saddle point equations
$
\delta\mathcal{S}^{\,}_{\mathrm{eff}}[\bm{\phi}]/\delta\bm{\phi}
|^{\,}_{\bm{\phi}=\bar{\bm{\phi}}}=0
$
are
\begin{equation*}
\int\mathrm{d}\omega 
\int\mathrm{d}^{d-1}\bm{k}\, 
\left(
\frac{
2\bar\phi^{\,}_{\beta}
     }
     {
\omega^{2}
+
|\bm{k}|^{2}
-
4\,\bar{\bm{\phi}}^{2} 
     }
\right)
= 
\frac{1}{\lambda\,r}\, 
\bar\phi^{\,}_{\beta}.
\end{equation*}
We denote with $\Omega^{\,}_{d-1}$ the area of the unit sphere 
$S^{d-1}$,
with $k$ the length of the vector $(\omega,\bm{k})$,
and with $\Lambda$ the ultraviolet cutoff in $(\omega,\bm{k})$ space.
The saddle-point equations reduce to the equation
\begin{equation*}
\Omega^{\,}_{d-1}\, 
\int_{0}^{\Lambda}\mathrm{d}k k^{d}\, 
\frac{2}{k^{2}-4 \bar{\bm{\phi}}^{2}}=
\frac{1}{\lambda\,r}.
\end{equation*}
It has the solution
\begin{equation*}
|\bar\phi_{\beta}|=\mathrm{i}\phi^{\,}_{0}(\lambda\,r),
\qquad
\phi^{\,}_{0}(\lambda\,r)>0.
\end{equation*}
         }
Since fluctuations about this direction are soft, 
these are the Goldstone modes associated with the spontaneous
breaking of a continuous symmetry. 

The effective low-energy theory
governing the fluctuations of these Goldstone modes is obtained 
from a gradient expansion of the fermion determinant
\begin{equation}
\begin{split}
&
\mathrm{Det}\,
\left(
\partial^{\,}_{\tau}
+
\mathcal{H}^{(\mathrm{dyn})}_{\mathrm{bd}}
\right):=
\\
&\qquad
\int\mathcal{D}[\Psi,\Psi^{\dag}]\,
e^{
-
\int\mathrm{d}\tau
\int\mathrm{d}^{d-1}\bm{x}\,
\Psi^{\dag}
\left(
\partial^{\,}_{\tau}
+
\mathcal{H}^{(\mathrm{dyn})}_{\mathrm{bd}}
\right)
\Psi
  }.
\end{split}
\end{equation}  
It is captured by the partition function
\begin{equation}
Z^{\,}_{\mathrm{bd}}\approx
\int\mathcal{D}[\bm{\phi}]\,
\delta(\bm{\phi}^{2}-1)\,
e^{-S^{\,}_{\mathrm{QNLSM}}-S^{\,}_{\mathrm{top}}},
\end{equation}
after we have rescaled the vector $\bm{\phi}$
so that it squares to one.
The Euclidean action
\begin{equation}
S^{\,}_{\mathrm{QNLSM}}=
\frac{1}{2\,g}
\int\mathrm{d}\tau
\int\mathrm{d}^{d-1}\bm{x}\
(\partial^{\,}_{i}\bm{\phi})^{2}
\label{eq: def action QNLSM O(N)}
\end{equation}
is the action of the quantum nonlinear sigma model (QNLSM)
with the base space
$\mathbb{R}^{(d-1)+1}$ in space and time and the target space 
\begin{equation}
S^{N(\nu)-1}
\end{equation}
with the integer $N(\nu)$ counting the pairwise anticommuting 
Dirac masses that have been retained in the set $\{\beta\}$.
The effective coupling constant $g$ is positive.
The topological term $S^{\,}_{\mathrm{top}}$ is present whenever
any one of the homotopy groups
\begin{equation}
\begin{split}
&
\pi^{\,}_{0}\left(S^{N(\nu)-1}\right),
\\
&\qquad
\pi^{\,}_{1}\left(S^{N(\nu)-1}\right),
\\
&\qquad\qquad
\ddots
\\
&\qquad\qquad\qquad
\pi^{\,}_{d}\left(S^{N(\nu)-1}\right),
\\
&\qquad\qquad\qquad\qquad
\pi^{\,}_{d+1}\left(S^{N(\nu)-1}\right),
\end{split}
\label{eq: family of homotopy groups from 0 to d+1 if any nu}
\end{equation}
is non-vanishing.%
~\cite{Abanov00}
(The reason why we ignore all topological terms associated with
non-vanishing homotopy group of order larger than $d+1$ 
is that such topological terms would modify the 
local equations of motion derived from $S^{\,}_{\mathrm{QNLSM}}$
in a non-local way.)
It signals the existence of zero modes
of the Dirac Hamiltonian (\ref{eq: partition fct after HS d})
in the presence of topological defects in the order parameter
$\bm{\phi}$.
We expect that these zero modes prevent the 
gapping of the boundary Dirac fermions. 
We define the smallest value
$\nu^{\,}_{\mathrm{min}}$ for the dimension
$\nu$ of the unit matrix $\openone$
in Eq.\ (\ref{eq: def mathcal{H}^{(0)}{mathrm{bd}}}) 
for which 
\begin{equation}
\begin{split}
&
\pi^{\,}_{0}\left(S^{N(\nu^{\,}_{\mathrm{min}})-1}\right)=0,
\\
&\qquad
\pi^{\,}_{1}\left(S^{N(\nu^{\,}_{\mathrm{min}})-1}\right)=0,
\\
&\qquad\qquad
\ddots
\\
&\qquad\qquad\qquad
\pi^{\,}_{d}\left(S^{N(\nu^{\,}_{\mathrm{min}})-1}\right)=0,
\\
&\qquad\qquad\qquad\qquad
\pi^{\,}_{d+1}\left(S^{N(\nu^{\,}_{\mathrm{min}})-1}\right)=0.
\end{split}
\label{eq: family of homotopy groups from 0 to d+1 if any nu min}
\end{equation}
As all homotopy groups of the spheres are known, one may verify that
\begin{equation}
d+1<N(\nu^{\,}_{\mathrm{min}})-1.
\end{equation}

When Eq.\ (\ref{eq: family of homotopy groups from 0 to d+1 if any nu min}) 
holds, the topological term $S^{\,}_{\mathrm{top}}$ is absent, and
the effective action in the
partition function is simply the action 
(\ref{eq: def action QNLSM O(N)})
for a QNLSM on a sphere.%
~\footnote{
When $d=2$ and $N(\nu^{\,}_{\mathrm{min}})>2$, 
the Mermin-Wagner theorem 
applied to the QNLSM describing the one-dimensional boundary prevents
the spontaneous symmetry breaking on the target space 
$S^{N(\nu^{\,}_{\mathrm{min}})-1}$.
The coupling constant $g$ always flows to strong coupling,
the quantum-disordered phase at $g\to\infty$. 
When $d>2$, the fixed point at $g=0$ 
of the  QNLSM describing the $(d-1)$-dimensional boundary
is stable.
At this fixed point, one linear combination of the
bilinears
$\Psi^\dagger \beta\, \Psi$
acquires an expectation value. It thereby
breaks spontaneously one of the protecting symmetries.
In this case, interactions remove the noninteracting topological attributes
by spontaneously breaking one of the protecting symmetries.
The transition between the fixed point at $g=0$ and 
$g=\infty$ occurs
at $g=g^{\,}_{\star}\sim1$. 
Microscopics determine if the bare value of $g$ is smaller
or larger than the unstable quantum-critical point at $g^{\,}_{\star}$.
        }
In this case, the quantum-disordered phase at
the strong-coupling fixed point $g\to\infty$ is stable.
In this strongly interacting phase
and when $\mathfrak{G}=\mathbb{Z}$,
quantum fluctuations restore dynamically and non-perturbatively all 
the symmetries broken by the saddle-point,
including any protecting symmetries. If so,
all boundary Dirac fermions are gapped out.
We then conclude that 
\begin{equation}
\mathfrak{G}^{\,}_{\mathrm{int}}=
\mathbb{Z}^{\,}_{\nu^{\,}_{\mathrm{min}}}.
\end{equation}
The stability 
\begin{equation}
\mathfrak{G}^{\,}_{\mathrm{int}}=
\mathfrak{G}
\end{equation}
when 
$\mathfrak{G}=\mathbb{Z}^{\,}_{2}$
follows from the fact that one of the homotopy groups 
$\pi^{\,}_{D}(S^{N(\nu=1)-1})$ with $D \le d+1$ is always non-trivial
when $\mathfrak{G}=\mathbb{Z}^{\,}_{2}$ 
(see Sec.~\ref{sec: higher dims}).

As an illustration of this method, we give in Table
\ref{table: topo classification short-rnaged entangles AZ classes}
the equivalence classes of topological insulators and superconductors
belonging to the ten AZ symmetry classes in the presence
of interactions that select a short-ranged entangled many-body ground state.
It becomes apparent that the Bott periodicity of the tenfold way,
i.e., the periodicity of the (zeroth) homotopy groups of the 
classifying spaces with respect to $d$,
is lost. 
It also becomes apparent that the reduction of the
topologically distinct equivalence classes of noninteracting fermions
for any given AZ symmetry class occurs only in odd dimensions of space.
Finally, two of the
AZ symmetry classes, namely the chiral symmetry classes BDI and CII,
have the particularity that they may be interpreted either as a 
superconductor or an insulator. Correspondingly, the reduction
of their classification
$\mathbb{Z}\to \mathbb{Z}^{\,}_{m}$ for the superconductor interpretation
and
$\mathbb{Z}\to \mathbb{Z}^{\,}_{n}$ for the insulator interpretation
of these symmetry classes obeys
\begin{subequations}
\begin{align}
m&=2n, &(\t{class BDI})
\\ 
m&=n, &(\t{class CII})
\end{align}
\end{subequations}
when the dimensionality of space is $d=1$ mod 4.

\begin{table*}[tb]
\begin{center}
\caption{
The ten Altland-Zirnbauer (AZ) symmetry classes and their
topological classification when 
(i) fermion-fermion interactions neither break 
explicitly their defining symmetries nor spontaneously, 
(ii)
and the many-body ground state is short-ranged entangled.
Two complex and eight real symmetry classes are characterized
by the presence or the absence of 
time-reversal symmetry ($T$), 
particle-hole symmetry ($C$), 
and chiral symmetry ($\Gamma^{\,}_{5}$).
Their presence is complemented
by the sign multiplying the identity 
in $T^{2}=\pm1$ or $C^{2}=\pm1$, 
and by $1$ for $\Gamma^{\,}_{5}$.
Their absence is indicated by 0.
For each symmetry class and for any dimension 
$d=0,1,2,\ldots$ of space,
the classifying space $V^{\,}_{d}$, 
the space of normalized Dirac masses allowed by symmetry,
is given in the fifth column.
Explicit forms of the classifying spaces 
$C^{\,}_{q}$ and $R^{\,}_{q}$ and their stable homotopy groups are found 
in Table~\ref{table: all homotopies classifying spaces}
from Appendix~\ref{app: tenfold way}.
The reduction, if any, that arises from the effects of interactions
on the topological classification of noninteracting fermions
for $d=1,\ldots,8$ is given in the last eight columns.
Each entry with a non-trivial Abelian group
defines equivalence classes of interacting 
topological insulators (superconductors) 
with a short-ranged entangled many-body ground state.
We color in blue the entry corresponding to a given symmetry class
and a given column of odd dimensionality $d$ to indicate that
this entry is a quotient group $\mathfrak{G}^{\,}_{\t{int}}$
of $\mathfrak{G}^{\,}=\mathbb{Z}$.
The reduction $\mathbb{Z}\to\mathfrak{G}^{\,}_{\t{int}}$
results from an instability of the noninteracting topological classification
to fermion-fermion interactions.
The four entries corresponding to the symmetry classes BDI and CII 
and the dimensions $d=1$ and $d=5$ occur
in pairs depending on whether these two classes are interpreted
as describing superconductors (i.e., interacting Majorana fermions)
or insulators (i.e., interacting complex fermions), respectively.
\label{table: topo classification short-rnaged entangles AZ classes}
        }
\begin{tabular}[t]{ c c c c c c c c c c c c c c}
\hline \hline
Class 
& 
~$T$~
& 
~$C$~
&
~$\Gamma^{\,}_{5}$~
& 
~$V^{\,}_{d}$~
& 
~~$d=1$~~
&
~~$d=2$~~
&
~~$d=3$~~
&
~~$d=4$~~
&
~~$d=5$~~
&
~~$d=6$~~
&
~~$d=7$~~
&
~~$d=8$~~
\\
\hline
A   
& 
0 
& 
0    
&  
0     
& 
$C^{\,}_{0+d}$ 
& 
0
&
$\mathbb{Z}$   
&
0
& 
$\mathbb{Z}$ 
& 
0
&
$\mathbb{Z}$   
&
0
& 
$\mathbb{Z}$ 
\\
AIII  
& 
0 
& 
0    
&  
1     
& 
$C^{\,}_{1+d}$ 
&
$\blue{\mathbb{Z}^{\,}_{4}}$   
&
0
&
$\blue{\mathbb{Z}^{\,}_{8}}$   
& 
0 
&
$\blue{\mathbb{Z}^{\,}_{16}}$   
&
0
&
$\blue{\mathbb{Z}^{\,}_{32}}$   
& 
0 
\\
\hline
AI   
& 
$+1$ 
& 
0    
&  
0     
& 
$R^{\,}_{0-d}$ 
&
0
&
0
&
0
&
$\mathbb{Z}$   
&
0
&
$\mathbb{Z}^{\,}_{2}$
&
$\mathbb{Z}^{\,}_{2}$
&  
$\mathbb{Z}$  
\\
BDI  
& 
$+1$ 
& 
$+1$ 
&  
1     
& 
$R^{\,}_{1-d}$ 
&
$\blue{\mathbb{Z}^{\,}_{8}},\blue{\mathbb{Z}^{\,}_{4}}$
&
0
&
0
&
0
&
$\blue{\mathbb{Z}^{\,}_{16}},\blue{\mathbb{Z}^{\,}_{8}}$
&
0
&
$\mathbb{Z}^{\,}_{2}$
& 
$\mathbb{Z}^{\,}_{2}$ 
\\
D    
& 
0    
& 
$+1$ 
&  
0     
& 
$R^{\,}_{2-d}$ 
&
$\mathbb{Z}^{\,}_{2}$ 
&
$\mathbb{Z}$
&
0
&
0
&
0
&
$\mathbb{Z}$
&
0
& 
$\mathbb{Z}^{\,}_{2}$ 
\\
DIII 
& 
$-1$ 
& 
$+1$ 
&  
1     
& 
$R^{\,}_{3-d}$ 
&
$\mathbb{Z}^{\,}_{2}$ 
&
$\mathbb{Z}^{\,}_{2}$ 
&
$\blue{\mathbb{Z}^{\,}_{16}}$
&
0
&
0
&
0
&
$\blue{\mathbb{Z}^{\,}_{32}}$
& 
0              
\\
AII  
& 
$-1$ 
& 
0    
&  
0     
& 
$R^{\,}_{4-d}$ 
&
0
&
$\mathbb{Z}^{\,}_{2}$ 
&
$\mathbb{Z}^{\,}_{2}$ 
&
$\mathbb{Z}$ 
&
0
&
0
&
0
& 
$\mathbb{Z}$   
\\
CII  
& 
$-1$ 
& 
$-1$ 
&  
1     
& 
$R^{\,}_{5-d}$ 
&
$\blue{\mathbb{Z}^{\,}_{2}},\blue{\mathbb{Z}^{\,}_{2}}$
&
0
&
$\mathbb{Z}^{\,}_{2}$
&
$\mathbb{Z}^{\,}_{2}$ 
&
$\blue{\mathbb{Z}^{\,}_{16}},\blue{\mathbb{Z}^{\,}_{16}}$
&
0
&
0
& 
0              
\\
C    
& 
0    
& 
$-1$ 
&  
0     
& 
$R^{\,}_{6-d}$ 
&
0
&
$\mathbb{Z}$
&
0
&
$\mathbb{Z}^{\,}_{2}$
&
$\mathbb{Z}^{\,}_{2}$
&
$\mathbb{Z}$
&
0 
& 
0              
\\
CI   
& 
$+1$
& 
$-1$ 
&  
1     
& 
$R^{\,}_{7-d}$ 
&
0
&
0
&
$\blue{\mathbb{Z}^{\,}_{4}}$
&
0
&
$\mathbb{Z}^{\,}_{2}$ 
&
$\mathbb{Z}^{\,}_{2}$ 
&
$\blue{\mathbb{Z}^{\,}_{32}}$
& 
0              
\\
\hline \hline
\end{tabular}
\end{center}
\end{table*}

\section{Reduction of the periodic table for strong TI and TS}
\label{sec: Reduction of the periodic table for strong TI and TS}

In this section,
we apply the strategy explained above to study the breakdown
of the tenfold way in the presence of quartic contact interactions
in the ascending order of the spatial dimension $d$,
i.e., $d=1,2,3$, and higher dimensions.

We will use the following conventions.
The operation of complex conjugation will be denoted by $\mathsf{K}$.
Linear maps of two-dimensional vector space $\mathbb{C}^{2}$
shall be represented by $2\times2$ matrices that we expand
in terms of the unit matrix $\tau^{\,}_{0}$ and the three Pauli matrices
$\tau^{\,}_{1}$,
$\tau^{\,}_{2}$,
and 
$\tau^{\,}_{3}$.
Linear maps of the four-dimensional vector space 
$\mathbb{C}^{4}=\mathbb{C}^{2}\otimes\mathbb{C}^{2}$
will be represented by $4\times4$ matrices that we expand
in terms of the 16 Hermitian matrices
\begin{equation}
X^{\,}_{\mu\mu'}\equiv
\tau^{\,}_{\mu}\otimes\sigma^{\,}_{\mu'},
\qquad
\mu,\mu'=0,1,2,3,
\end{equation}
where $\sigma^{\,}_{\nu}$ is a second set comprised of the unit matrix
and the three Pauli matrices. Linear maps of the $2^{n}$-dimensional vector space 
$\mathbb{C}^{2^{n}}=\mathbb{C}^{2}\otimes\cdots\otimes\mathbb{C}^{2}$
will be represented by $2^{n}\times2^{n}$ matrices that we expand
in terms of the $4^{n}$ Hermitian matrices
\begin{equation}
X^{\,}_{\mu^{\,}_{1}\cdots\mu^{\,}_{n}}\equiv
\tau^{\,}_{\mu^{\,}_{1}}
\otimes
\tau^{\,}_{\mu^{\,}_{2}}
\otimes
\cdots
\otimes
\tau^{\,}_{\mu^{\,}_{n}}
\end{equation}
where
$\mu^{\,}_{1},\cdots,\mu^{\,}_{n}=0,1,2,3$.

\subsection{The case of one-dimensional space}

Fidkowski and Kitaev showed in Ref.\
\onlinecite{Fidkowski10}
that, in one spatial dimension,
any pair of Hamiltonian in the symmetry class BDI whose noninteracting
topological indices differ by eight 
can be transformed into each other adiabatically
(i.e., without closing the spectral gap)
in the presence of a quartic contact interaction
that preserves TRS.
This work was followed up in Refs.\
\onlinecite{Fidkowski11,Turner11}
with the construction of a topological invariant
for interacting fermions from 
the matrix product representation of ground states.
This topological invariant establishes that the
reduction $\mathbb{Z}\to\mathbb{Z}^{\,}_{8}$
is exhaustive. The same approach with matrix product states
was used to obtain an exhaustive classification of one-dimensional
 gapped spin systems
in Ref.\ \onlinecite{Chen11}.

Here, we focus on the three chiral symmetry classes
that support the $\mathbb{Z}$ topological classification 
in the noninteracting limit. We shall reproduce the reduction
$\mathbb{Z}\to\mathbb{Z}^{\,}_{8}$
and
$\mathbb{Z}\to\mathbb{Z}^{\,}_{2}$
when the symmetry classes BDI and CII are interpreted as chains
of Majorana fermions, respectively.

The one-dimensional chiral symmetry classes can be also realized
as chains of complex fermions with sublattice symmetry and 
fermion-number conservation, e.g., polyacetylene.
For example, polyacetylene-like chains realize 
the symmetry class AIII when TRS is broken, 
the symmetry class BDI when both TRS and the $SU(2)$ spin-rotation symmetry
are present, 
and the symmetry class CII when TRS holds 
but not the $SU(2)$ spin-rotation symmetry
if spin-orbit coupling is sizable.
We show that the reduction of the noninteracting topological classification is 
$\mathbb{Z}\to\mathbb{Z}^{\,}_{4}$
for the symmetry classes AIII and BDI, while it is
$\mathbb{Z}\to\mathbb{Z}^{\,}_{2}$
for the symmetry class CII, 
provided conservation of the fermion number holds.

\subsubsection{The symmetry class BDI when $d=1$}
\label{subsec: The BDI symmetry class when d=1}

Consider the one-dimensional bulk single-particle Dirac Hamiltonian
(with Dirac matrices of dimension $r=2\equiv r^{\,}_{\mathrm{min}}$),
\begin{subequations}
\label{eq: def 1D mathcal H for BDI}
\begin{align}
\mathcal{H}^{(0)}(x):=
-\mathrm{i}
\partial^{\,}_{x}
\,
\tau^{\,}_{3}
+ 
m(x)\,
\tau^{\,}_{2}.
\label{eq: def 1D mathcal H for BDI a}
\end{align}
This single-particle Hamiltonian belongs to the symmetry class BDI, for
\begin{align}
&
\mathcal{T}\,\mathcal{H}^{(0)}(x)\,\mathcal{T}^{-1}=
+
\mathcal{H}^{(0)}(x),
\label{eq: def 1D mathcal H for BDI b}
\\
&
\mathcal{C}\,\mathcal{H}^{(0)}(x)\,\mathcal{C}^{-1}=
-
\mathcal{H}^{(0)}(x),
\label{eq: def 1D mathcal H for BDI c}
\end{align}
where
\begin{align}
\mathcal{T}&:=
\tau^{\,}_{1}\, \mathsf{K}, 
& 
\mathcal{C}&:=
\tau^{\,}_{0}\,
\mathsf{K}.
\label{eq: def 1D mathcal H for BDI d}
\end{align}
\end{subequations}

The Dirac mass matrix $\tau^{\,}_{2}$
is here the only one allowed for dimension two Dirac matrices
under the constraints (\ref{eq: def 1D mathcal H for BDI b})
and (\ref{eq: def 1D mathcal H for BDI c}).
As was shown by Jackiw and Rebbi,
if translation symmetry is broken by the mass term supporting the domain
wall 
\begin{subequations}
\label{eq: zero mode if nu=1 for BDI}
\begin{equation}
m(x)=
m^{\,}_{\infty}\,
\mathrm{sgn}(x),
\qquad
m^{\,}_{\infty}\in\mathbb{R},
\label{eq: def Jackiw Rebbi domain wall}
\end{equation}
at $x=0$, then the zero mode 
\begin{equation}
e^{
-\mathrm{i}
\tau^{\,}_{3}\,\tau^{\,}_{2}\,
\int\limits_{0}^{x}\mathrm{d}x'\,m(x')
  }\,
\chi\,
=
e^{
-|m^{\,}_{\infty}\,x|
  }\,
\chi,
\end{equation}
where
\begin{equation}
\tau^{\,}_{1}\,
\chi=
\mathrm{sgn}\,(m^{\,}_{\infty})\,
\chi,
\end{equation}
is the only normalizable state bound to this domain wall.
This boundary state is a zero mode. It is an eigenstate of the
single-particle boundary Hamiltonian
\begin{equation} 
\mathcal{H}^{(0)}_{\mathrm{bd}}=0.
\end{equation}
\end{subequations}

Suppose that we consider $\nu=1,2,\cdots$ identical copies of the 
single-particle Hamiltonian 
(\ref{eq: def 1D mathcal H for BDI})
by defining
\begin{subequations}
\begin{equation}
\mathcal{H}^{(0)}_{\nu}(x):=
\mathcal{H}^{(0)}(x)\otimes\openone,
\end{equation}
and
\begin{align}
\mathcal{T}&:=
\tau^{\,}_{1}\otimes\openone\,\mathsf{K}, 
& 
\mathcal{C}&:=
\tau^{\,}_{0}\otimes\openone\,\mathsf{K},
\end{align}
\end{subequations}
where $\openone$ is a $\nu\times\nu$ unit matrix.
Observe that $\mathcal{T}$ and $\mathcal{C}$ commute with 
$\tau^{\,}_{1}\otimes\openone$ and with each other. 
The domain wall (\ref{eq: def Jackiw Rebbi domain wall}) 
must then support $\nu$ linearly independent
boundary zero modes. They are annihilated by the boundary Hamiltonian
\begin{equation} 
\mathcal{H}^{(0)}_{\mathrm{bd}\,\nu}=
\mathcal{H}^{(0)}_{\mathrm{bd}}\otimes\openone=0.
\end{equation}
The topological sectors for noninteracting Hamiltonians
are thus labeled by the integer $\nu$ taking values in $\mathbb{Z}$
in the limit $\nu\to\infty$.

A generic local quartic interaction that respects the defining BDI
symmetries with the potential to gap out these boundary zero modes
reduces to a dynamical Dirac mass 
(that depends on imaginary time $\tau$ in addition to 
space $x$)  that belongs to the symmetry class D, upon performing 
a Hubbard-Stratonovich transformation. Hence, we must consider the dynamical 
bulk single-particle Hamiltonian
\begin{subequations}
\label{eq: dyanmical bulk 1d BDI}
\begin{equation}
\mathcal{H}^{(\mathrm{dyn})}_{\nu}(\tau,x):=
\left[
-\mathrm{i}
\partial^{\,}_{x}
\,
\tau^{\,}_{3}
+ 
m(x)\,
\tau^{\,}_{2}
\right]
\otimes\openone
+
\mathcal{V}(\tau,x).
\label{eq: dyanmical bulk 1d BDI a}
\end{equation}
The dynamical Dirac mass $\mathcal{V}(\tau,x)$ 
is here defined by the condition that
it anticommutes with 
$\mathcal{H}^{(0)}(x)\otimes\openone$, 
when independent of $x$,
and obeys the transformation laws
dictated by the symmetry class D, i.e., it is of the form
\begin{equation}
\mathcal{V}(\tau,x):=
\tau^{\,}_{1}\otimes
\gamma'(\tau,x),
\qquad
\gamma'(\tau,x):=
\mathrm{i}M(\tau,x),
\label{eq: dyanmical bulk 1d BDI b}
\end{equation}
where 
\begin{equation}
M(\tau,x)=
M^{*}(\tau,x),
\qquad
M(\tau,x)=
-
M^{\mathsf{T}}(\tau,x),
\end{equation}
is a real-valued antisymmetric $\nu\times\nu$ matrix.
Consequently, TRS is only retained for a given $\mathcal{V}(\tau,x)$
if
\begin{equation}
M(\tau,x)=
-
M(-\tau,x).
\end{equation}
\end{subequations}

On the boundary, the operations for reversal of time and charge conjugation
are now represented by
\begin{subequations}
\begin{equation}
\mathcal{T}^{\,}_{\mathrm{bd}}:=\mathsf{K},
\qquad
\mathcal{C}^{\,}_{\mathrm{bd}}:=\mathsf{K}.
\end{equation}
Hence, we must consider the dynamical single-particle boundary Hamiltonian
\begin{equation}
\mathcal{H}^{(\mathrm{dyn})}_{\mathrm{bd}\,\nu}(\tau)\equiv
\gamma'(\tau):=
\mathrm{i}M(\tau),
\end{equation}
\end{subequations}
where $M(\tau)$ is a real-valued antisymmetric $\nu\times\nu$ matrix.
The space of boundary normalized Dirac mass matrices 
obtained by demanding that $\gamma'$ square to the unit $\nu\times\nu$ matrix
is topologically equivalent to the space 
\begin{align}
V^{\,}_{\nu}=O(\nu)/U(\nu/2)
\label{eq: case d=1 BDI a supercond: B nu= O nu / U nu/2}
\end{align}
for the symmetry class D in zero-dimensional space,
provided the rank $\nu\ge 2$ and $\nu$ is even. 
The limit $\nu\to\infty$ of these spaces is the classifying space
$R^{\,}_{2}$.  
In order to gap out dynamically the boundary zero modes 
without breaking the defining symmetries of the symmetry class BDI,
we need to construct a (0+1)-dimensional
QNLSM for the (boundary) dynamical Dirac masses
from the zero-dimensional symmetry class D without topological 
obstructions. We construct explicitly the spaces 
for the relevant normalized boundary dynamical Dirac masses 
of dimension $\nu=2^{n}$ with $n=0,1,2,3$ in the following.%
~\footnote{
In order to study the topological obstructions in the target space of
the QNLSMs, it is sufficient to consider the dimensions $\nu=2^{n}$
with $n=0,1,2,3$ of the dynamical Dirac mass matrices. Indeed,
the target space of the QNLSM is a sphere generated by a maximum number of 
anticommuting dynamical Dirac masses.
The increase in the number of anticommuting dynamical Dirac masses $N(\nu)$
takes place if and only if the dimensions of the 
Dirac matrices are doubled. In other words, as
$N(\nu)$ remains the same for $\nu=2^{n},\ldots,2^{n+1}-1$,
the same topological obstruction for the QNLSM prevents gapping
out of the excitations at the boundary for $\nu=2^{n},\ldots,2^{n+1}-1$.
This is why, to study the breakdown of the noninteracting classification,
we only focus on the cases with $\nu=2^{n}$ in the following.
          }
The relevant homotopy groups are given in Table%
~\ref{tab: 1D BDI}.
\footnote{
The homotopy groups for the space of $\nu\times\nu$ 
normalized Dirac mass matrices
$V^{\,}_{\nu}$ for finite $\nu$ can be different from those for the space 
$R^{\,}_{2}$ (i.e., the limit $\nu\to\infinity$).
In fact, the latter obey the Bott periodicity, 
while the former do not.
However, we find by an explicit enumeration of the Dirac mass matrices 
in the following
that the non-trivial entries of the relevant homotopy groups 
$\pi^{\,}_{D}(V^{\,}_{\nu})$ 
appear when $\pi^{\,}_{D}(R^{\,}_{2})$ is non-trivial.
It turns out that this correspondence
between homotopy groups at finite 
$\nu$ and infinite $\nu$ always holds for any example that we worked out later.
While we do not rely on this fact for the analysis in one,
two, and three dimensions,
the analysis in higher dimensions made in 
Sec.~\ref{sec: higher dims}
assumes this correspondence.
          }

\textit{Case $\nu=1$:}
No Dirac mass is allowed on the boundary, because the boundary
is the end of a one-dimensional $\mathbb{Z}^{\,}_{2}$ 
topological superconductor in the topologically non-trivial phase
of the symmetry class D.

\textit{Case $\nu=2$:}
We use the representation 
$\openone=\sigma^{\,}_{0}$.
There is one dynamical normalized Dirac mass on the boundary 
that is proportional to
the matrix $\sigma^{\,}_{2}$. A domain wall in imaginary time such as
$m^{\,}_{2\,\infty}\,\mathrm{sign}(\tau)\,\sigma^{\,}_{2}$ 
prevents the dynamical generation of a spectral gap on the boundary.

\textit{Case $\nu=4$:}
We use the representation 
$\openone=\sigma^{\,}_{0}\otimes\rho^{\,}_{0}$.
A (maximum) set of pairwise anticommuting boundary dynamical 
Dirac mass matrices follows from the set
\begin{align}
\{
\sigma^{\,}_{2}\otimes\rho^{\,}_{0},
\sigma^{\,}_{1}\otimes\rho^{\,}_{2},
\sigma^{\,}_{3}\otimes\rho^{\,}_{2}
\}.
\end{align}
This set spans the space of normalized boundary dynamical Dirac masses
that is homeomorphic to $S^{2}$.
Even though $\pi^{\,}_{0+1}(S^{2})=0$, it is possible to add a topological term 
that is nonlocal, yet only modifies the equations of motion 
of the (0+1)-dimensional QNLSM on the boundary
by local terms as a consequence of the fact that
$\pi^{\,}_{0+1+1}(S^{2})=\mathbb{Z}$. Such a term is a (0+1)-dimensional 
example of a Wess-Zumino (WZ) term. In the presence of this WZ term,
the boundary theory remains gapless. It is nothing but a bosonic representation
of the gapless $S=1/2$ degrees of freedom at the end of a quantum spin-1
antiferromagnetic spin chain in the Haldane phase.%
~\cite{You14}

\textit{Case $\nu=8$:}
We use the representation 
$\openone=\sigma^{\,}_{0}\otimes\rho^{\,}_{0}\otimes\lambda^{\,}_{0}$.
One set of pairwise anticommuting boundary dynamical Dirac mass matrices
follows from the set
\begin{align}
\{
\sigma^{\,}_{2}\otimes\rho^{\,}_{0}\otimes\lambda_{0},
\sigma^{\,}_{3}\otimes\rho^{\,}_{2}\otimes\lambda_{0},
\sigma^{\,}_{3}\otimes\rho^{\,}_{3}\otimes\lambda_{2},
\sigma^{\,}_{1}\otimes\rho^{\,}_{0}\otimes\lambda_{2}
\}
\end{align}
This set spans a manifold homeomorphic to $S^3$
(we may find a set of pairwise anticommuting masses 
spanning $S^6$).
No topological term is admissible over this 
target manifold 
that delivers local equations of motion.
The QNLSM over this target space endows dynamically
the boundary Hamiltonian with a spectral gap.

We conclude that the effects of interactions on the one-dimensional
SPT phases in the symmetry class BDI are
to reduce the topological classification $\mathbb{Z}$
in the noninteracting limit down to $\mathbb{Z}^{\,}_{8}$ under the
assumption that a Hamiltonian from the
symmetry class BDI is interpreted as a mean-field description of
a superconductor. The logic used to reach this conclusion
is summarized by Table \ref{tab: 1D BDI} 
once the line corresponding to
$\nu=2$ has been identified.
It is given by the smallest $D$ that
accommodates a non-trivial entry for the corresponding homotopy group.
The line for $\nu=4$ is then identified with the next smallest
$D$ with $\pi^{\,}_{D}(R^{\,}_{2})\ne0$, and so on.

\begin{table}[tb]
\caption{
Reduction from $\mathbb{Z}$ 
to
$\mathbb{Z}^{\,}_{8}$ 
for the topologically equivalent classes of the
one-dimensional SPT phases in the symmetry class BDI
that arises from interactions. 
We denote by $V^{\,}_{\nu}$ the space of
$\nu\times\nu$ normalized Dirac mass matrices
in zero-dimensional Hamiltonians belonging to the symmetry class D. 
The limit $\nu\to\infty$ of these spaces is the classifying space 
$R^{\,}_{2}$.
The second column shows the stable $D$-th homotopy groups 
of the classifying space $R^{\,}_{2}$. 
The third column gives the number $\nu$ of copies of 
boundary (Dirac) fermions for which a topological obstruction is permissible.
The fourth column gives the type of topological obstruction
that prevents the gapping of the boundary (Dirac) fermions.
\label{tab: 1D BDI}
        }
\begin{tabular}{ccccccc}
\hline \hline
$D$ 
&\qquad\qquad& 
$
\pi^{\,}_{D}(R^{\,}_{2})$
&\qquad\qquad& 
$\nu$ 
&\qquad\qquad&
Topological obstruction 
\\
\hline
0&&$\mathbb{Z}^{\,}_{2}$ && $2$  && Domain wall \\
1&&0                    &&     &&              \\
2&&$\mathbb{Z}$         && $4$ && WZ term      \\
3&&0                    &&     &&              \\
4&&0                    &&     &&              \\
5&&0                    &&     &&              \\
6&&$\mathbb{Z}$         && $8$ &&         None \\
7&&$\mathbb{Z}^{\,}_{2}$  &&    &&               \\
\hline \hline
\end{tabular}
\end{table}

\subsubsection{The symmetry class CII when $d=1$}
\label{subsec: The CII symmetry class when d=1}

Consider the one-dimensional bulk single-particle Dirac Hamiltonian
(with Dirac matrices of dimension $r=4\equiv r^{\,}_{\mathrm{min}}$),
\begin{subequations}
\label{eq: def 1D mathcal H for CII}
\begin{align}
\mathcal{H}^{(0)}(x):=
-\mathrm{i}
\partial^{\,}_{x}\,
X^{\,}_{30}
+ 
m(x)\,
X^{\,}_{20}.
\label{eq: def 1D mathcal H for CII a}
\end{align}
This single-particle Hamiltonian belongs to the symmetry class CII, for
\begin{align}
&
\mathcal{T}\,\mathcal{H}^{(0)}(x)\,\mathcal{T}^{-1}=
+
\mathcal{H}^{(0)}(x),
\label{eq: def 1D mathcal H for CII b}
\\
&
\mathcal{C}\,\mathcal{H}^{(0)}(x)\,\mathcal{C}^{-1}=
-
\mathcal{H}^{(0)}(x),
\label{eq: def 1D mathcal H for CII c}
\end{align}
where
\begin{align}
\mathcal{T}&:=
\mathrm{i}
X^{\,}_{12}\,\mathsf{K}, 
& 
\mathcal{C}&:=
\mathrm{i}
X^{\,}_{02}\,
\mathsf{K}.
\label{eq: def 1D mathcal H for CII d}
\end{align}
\end{subequations}

The Dirac mass matrix $X^{\,}_{20}$
is here the only one allowed for dimension four Dirac matrices
in the symmetry class CII. 
If translation symmetry is broken by the Dirac mass term supporting the domain
wall 
\begin{subequations}
\begin{equation}
m(x)=
m^{\,}_{\infty}\,
\mathrm{sgn}(x),
\qquad
m^{\,}_{\infty}\in\mathbb{R},
\label{eq: def Jackiw Rebbi domain wall CII}
\end{equation}
at $x=0$, then the zero mode 
\begin{equation}
e^{
- 
\mathrm{i}X^{\,}_{30}\,X^{\,}_{20}\,
\int\limits_{0}^{x}\mathrm{d}x'\,m(x')
  }\,
\chi
=
e^{
-|m^{\,}_{\infty}\,x|
  }\,
\chi,
\end{equation}
where
\begin{equation}
X^{\,}_{10}\,
\chi=
\mathrm{sgn}\,(m^{\,}_{\infty})\,
\chi,
\end{equation}
is the only normalizable state bound to this domain wall.
This boundary state is a zero mode. It is an eigenstate of the
single-particle boundary Hamiltonian
\begin{equation} 
\mathcal{H}^{(0)}_{\mathrm{bd}}=0.
\end{equation}
\end{subequations}

Suppose that we consider $\nu=1,2,\cdots$ identical copies of the 
single-particle Hamiltonian 
(\ref{eq: def 1D mathcal H for BDI})
by defining
\begin{subequations}
\begin{equation}
\mathcal{H}^{(0)}_{\nu}(x):=
\mathcal{H}^{(0)}(x)\otimes\openone,
\end{equation}
and
\begin{align}
\mathcal{T}&:=
\mathrm{i}
X^{\,}_{12}\otimes\openone\,\mathsf{K}, 
& 
\mathcal{C}&:=
\mathrm{i}
X^{\,}_{02}\otimes\openone\,\mathsf{K},
\end{align}
\end{subequations}
where $\openone$ is a $\nu\times\nu$ unit matrix.
Observe that $\mathcal{T}$ and $\mathcal{C}$ commute with 
$X^{\,}_{10}\otimes\openone$ and with each other. 
The domain wall (\ref{eq: def Jackiw Rebbi domain wall CII}) 
must then support $\nu$ linearly independent boundary zero modes. 
They are annihilated by the boundary Hamiltonian
\begin{equation} 
\mathcal{H}^{(0)}_{\mathrm{bd}\,\nu}=
\mathcal{H}^{(0)}_{\mathrm{bd}}\otimes\openone=0.
\end{equation}
The topological sectors for noninteracting Hamiltonians
are thus labeled by the integer $\nu$ taking values in $\mathbb{Z}$
in the limit $\nu\to\infty$.

A generic local quartic interaction that respects the defining CII
symmetries with the potential to gap out the boundary zero modes
reduces to a dynamical Dirac mass 
(that depends on imaginary time $\tau$ in addition to 
space $x$)  that belongs to the symmetry class C, upon performing 
a Hubbard-Stratonovich transformation. Hence, we must consider the dynamical 
bulk single-particle Hamiltonian
\begin{subequations}
\label{eq: mathcal H dyb nu d=1 CII}
\begin{equation}
\mathcal{H}^{(\mathrm{dyn})}_{\nu}(\tau,x):=
\left[
-\mathrm{i}
\partial^{\,}_{x}\,
X^{\,}_{30}
+ 
m(x)\,
X^{\,}_{20}
\right]
\otimes\openone
+
\mathcal{V}(\tau,x),
\label{eq: mathcal H dyb nu d=1 CII a}
\end{equation}
where the dynamical Dirac mass $\mathcal{V}(\tau,x)$ 
is defined by the condition that
it anticommutes with $\mathcal{H}^{(0)}(x)\otimes\openone$, 
when independent of $x$,
and obeys the transformation laws dictated by the symmetry class C, 
i.e., it must obey
\begin{equation}
\mathcal{C}\,
\mathcal{V}(\tau,x)\,
\mathcal{C}^{-1}=
-\mathcal{V}(\tau,x).
\label{eq: mathcal H dyb nu d=1 CII b}
\end{equation} 
\end{subequations}

On the boundary, the operations for reversal of time and charge conjugation
are now represented by
\begin{subequations}
\begin{equation}
\mathcal{T}^{\,}_{\mathrm{bd}}:=\sigma^{\,}_{2}\otimes\openone\,\mathsf{K},
\qquad
\mathcal{C}^{\,}_{\mathrm{bd}}:=\sigma^{\,}_{2}\otimes\openone\,\mathsf{K}.
\end{equation}
Hence, we must consider the dynamical single-particle boundary Hamiltonian
\begin{equation}
\mathcal{H}^{(\mathrm{dyn})}_{\mathrm{bd}\,\nu}(\tau):=
\gamma'(\tau),
\end{equation}
where
\begin{equation}
\mathcal{C}^{\,}_{\mathrm{bd}}\,
\gamma^{\prime}(\tau)\,
\mathcal{C}^{-1}_{\mathrm{bd}}=
-
\gamma^{\prime}(\tau).
\end{equation}
\end{subequations}
The space of normalized Dirac mass matrices obtained by demanding
that $\gamma'(\tau)$ squares to the unit $2\nu\times2\nu$ matrix
for all imaginary times is the space 
\begin{align}
V^{\,}_{\nu}=Sp(\nu)/U(\nu)
\end{align}
for the symmetry class C in zero-dimensional space. 
The limit $\nu\to\infty$ of these spaces is the classifying space
$R^{\,}_{6}$. In order to gap out dynamically the boundary zero modes 
without breaking the defining symmetries of the symmetry class CII,
we need to construct a (0+1)-dimensional
QNLSM for the (boundary) dynamical Dirac masses
from the zero-dimensional symmetry class C without topological 
obstructions. We construct explicitly the spaces 
for the relevant normalized boundary dynamical Dirac masses 
of dimension $\nu=2^{n}$ with $n=0,1$ in the following.
The relevant homotopy groups are given in Table~\ref{tab: 1D CII}.

\textit{Case $\nu=1$:}
The three Dirac mass matrices 
$\sigma^{\,}_{1}$,
$\sigma^{\,}_{2}$,
and
$\sigma^{\,}_{3}$,
are allowed on the boundary.
They all anticommute pairwise.
A WZ term is permissible as
$\pi^{\,}_{0+1+1}(S^{2})=\mathbb{Z}$. 
In the presence of this WZ term,
the boundary theory remains gapless.

\textit{Case $\nu=2$:}
The minimum number of anticommuting mass matrices is larger than three.
Hence,  the zeroth, first, and second homotopy groups over the 
boundary normalized dynamical Dirac masses all vanish.
No topological term is possible.
The (0+1)-dimensional QNLSM over this target space endows dynamically
the boundary Hamiltonian with a spectral gap.

We conclude that the effects of interactions on the one-dimensional
SPT phases in the symmetry class CII are
to reduce the topological classification $\mathbb{Z}$
in the noninteracting limit down to $\mathbb{Z}^{\,}_{2}$ under the
assumption that a Hamiltonian from the symmetry class CII 
is interpreted as a mean-field description of a superconductor.
The logic used to reach this conclusion
is summarized by Table \ref{tab: 1D CII} once the line corresponding to
$\nu=1$ has been identified. It is given by the smallest $D$ that
accommodates a non-trivial entry for the corresponding homotopy group.

\begin{table}[tb]
\caption{
Reduction from $\mathbb{Z}$ 
to
$\mathbb{Z}^{\,}_{2}$ 
for the topologically equivalent classes of the
one-dimensional SPT phases in the symmetry class CII
that arises from interactions. 
We denote by $V^{\,}_{\nu}$ the space of
$\nu\times\nu$ normalized Dirac mass matrices
in zero-dimensional Hamiltonians belonging to the
symmetry class C. 
The limit $\nu\to\infty$ of these spaces is the classifying spaces
$R^{\,}_{6}$. 
The second column shows the stable $D$-th homotopy groups 
of the classifying space $R^{\,}_{6}$. 
The third column gives the number $\nu$ of copies of 
boundary (Dirac) fermions for which a topological obstruction is permissible.
The fourth column gives the type of topological obstruction
that prevents the gapping of the boundary (Dirac) fermions.
\label{tab: 1D CII}
        }
\begin{tabular}{ccccccc}
\hline \hline
$D$ 
&\qquad\qquad& 
$
\pi^{\,}_{D}(R^{\,}_{6})$
&\qquad\qquad& 
$\nu$ 
&\qquad\qquad&
Topological obstruction 
\\
\hline
0&&0                   &&           &&        \\
1&&0                   &&           &&        \\
2&&$\mathbb{Z}$        &&  $1$      && WZ term\\
3&&$\mathbb{Z}^{\,}_{2}$ &&  $2$      && None   \\
4&&$\mathbb{Z}^{\,}_{2}$ &&           &&        \\
5&&0                   &&            &&        \\
6&&$\mathbb{Z}$        &&            &&        \\
7&&0                   &&            &&        \\
\hline \hline
\end{tabular}
\end{table}

\subsubsection{The chiral symmetry classes as one-dimensional insulators}
\label{subsec: The chiral symmetry classes when d=1}

So far we have interpreted the symmetry classes BDI and CII
as examples of topological superconductors by focusing on the fact that their
second-quantized Hamiltonian respects a unitary charge-conjugation symmetry,
a PHS
(see Appendix 
\ref{appsec: Defining symmetries of strong topological insulators}).
As the  symmetry classes BDI and CII also preserve TRS, the composition of
reversal of time with charge conjugation delivers a non-unitary symmetry
of their second-quantized Hamiltonian, namely a CHS
(see Appendix 
\ref{appsec: Defining symmetries of strong topological insulators}).
The third chiral symmetry class AIII is defined by demanding that it
preserves CHS, no more and no less. 
Hence, any representative gapped Hamiltonian
from the chiral symmetry class AIII can always be
interpreted as a topological insulator with fermion number conservation.

The CHS can be implemented at the single-particle level by
a unitary sublattice spectral symmetry for (complex) electrons hopping
between two sublattices. From this point of view, the three chiral
symmetry classes AIII, BDI, and CII, 
when interpreted as metals or as insulators,
can be treated on equal footing. 
The fermion number is conserved in a metal or in an insulator, 
unlike in the mean-field treatment of a superconductor. 
This is the case when the  symmetry class
BDI is interpreted as an effective theory for polyacetylene, in which
case the Dirac gap is induced by coupling the electrons to phonons,
i.e., it realizes a Peierls or bond-density wave instability.%
~\cite{Su79}
In this interpretation of the chiral classes, it is necessary to
introduce an additional particle-hole grading in order to include
through dynamical Dirac masses the effects of superconducting fluctuations
induced by any quartic interaction. Failure to do so can produce a
distinct reduction pattern of the noninteracting
topological equivalence classes arising from interactions,
for it can matter whether the boundary dynamical masses belong to the 
classifying space associated with the symmetry classes D
or to the symmetry class A.

\paragraph{Symmetry class AIII}

Consider the one-dimensional bulk single-particle Dirac Hamiltonian
in the  symmetry class AIII
\begin{subequations}
\label{eq: def 1D mathcal H for AIII}
\begin{align}
\mathcal{H}^{(0)}(x):=
-\mathrm{i}
\partial^{\,}_{x}
\tau^{\,}_{3}
+ 
m(x)
\tau^{\,}_{2}.
\label{eq: def 1D mathcal H for AIII a}
\end{align}
It anticommutes with the unitary operator
\begin{equation}
\Gamma^{\,}_{5}:=
\tau^{\,}_{1}.
\label{eq: def 1D mathcal H for AIII b}
\end{equation}
It supports the zero mode
(\ref{eq: zero mode if nu=1 for BDI})
at the boundary where it identically vanishes,
\begin{equation}
\mathcal{H}^{(0)}_{\mathrm{bd}}(x)=0.
\label{eq: def 1D mathcal H for AIII c}
\end{equation}
\end{subequations}

Upon tensoring Hamiltonians
(\ref{eq: def 1D mathcal H for AIII a})
and
(\ref{eq: def 1D mathcal H for AIII c})
together with the Dirac $\Gamma^{\,}_{5}$ matrix by the 
$\nu\times\nu$ unit matrix $\openone$,
there follows $\nu=1,2,3,\cdots$ boundary zero modes.

The dynamical single-particle Hamiltonian that encodes
those non-superconducting fluctuations arising from a local
quartic interactions after a Hubbard-Stratonovich transformation
takes the form (\ref{eq: dyanmical bulk 1d BDI a})
with 
\begin{equation}
\mathcal{V}(\tau,x):=
\tau^{\,}_{1}
\otimes
\gamma'(\tau,x)
\end{equation}
and $\gamma'(\tau,x)$ a $\nu\times\nu$ Hermitian matrix.

On the boundary, we must consider the dynamical single-particle boundary 
Hamiltonian 
\begin{equation}
\mathcal{H}^{(\mathrm{dyn})}_{\mathrm{bd}\,\nu}(\tau)=
\gamma'(\tau).
\label{eq: dynamical boundary mathcal{H} d=1 AIII} 
\end{equation}
The $\nu\times\nu$ Hermitian matrix $\gamma'(\tau)$ 
belongs to the zero-dimensional symmetry class A.
Consequently, it is assigned the classifying space $C^{\,}_{0}$
in the limit $\nu\to\infty$ with the zeroth-homotopy group
$\pi^{\,}_{0}(C^{\,}_{0})=\mathbb{Z}$. 
When $\nu=1$, $\gamma'$ is a real number, and
the domain wall $\gamma'(\tau)\propto\mathrm{sgn}(\tau)$
binds a zero mode at $\tau=0$ 
[i.e., a normalizable zero mode of the operator 
$\partial^{\,}_{\tau}+\mathcal{H}^{(\mathrm{dyn})}_{\mathrm{bd}\,\nu}(\tau)$].
When $\nu=2$, we can write
$\gamma'(\tau)$
as a linear combination of the matrices
$\sigma^{\,}_{\mu}$ 
with the real-valued functions $m^{\,}_{\mu}(\tau)$
as coefficients for $\mu=0,1,2,3$, respectively.
Any one of the three Pauli matrices 
($\sigma^{\,}_{1},\sigma^{\,}_{2},\sigma^{\,}_{3}$)
anticommutes with the other two
Pauli matrices.
Hence, the space of normalized boundary dynamical Dirac masses 
that anticommute pairwise is homeomorphic to $S^{2}$ with the homotopy group 
$\pi^{\,}_{0+1+1}(S^{2})=\mathbb{Z}$ when $\nu=2$. 
A (0+1)-dimensional QNLSM for the (boundary) dynamical Dirac masses
is augmented by a WZ term.

We conclude that the effects of interactions on the one-dimensional
SPT phases in the symmetry class AIII are to reduce the topological
classification $\mathbb{Z}$
in the noninteracting limit down to $\mathbb{Z}^{\,}_{4}$ under the
assumption that only fermion-number-conserving 
dynamical Dirac masses are included in the 
single-particle Hamiltonian.
The logic used to reach this conclusion
is summarized by Table \ref{tab: 1D AIII} once the line corresponding to
$\nu=1$ has been identified. It is given by the smallest $D$ that
accommodates a non-trivial entry for the corresponding homotopy group.

\begin{table}[tb]
\caption{
Reduction from $\mathbb{Z}$ 
to
$\mathbb{Z}^{\,}_{4}$ 
for the topologically equivalent classes of the
one-dimensional SPT phases in symmetry classes AIII or BDI
that arises from 
the fermion-number-conserving interacting channels. 
We denote by $V^{\,}_{\nu}$ the space of $\nu\times\nu$ 
normalized Dirac mass matrices
in zero-dimensional Hamiltonians belonging to the symmetry class A. 
The limit $\nu\to\infty$ of these spaces is the classifying spaces
$C^{\,}_{0}$.
The second column shows the stable $D$-th homotopy groups 
of the classifying space $C^{\,}_{0}$. 
The third column gives the number $\nu$ of copies of 
boundary (Dirac) fermions for which a topological obstruction is permissible.
The fourth column gives the type of topological obstruction
that prevents the gapping of the boundary (Dirac) fermions.
\label{tab: 1D AIII}
        }
\begin{tabular}{ccccccc}
\hline \hline
$D$ 
&\qquad\qquad& 
$
\pi^{\,}_{D}(C^{\,}_{0})$
&\qquad\qquad& 
$\nu$ 
&\qquad\qquad&
Topological obstruction 
\\
\hline
0&&$\mathbb{Z}$ && $1$  && Domain wall \\
1&&0                    &&     &&      \\
2&&$\mathbb{Z}$ && $2$  && WZ term     \\
3&&0            &&      &&             \\
4&&$\mathbb{Z}$ && $4$  && None        \\
5&&0            &&      &&             \\
6&&$\mathbb{Z}$ &&      &&             \\
7&&0                    &&     &&      \\
\hline \hline
\end{tabular}
\end{table}

To include the effects of superconducting fluctuations
in the single-particle Hamiltonian 
after a Hubbard-Stratonovich transformation, 
we need to consider the direct sum
\begin{subequations}
\label{eq: def 1D mathcal H BdG for AIII}
\begin{align}
\mathcal{H}^{(0)}_{\mathrm{BdG}}(x):=&
\left[\mathcal{H}^{(0)}(x)\otimes\openone\right]
\oplus
\left[-\mathcal{H}^{(0)\,*}(x)\otimes\openone\right]
\nonumber\\
\equiv&\,
\mathcal{H}^{(0)}(x)\otimes\openone\otimes\rho^{\,}_{0}.
\label{eq: def 1D mathcal H BdG for AIII a}
\end{align}
This Bogoliubov-de-Gennes (BdG)
single-particle Hamiltonian anticommutes with the operation for
charge conjugation
\begin{equation}
\mathcal{C}:=
\tau^{\,}_{0}\otimes\openone\otimes\rho^{\,}_{1}\,\mathsf{K},
\label{eq: def 1D mathcal H BdG for AIII b}
\end{equation}
\end{subequations}
in addition to anticommuting with 
$\tau^{\,}_{1}\otimes\openone\otimes\rho^{\,}_{0}$.
Hence, it belongs to the  symmetry class BDI.
The dynamical single-particle Hamiltonian that accounts for
superconducting fluctuations 
(dynamical Dirac masses from the symmetry class D)
takes the form 
\begin{subequations}
\begin{equation}
\mathcal{V}(\tau,x):=
\tau^{\,}_{1}\otimes
\gamma'(\tau,x),
\end{equation}
where the $2\nu\times2\nu$ Hermitian matrix $\gamma'(\tau,x)$
must obey
\begin{equation}
\gamma'(\tau,x)=
-
(\tau^{\,}_{0}\otimes\openone\otimes\rho^{\,}_{1})\,
\gamma^{\prime\,*}(\tau,x)\,
(\tau^{\,}_{0}\otimes\openone\otimes\rho^{\,}_{1}).
\end{equation}
\end{subequations}
The stability analysis of the boundary zero modes is similar to the
one performed below Eq.\ 
(\ref{eq: case d=1 BDI a supercond: B nu= O nu / U nu/2})
except that one must replace $\nu$ in Eq.\ 
(\ref{eq: case d=1 BDI a supercond: B nu= O nu / U nu/2})
by $\nu^{\,}_{\mathrm{BdG}}=2\nu$ 
and that we have a different representation of the PHS. 
When $\nu=1$,
\begin{equation}
\gamma'(\tau)=M(\tau)\,\rho^{\,}_{3},
\qquad
M(\tau)\in\mathbb{R},
\label{eq: d=1 AIII nu=1 dynamical mass}
\end{equation}
supports a domain wall in imaginary time on the boundary.
When $\nu=2$, we use the representation
$\openone=\sigma^{\,}_{0}$ and introduce the notation
$X^{\,}_{\mu\mu'}\equiv\sigma^{\,}_{\mu}\otimes\rho^{\,}_{\mu'}$
with $\mu,\mu'=0,1,2,3$. Now,
\begin{equation}
\gamma'(\tau)= 
\sum_{\{(\mu,\mu')\}}M^{\,}_{\mu\mu'}(\tau)\,X^{\,}_{\mu\mu'},
\qquad
M^{\,}_{\mu\mu'}(\tau)\in\mathbb{R},
\label{eq: d=1 AIII nu=2 dynamical mass}
\end{equation} 
where the sum on the right-hand side
is to be performed over
the six matrices
$X^{\,}_{21}$,$X^{\,}_{22}$,$X^{\,}_{03}$,$X^{\,}_{13}$,$X^{\,}_{20}$,$X^{\,}_{33}$.
This set of six matrices decomposes into 
two triplets of pairwise anticommuting matrices. 
The first triplet is given by
$X^{\,}_{21}$,$X^{\,}_{22}$,$X^{\,}_{03}$.
The second triplet is given by
$X^{\,}_{13}$,$X^{\,}_{20}$,$X^{\,}_{33}$.
Each triplet defines a two-sphere $S^{2}$.
Hence, each triplet has the potential to accommodate a WZ term.
However, we must make sure that the integrity of any one $S^{2}$
entering the decomposition $S^{2}\cup S^{2}$
of the normalized dynamical masses on the boundary is compatible with
maintaining the global $U(1)$ symmetry associated with the conservation of the
fermion number.

To address the fate
of the fermion-number conservation, observe that
Hamiltonian (\ref{eq: def 1D mathcal H for AIII a})
is invariant under the global $U(1)$ transformation
\begin{equation}
\mathcal{H}^{(0)}\mapsto
\mathcal{U}(\alpha)\,
\mathcal{H}^{(0)}\,
\mathcal{U}^{-1}(\alpha),
\qquad
\mathcal{U}(\alpha):= e^{+\mathrm{i}\alpha\,\tau^{\,}_{0}},
\end{equation}
with $0\leq\alpha<2\pi$ independent of $x$.
This symmetry is to be preserved when treating superconducting
fluctuations. In the BdG representation 
(\ref{eq: def 1D mathcal H BdG for AIII a}),
this symmetry becomes the symmetry under the global $U(1)$ transformation
\begin{subequations}
\label{eq: def global U(1) gauge trsf d=1 AIII}
\begin{equation}
\mathcal{H}^{(0)}_{\mathrm{BdG}}\mapsto
\mathcal{U}^{\,}_{\mathrm{BdG}}(\alpha)\,
\mathcal{H}^{(0)}_{\mathrm{BdG}}\,
\mathcal{U}^{-1}_{\mathrm{BdG}}(\alpha),
\end{equation}
where
\begin{equation}
\mathcal{U}^{\,}_{\mathrm{BdG}}(\alpha):= 
e^{+\mathrm{i}\alpha\,\tau^{\,}_{0}\otimes\openone\otimes\rho^{\,}_{3}}.
\end{equation}
\end{subequations}
When $\nu=1$, the boundary dynamical mass 
(\ref{eq: d=1 AIII nu=1 dynamical mass})
is invariant under multiplication from the left with 
$\exp(+\mathrm{i}\alpha\,\rho^{\,}_{3})$
and multiplication from the right with 
$\exp(-\mathrm{i}\alpha\,\rho^{\,}_{3})$.
When $\nu=2$,  if we normalize the boundary dynamical mass 
(\ref{eq: d=1 AIII nu=2 dynamical mass})
by demanding that
\begin{subequations}
\begin{align}
&
1=
M^{2}_{21}
+
M^{2}_{22}
+
M^{2}_{03},
\label{eq: d=1 AIII first sphere S2}
\\
&
1=
M^{2}_{13}
+
M^{2}_{20}
+
M^{2}_{33},
\label{eq: d=1 AIII second sphere S2}
\end{align}
\end{subequations}
we may then identify these boundary dynamical masses 
as the union of two two-spheres $S^{2}$. 
The global $U(1)$ transformation defined by
multiplication from the left with
$\exp(+\mathrm{i}\alpha\,X^{\,}_{03})$
and multiplication from the right with 
$\exp(-\mathrm{i}\alpha\,X^{\,}_{03})$
leaves the two-sphere (\ref{eq: d=1 AIII first sphere S2})
invariant as a set, for it is represented by a rotation about the north pole
$X^{\,}_{03}$ that rotates the equator spanned by 
$X^{\,}_{21}$ and $X^{\,}_{22}$ with the angle $2\alpha$.
The same transformation leaves the two-sphere
(\ref{eq: d=1 AIII second sphere S2})
invariant point-wise. Hence, each $S^{2}$ in $S^{2}\cup S^{2}$
is compatible with the conservation of the global fermion number.
Because $\pi^{\,}_{0+1+1}(S^{2})=\mathbb{Z}$, 
a WZ topological term in the QNLSM for the boundary is permissible.

We may then safely conclude that the effects of interactions 
on the one-dimensional SPT phases in the symmetry class AIII are
also to reduce the topological 
classification $\mathbb{Z}$
in the noninteracting limit down to $\mathbb{Z}^{\,}_{4}$ under the
assumption that superconducting fluctuation
channels are included in the stability analysis.
The $\mathbb{Z}^{\,}_{4}$ classification 
for one-dimensional SPT phases in the symmetry class AIII 
agrees with the one derived using
group cohomology in Ref.\ \onlinecite{Tang-Wen12}
[with AIII interpreted in Ref.\ \onlinecite{Tang-Wen12}
as a time-reversal-symmetric superconductor with the full
spin-1/2 rotation symmetry broken down to a $U(1)$ subgroup].

\paragraph{Symmetry class BDI}
Dirac Hamiltonians in the symmetry class BDI are obtained
from those in the symmetry class AIII by imposing the constraint of TRS,
Eqs.\ (\ref{eq: def 1D mathcal H for BDI b}) and
(\ref{eq: def 1D mathcal H for BDI d})
(note that $r^{\,}_{\min}=2$ for both AIII and BDI classes in $d=1$).
Since the TRS is not relevant for dynamical Dirac masses
in the single-particle Dirac Hamiltonian after a Hubbard-Stratonovich
transformation, the stability analysis of gapless boundary states in the
symmetry class BDI in $d=1$
follows from that of the symmetry class AIII.
Consequently, the effects of interactions in the 
symmetry class BDI, when interpreted as realizing complex fermions
as opposed to Majorana fermions,
is to reduce the topological classification $\mathbb{Z}$
in the noninteracting limit down to $\mathbb{Z}^{\,}_{4}$ under the
assumption that only fermion-number preserving dynamical
Dirac masses taken from the symmetry class A
are included in the stability analysis.
Furthermore, if dynamical superconducting fluctuations
are allowed by introducing an additional particle-hole grading
and dynamical Dirac masses from the symmetry class D, then
the same reduction pattern $\mathbb{Z}\to\mathbb{Z}^{\,}_{4}$ follows.
The topological classification $\mathbb{Z}^{\,}_{8}$ of the symmetry class BDI
when interpreted as describing Majorana fermions is thus finer than the
classification $\mathbb{Z}^{\,}_{4}$ of the symmetry class BDI when
interpreted as describing complex fermions.

\paragraph{Symmetry class CII}

We interpret the single-particle Hamiltonian
(\ref{eq: def 1D mathcal H for CII a})
as describing an insulator, not a superconductor.
This is to say that the defining symmetries are TRS
(\ref{eq: def 1D mathcal H for CII b})
and the CHS
\begin{equation}
\Gamma^{\,}_{5}\,\mathcal{H}^{(0)}(x)\,\Gamma^{-1}_{5}=
-
\mathcal{H}^{(0)}(x),
\qquad
\Gamma^{\,}_{5}:=
X^{\,}_{10}.
\end{equation}
If we are after the dynamical effects of interactions that preserve
the (complex) fermion number, we may use
Eq.\ (\ref{eq: mathcal H dyb nu d=1 CII})
with the only caveat that the dynamical mass matrix is now required to belong
to the symmetry class A instead of the symmetry class C. 
The boundary dynamical Hamiltonian
is then the same as for the symmetry classes AIII and BDI,
i.e., Eq.\ (\ref{eq: dynamical boundary mathcal{H} d=1 AIII}),
except for its rank being twice as large as compared to the symmetry classes
AIII and BDI, for an additional grading 
(that of the spin-1/2 degrees of freedom)
has been accounted for. 
This larger rank implies that the WZ term is already permissible
at $\nu=1$, i.e., the reduction pattern for the noninteracting
topological classification
is $\mathbb{Z}\to\mathbb{Z}^{\,}_{2}$. 
It remains to verify that 
the same reduction pattern is also obtained if superconducting
fluctuations are included, as was the case for the symmetry classes AIII
and BDI. To this end, we must tensor
(\ref{eq: def 1D mathcal H for CII a})
with $\rho^{\,}_{0}$, in which case the charge conjugation symmetry
is realized by 
$\tau^{\,}_{0}\otimes\sigma^{\,}_{0}\otimes\rho^{\,}_{1}\,\mathsf{K}$.
We can borrow the stability analysis with respect to superconducting
interacting channels from the symmetry class D that we performed for
the symmetry classes AIII and BDI, again with the caveat that the rank
of the boundary dynamical Hamiltonian is twice as large as it was.
This larger rank implies again that the WZ term is already permissible
for $\nu=1$, i.e., the reduction pattern for the noninteracting
topological classification
is again $\mathbb{Z}\to\mathbb{Z}^{\,}_{2}$. 
Thus we obtain the same topological
classification $\mathbb{Z}^{\,}_{2}$ of the symmetry class CII 
in $d=1$ both when
interpreted as describing Majorana fermions (superconductors)
and when
interpreted as describing complex fermions (insulators).

\subsection{The case of two-dimensional space}

The notion that the chiral edge modes in the IQHE are immune to
local interactions is rather intuitive. Neither backscattering nor
umklapp scattering is permissible. 
An operational and quantitative validation for this intuition goes
back to Niu and Thouless in
Ref.\ \onlinecite{Niu85},
who proposed to average the Kubo Hall conductivity
over all twisted boundary conditions of the many-body ground state
as a signature of both the IQHE and FQHE.
A mathematically rigorous proof of this intuition 
can be found in
Refs.\ \onlinecite{Hastings15} and \onlinecite{Koma15}.
This intuition readily extends to the symmetry class D and C
as they realize quantized thermal Hall effects.
The robustness of chiral edge modes in the symmetry classes 
D, C, and A to quartic contact interactions
will be derived using the method of Sec.\ 
\ref{sec: Definition and strategy}.

Let $\bm{x}=(x^{\,}_{1},x^{\,}_{2})$ denote a point in two-dimensional space.
The single-particle Dirac Hamiltonian with the smallest rank
$r^{\,}_{\mathrm{min}}=2$ 
that admits a Dirac mass can be chosen to be represented by
\begin{align}
\mathcal{H}^{(0)}_{\mathrm{A}}(\bm{x}):=&\,
\left[
-\mathrm{i}\partial^{\,}_{1}
+
A^{\,}_{1}(\bm{x})
\right]
\sigma^{\,}_{3}
+
\left[
-\mathrm{i}\partial^{\,}_{2}
+
A^{\,}_{2}(\bm{x})
\right]
\sigma^{\,}_{1}
\nonumber\\
&\,
+
A^{\,}_{0}(\bm{x})\,
\sigma^{\,}_{0}
+
m(\bm{x})\,
\sigma^{\,}_{2}.
\end{align}
It belongs to the  symmetry class A for arbitrarily
chosen vector potentials $\bm{A}(\bm{x})$,
scalar potential $A^{\,}_{0}(\bm{x})$,
and mass $m(\bm{x})$.
When the gauge fields are vanishing,
\begin{align}
\mathcal{H}^{(0)}_{\mathrm{D}}(\bm{x}):=&\,
-\mathrm{i}\partial^{\,}_{1}\,\sigma^{\,}_{3}
-\mathrm{i}\partial^{\,}_{2}\,\sigma^{\,}_{1}
+
m(\bm{x})\,\sigma^{\,}_{2}
\nonumber\\
=&\,
-
\left[\mathcal{H}^{(0)}(\bm{x})\right]^{*}
\end{align}
belongs to the  symmetry class D. 
Finally, the single-particle
Hamiltonian with the smallest rank $r^{\,}_{\min}=4$ that belongs to
the symmetry class C can be chosen to be represented by
\begin{equation}
\begin{split}
\mathcal{H}^{(0)}_{\mathrm{C}}(\bm{x}):=&\,
-\mathrm{i}\partial^{\,}_{1}\,X^{\,}_{30}
+
\sum_{j=1}^{3}
A^{\,}_{1j}(\bm{x})\,
X^{\,}_{3j}
\\
&\,
-\mathrm{i}\partial^{\,}_{2}\,X^{\,}_{10}
+
\sum_{j=1}^{3}
A^{\,}_{2j}(\bm{x})\,X^{\,}_{1j}
\\
&\,
+
\sum_{j=1}^{3}
A^{\,}_{0j}(\bm{x})\,
X^{\,}_{0j}
+
m(\bm{x})\,
X^{\,}_{20}
\\
=&\,
-
X^{\,}_{02}
\left[\mathcal{H}^{(0)}(\bm{x})\right]^{*}
X^{\,}_{02}.
\end{split}
\end{equation}

In two spatial dimensions the symmetry classes A, D, and C
realize noninteracting topological insulators and
superconductors with the Grassmannian manifolds
$C^{\,}_{0}\equiv
\lim_{N\to\infty} 
\cup_{n=0}^{N}\,U(N)/[U(n)\times U(N-n)]$,
$R^{\,}_{0}\equiv
\lim_{N\to\infty} 
\cup_{n=0}^{N}\,
O(N)/[O(n)\times )(N-n)]$,
and
$R^{\,}_{4}\equiv
\lim_{N\to\infty} 
\cup_{n=0}^{N}\,
Sp(N)/[Sp(n)\times )(N-n)]$
as classifying spaces, respectively. They share the same zeroth-order
homotopy group $\mathbb{Z}$. This group also serves as defining the
topological attributes of 
noninteracting topological insulators and superconductors
in the  symmetry classes A, D, and C.

\subsubsection{The symmetry class D when $d=2$}

Let $\openone$ denote a $\nu\times\nu$ unit matrix with $\nu=1,2,\cdots$.
Consider the two-dimensional bulk single-particle Dirac Hamiltonian
\begin{equation}
\mathcal{H}^{(0)}(\bm{x}):=
-\mathrm{i}\partial^{\,}_{1}\,
\sigma^{\,}_{3}\otimes\openone
-\mathrm{i}\partial^{\,}_{2}\,
\sigma^{\,}_{1}\otimes\openone
+m(\bm{x})\,
\sigma^{\,}_{2}\otimes\openone
\label{eq: def Dirac Hamiltonian 2d symmetry class D}
\end{equation}
of rank $2\nu$.
There is no Hermitian 
$(2\nu)\times(2\nu)$ matrix that anticommutes with
$\mathcal{H}^{(0)}_{\mathrm{A}}(\bm{x})$. If so,
the set $\{\beta\}$ in Eq.\ 
(\ref{eq: def interacting boundary massless Dirac fermions})
is empty. In other words, no dynamical mass is available to induce
a dynamical instability of the $\nu$ boundary zero modes.

\subsubsection{The symmetry class C when $d=2$}

The same reasoning applies to the bulk single-particle
Hamiltonian 
\begin{subequations}
\begin{equation}
\mathcal{H}^{(0)}(\bm{x}):=
\left[
-\mathrm{i}\partial^{\,}_{1}\,
X^{\,}_{30}
-\mathrm{i}\partial^{\,}_{2}\,
X^{\,}_{10}\,
+m(\bm{x})\,
X^{\,}_{20}
\right]
\otimes\openone,
\label{eq: def Dirac Hamiltonian 2d symmetry class C a}
\end{equation}
of rank $4\nu$ that realizes a 
topological superconductor in the  
symmetry class C, 
\begin{equation}
\mathcal{H}^{(0)}(\bm{x})=
-
\left(X^{\,}_{02}\otimes\openone\right)
\left[\mathcal{H}^{(0)}(\bm{x})\right]^{*}
\left(X^{\,}_{02}\otimes\openone\right).
\label{eq: def Dirac Hamiltonian 2d symmetry class C b}
\end{equation}
\end{subequations}
No dynamical mass is available to induce a dynamical instability
of the $\nu$ boundary zero modes.

\subsubsection{The symmetry class A when $d=2$}

If the single-particle Dirac Hamiltonian
(\ref{eq: def Dirac Hamiltonian 2d symmetry class D})
is interpreted as describing an insulator with fermion-number
conservation, then no dynamical mass that preserves the fermion number
and anticommutes with $\sigma^{\,}_{2}\otimes\openone$ is permissible.
The same remains true if we account for superconducting fluctuations, for
the BdG extension of
(\ref{eq: def Dirac Hamiltonian 2d symmetry class D})
that is given by
Eq.\ (\ref{eq: def Dirac Hamiltonian 2d symmetry class C a}),
whereby charge conjugation is defined by
[and not by Eq.\
(\ref{eq: def Dirac Hamiltonian 2d symmetry class C b})]
\begin{equation}
\mathcal{C}:=
X^{\,}_{01}\otimes\openone\,\mathsf{K},
\label{eq: def C for 2d A}
\end{equation}
fails to anticommute with any $(4\nu)\times(4\nu)$
Hermitian matrix allowed by the PHS generated by
the operation of charge conjugation (\ref{eq: def C for 2d A}).

\subsection{The case of three-dimensional space}

The reduction $\mathbb{Z}\to\mathbb{Z}^{\,}_{16}$ for the 
three-dimensional interacting topological superconductors belonging to 
the symmetry class DIII 
has been understood in the following ways
after a conjecture by Kitaev from Ref.%
~\onlinecite{kitaev-topomat11}.
One approach is to enumerate the distinct topological orders at the 
two-dimensional surface of the three-dimensional bulk 
that cannot be realized with any bulk two-dimensional Hamiltonian.%
~\cite{Fidkowski13,Metlitski14,Wang14}
In this approach, the breakdown of $\mathbb{Z}$ takes place when
vortices (point-like defects of a symmetry-broken phase) 
at the surface proliferate (deconfine) so as to stabilize
a gapped and fully symmetric surface phase.
Another approach advocated by You and Xu consists in
relating fermionic short-ranged entangled ground states
to bosonic short-ranged entangled ground states.%
~\cite{You14}
They also applied their approach to systems with inversion symmetry.
The reductions
$\mathbb{Z}\to\mathbb{Z}^{\,}_{4}$
and
$\mathbb{Z}\to\mathbb{Z}^{\,}_{8}$
for the symmetry classes CI and AIII was obtained 
in Ref.\ \onlinecite{Wang14}.
Finally, Kapustin has proposed to classify
symmetry protected topological phases for interacting bosons or fermions
by considering low-energy effective actions that are invariant 
under cobordism (a certain type of equivalence relation between
manifolds).%
~\cite{kapustin2014symmetry,kapustin2014bosonic,Kapustin14d}

We apply the method of Sec.\ 
\ref{sec: Definition and strategy}
to the symmetry classes DIII, CI, and AIII
in the presence of quartic contact interactions.
We recover the reductions
$\mathbb{Z}\to\mathbb{Z}^{\,}_{16}$,
$\mathbb{Z}\to\mathbb{Z}^{\,}_{4}$,
and
$\mathbb{Z}\to\mathbb{Z}^{\,}_{8}$
for the symmetry class DIII, CI, and AIII, respectively.
We also verify that the topological classification $\mathbb{Z}^{\,}_{2}$ 
of the symmetry class AII is stable to quartic contact interactions.

We shall denote with 
$\bm{x}\equiv(x,y,z)\equiv(x^{\,}_{1},x^{\,}_{2},x^{\,}_{3})$ 
a point in three-dimensional space.

\subsubsection{The symmetry class DIII when $d=3$}

Let $X^{\,}_{\mu\mu'}\equiv\tau^{\,}_{\mu}\otimes\rho^{\,}_{\mu'}$ 
with $\mu,\mu'=0,1,2,3$,
Consider the three-dimensional bulk single-particle Dirac Hamiltonian
(with Dirac matrices of dimension $r=4\equiv r^{\,}_{\mathrm{min}}$),
\begin{subequations}
\label{eq: def 3D H(0) DIII} 
\begin{equation}
\mathcal{H}^{(0)}(\bm{x}):=
-\mathrm{i}\partial^{\,}_{1}\,
X^{\,}_{31} 
-\mathrm{i}\partial^{\,}_{2}\, 
X^{\,}_{02} 
-
\mathrm{i}\partial^{\,}_{3}\, 
X^{\,}_{11} 
+ 
m(\bm{x})\,
X^{\,}_{03}.
\label{eq: def 3D H(0) DIII a} 
\end{equation}
This single-particle Hamiltonian belongs to the 
three-dimensional symmetry class DIII, for
\begin{align}
&
\mathcal{T}\,
\mathcal{H}^{(0)}(\bm{x})\,
\mathcal{T}^{-1}
=
+
\mathcal{H}^{(0)}(\bm{x}),
\label{eq: def 3D H(0) DIII b}
\\
&
\mathcal{C}\,
\mathcal{H}^{(0)}(\bm{x})\,
\mathcal{C}^{-1}
=
-
\mathcal{H}^{(0)}(\bm{x}),
\label{eq: def 3D H(0) DIII c} 
\end{align}
where
\begin{align}
\mathcal{T}&:=
\mathrm{i}X^{\,}_{20}\,\mathsf{K},
&
\mathcal{C}&:=
X^{\,}_{01}\,\mathsf{K}.
\label{eq: def 3D H(0) DIII d} 
\end{align}
\end{subequations}
The multiplicative factor $\mathrm{i}$ in the definition of
$\mathcal{T}$ is needed for $\mathcal{T}$ 
to commute with $\mathcal{C}$.

The Dirac mass matrix 
$X^{\,}_{03}$
is here the only one allowed for dimension four Dirac matrices 
under the constraints (\ref{eq: def 3D H(0) DIII b})
and (\ref{eq: def 3D H(0) DIII c}).
Consequently,
the domain wall
\begin{subequations}
\label{eq: def domain wall d=3 DIII}
\begin{equation}
m(\bm{x})\equiv m(y):=
m^{\,}_{\infty}\,
\mathrm{sgn}(y),
\qquad
m^{\,}_{\infty}\in\mathbb{R},
\label{eq: def domain wall d=3 DIII a}
\end{equation}
at $y=0$, binds the zero mode
\begin{equation}
e^{
-\mathrm{i}
X^{\,}_{02}\,X^{\,}_{03}\,
\int\limits_{0}^{y}\mathrm{d}y'\,m(y')
  }\,\chi
=
e^{
-|m^{\,}_{\infty}y|
  }\,
\chi,
\label{eq: def domain wall d=3 DIII b}
\end{equation}
where
\begin{equation}
X^{\,}_{01}\,
\chi=
-\mathrm{sgn}\,(m^{\,}_{\infty})\,
\chi
\label{eq: def domain wall d=3 DIII c}
\end{equation}
\end{subequations}
with $\chi$ independent of $x$ and $z$.
The kinetics of the gapless boundary states is governed by
the Dirac Hamiltonian
\begin{equation}
\mathcal{H}^{(0)}_{\mathrm{bd}}(x,z)=
-\mathrm{i}\partial^{\,}_{x}
\tau^{\,}_{3}
-
\mathrm{i}\partial^{\,}_{z}
\tau^{\,}_{1},
\end{equation}
where we have chosen $m^{\,}_{\infty}<0$.

On the boundary, the operations for reversal of time and charge conjugation
are now represented by
\begin{subequations}
\label{eq: boundary dyn H d=3 class DIII}
\begin{align}
\mathcal{T}^{\,}_{\mathrm{bd}\,\nu}&:=
\mathrm{i}
\tau^{\,}_{2}
\otimes
\openone\,
\mathsf{K},
&
\mathcal{C}^{\,}_{\mathrm{bd}\,\nu}&:=
\tau^{\,}_{0}\otimes\openone\,\mathsf{K},
\end{align}
where $\openone$ is the $\nu\times\nu$ unit matrix.
We seek the single-particle Hamiltonian on the boundary
that encodes the fluctuations arising from the Hubbard-Stratonovich
decoupling of quartic interactions through a generic
dynamical mass that respects the PHS on the boundary.
It is given by
\begin{align}
\mathcal{H}^{(\mathrm{dyn})}_{\mathrm{bd}\,\nu}(\tau,x,z):=&\,
-\mathrm{i}\partial^{\,}_{x}
\tau^{\,}_{3}\otimes\openone
-
\mathrm{i}\partial^{\,}_{z}
\tau^{\,}_{1}\otimes\openone  
\nonumber\\
&\,
+\tau^{\,}_{2}\otimes M(\tau,x,z)
\label{eq: boundary dyn H d=3 class DIII a}
\end{align}
with the $\nu\times\nu$ real-valued and symmetric matrix 
\begin{equation}
M(\tau,x,z)=
M^{*}(\tau,x,z)=
M^{\mathsf{T}}(\tau,x,z).
\label{eq: boundary dyn H d=3 class DIII b}
\end{equation}
\end{subequations}

The space of normalized Dirac mass matrices of the form 
(\ref{eq: boundary dyn H d=3 class DIII b})
is topologically equivalent to the space 
\begin{equation}
V^{\,}_{\nu}:=
\bigcup_{k=1}^{\nu} O(\nu)/\left[O(k)\times O(\nu-k)\right]
\end{equation}
for the symmetry class D in two-dimensional space.
The limit $\nu\to\infty$ of these spaces is the classifying space
$R^{\,}_{0}$.
In order to gap out dynamically the boundary zero modes 
without breaking the defining symmetries of the symmetry class DIII,
we need to construct a (2+1)-dimensional
QNLSM for the (boundary) dynamical Dirac masses
from the two-dimensional symmetry class D without topological 
obstructions. We construct explicitly the spaces 
for the relevant normalized boundary dynamical Dirac mass matrices
of dimension $\nu=2^{n}$ with $n=0,1,2,3,4$ in the following.
The relevant homotopy groups are given in Table%
~\ref{tab: 3D DIII}.

\begin{table}[tb]
\caption{
Reduction from $\mathbb{Z}$ 
to
$\mathbb{Z}^{\,}_{16}$ 
for the topologically equivalent classes of the
three-dimensional SPT phases in the symmetry class DIII
that arises from interactions. 
We denote by $V^{\,}_{\nu}$ the space of
$\nu\times\nu$  normalized Dirac mass matrices
in boundary ($d=2$) Dirac Hamiltonians belonging to the
symmetry class D. 
The limit $\nu\to\infty$ of these spaces is the classifying space
$R^{\,}_{0}$.
The second column shows the stable $D$-th homotopy groups 
of the classifying space $R^{\,}_{0}$. 
The third column gives the number $\nu$ of copies of 
boundary (Dirac) fermions for which a topological obstruction is permissible.
The fourth column gives the type of topological obstruction
that prevents the gapping of the boundary (Dirac) fermions.
\label{tab: 3D DIII}
        }
\begin{tabular}{ccccccc}
\hline \hline
$D$ 
&\qquad\qquad& 
$\pi^{\,}_{D}(R^{\,}_{0})$ 
&\qquad\qquad& 
$\nu$ 
&\qquad\qquad& 
Topological obstruction 
\\
\hline
0
&\qquad\qquad& 
$\mathbb{Z}$ 
&\qquad\qquad&  
$1$ 
&\qquad\qquad& 
Domain wall  
\\
1
&\qquad\qquad& 
$\mathbb{Z}^{\,}_{2}$ 
&\qquad\qquad& 
$2$  
&\qquad\qquad& 
Vortex line  
\\
2
&\qquad\qquad& 
$\mathbb{Z}^{\,}_{2}$ 
&\qquad\qquad& 
$4$  
&\qquad\qquad& 
Monopole 
\\
3
&\qquad\qquad& 
0  
&\qquad\qquad&   
\\
4
&\qquad\qquad& 
$\mathbb{Z}$ 
&\qquad\qquad&  
$8$ 
&\qquad\qquad& 
WZ term 
\\
5
&\qquad\qquad& 
0  
&\qquad\qquad& 
\qquad\qquad&   
\\
6
&\qquad\qquad& 
0  
&\qquad\qquad& 
\qquad\qquad& 
\\
7
&\qquad\qquad& 
0  
&\qquad\qquad& 
\qquad\qquad&   
\\
8
&\qquad\qquad&
$\mathbb{Z}$
&\qquad\qquad&
16
&\qquad\qquad&
None
\\
\hline \hline
\end{tabular}
\end{table}

\textit{Case $\nu=1$:}
There is one boundary dynamical Dirac mass matrix $\gamma'(\tau,x,z)$
on the boundary that is proportional to $\tau^{\,}_{2}$.
A domain wall in imaginary time such as
$m^{\,}_{2\,\infty}\,\mathrm{sign}(\tau)\,\tau^{\,}_{2}$ 
prevents the dynamical generation of a spectral gap on the boundary.

\textit{Case $\nu=2$:} 
We use the representation 
$\openone=\sigma^{\,}_{0}$.
The $2\times2$ real-valued and symmetric matrix $M(\tau,x,z)$ is a linear
combination with real-valued coefficients of the pair of anticommuting matrices
$\sigma^{\,}_{x}$ and $\sigma^{\,}_{z}$.
If $M(\tau,x,z)$ is normalized by demanding that it squares to $\sigma^{\,}_{0}$, 
then the set spanned by $M(\tau,x,z)$ is homeomorphic to the one-sphere $S^{1}$.
As $\pi^{\,}_{1}(S^{1})=\mathbb{Z}$, it follows that
$M(\tau,x,z)$ supports vortex lines 
that bind zero modes in (2+1)-dimensional space and time and thus prevent 
the gapping of the boundary states.~%
\footnote{
Observe that $\pi^{\,}_{1}(S^1)=\mathbb{Z}$
whereas $\pi^{\,}_{1}(R^{\,}_{0})=\mathbb{Z}^{\,}_{2}$.
This discrepancy arises because we enter the
stable homotopy group $\pi^{\,}_{D}(R^{\,}_{0})=\mathbb{Z}^{\,}_{2}$
by taking the limit $R^{\,}_{0}:=\lim_{\nu\to\infty}V^{\,}_{\nu}$ 
in the second column of Table~\ref{tab: 3D DIII}.
         }

\textit{Case $\nu=4$:}
We use the representation
$\openone=\sigma^{\,}_{0}\tensor\sigma^{\prime}_{0}$.
The $4\times4$ real-valued and symmetric matrix $M(\tau,x,z)$ is a linear
combination with real coefficients of 
$X^{\,}_{\sigma^{\,}_{\mu}\sigma^{\prime}_{\mu'}}\equiv
\sigma^{\,}_{\mu}\tensor\sigma^{\prime}_{\mu'}$
with $\mu,\mu'=0,1,2,3$ such that either none or two
of $\mu$ and $\mu'$ equal the number 2.
Of these the three matrices 
$X^{\,}_{13}$,
$X^{\,}_{33}$,
and
$X^{\,}_{01}$
anticommute pairwise.
If $M(\tau,x,z)$ is a linear combinations with real-valued coefficients
of these three matrices and if $M$ is
normalized by demanding that it squares to $X^{\,}_{00}$,
then the set spanned by $M(\tau,x,z)$ is homeomorphic to 
the two-sphere $S^{2}$.
As $\pi^{\,}_{2}(S^{2})=\mathbb{Z}$, $M(\tau,x,z)$ supports
point-like defects of the monopole type that bind zero modes 
in (2+1)-dimensional space and time and thus prevent 
the gapping of the boundary states.~%
\footnote{
Observe that $\pi^{\,}_{2}(S^2)=\mathbb{Z}$
whereas $\pi^{\,}_{2}(R^{\,}_{0})=\mathbb{Z}^{\,}_{2}$.
This discrepancy arises because we enter the
stable homotopy group $\pi^{\,}_{D}(R^{\,}_{0})=\mathbb{Z}^{\,}_{2}$
by taking the limit $R^{\,}_{0}:=\lim_{\nu\to\infty}V^{\,}_{\nu}$ 
in the second column of Table~\ref{tab: 3D DIII}.
         }

\textit{Case $\nu=8$:}
We use the representation
$\openone=\sigma^{\,}_{0}\tensor\sigma^{\prime}_{0}\tensor\sigma^{\prime\prime}_{0}$.
The $8\times8$ real-valued and symmetric matrix $M(\tau,x,z)$ 
is a linear combination with real-valued coefficients of the matrices
$X^{\,}_{\mu\mu'\mu''}\equiv
\sigma^{\,}_{\mu}\otimes
\sigma^{\prime}_{\mu'}\otimes
\sigma^{\prime\prime}_{\mu''}$
where either none or two of $\mu,\mu',\mu''=0,1,2,3$
equal the number 2. Of these, one finds the five pairwise anticommuting
matrices 
$X^{\,}_{333}$,
$X^{\,}_{133}$,
$X^{\,}_{013}$,
$X^{\,}_{001}$,
and
$X^{\,}_{212}$.
If $M(\tau,x,z)$ is 
a linear combination with real-valued coefficients of these five matrices
and if $M$ is normalized by demanding that it squares to $X^{\,}_{000}$,
then the set spanned by $M(\tau,x,z)$ is homeomorphic to 
the four-sphere $S^{4}$.
As $\pi^{\,}_{4}(S^{4})=\mathbb{Z}$, it is possible to add
a topological term to the QNLSM on the boundary that is of the WZ type.
This term is conjectured to prevent the gapping of the boundary states.

\textit{Case $\nu=16$:}
We use the representation
$\openone=
\sigma^{\,}_{0}\tensor
\sigma^{\prime}_{0}\tensor
\sigma^{\prime\prime}_{0}\tensor
\sigma^{\prime\prime\prime}_{0}$.
The $16\times16$ real-valued and symmetric matrix $M(\tau,x,z)$ 
is a linear combination with real-valued coefficients of the matrices
$X^{\,}_{\mu\mu'\mu''\mu'''}=
\sigma^{\,}_{\mu}\otimes
\sigma^{\prime}_{\mu'}\otimes
\sigma^{\prime\prime}_{\mu''}\otimes
\sigma^{\prime\prime\prime}_{\mu'''}
$
where none, two, or four of 
$\mu,\mu',\mu'',\mu'''=0,1,2,3$
equal the number 2. 
Of these, one finds the nine pairwise anticommuting matrices 
$X^{\,}_{2222}$, $X^{\,}_{0122}$, $X^{\,}_{0322}$, $X^{\,}_{2012}$, $X^{\,}_{2032}$,
$X^{\,}_{1202}$, $X^{\,}_{3202}$, $X^{\,}_{0001}$, and $X^{\,}_{0003}$.
If $M(\tau,x,z)$ is 
a linear combination with real-valued coefficients of these nine matrices
and if $M(\tau,x,z)$ is normalized by demanding that it squares to 
$X^{\,}_{0000}$, 
then the set spanned by $M(\tau,x,z)$ is homeomorphic to 
the eight-sphere $S^{8}$.
It is then impossible to add a topological term to the QNLSM on the boundary.
The boundary zero modes can be gapped out.

We conclude that the effects of interactions on the three-dimensional
SPT phases in the symmetry class DIII are
to reduce the topological classification $\mathbb{Z}$
in the noninteracting limit down to $\mathbb{Z}^{\,}_{16}$.
The logic used to reach this conclusion
is summarized by Table \ref{tab: 3D DIII} once the line corresponding to
$\nu=1$ has been identified. It is given by the smallest $D$ that
accommodates a non-trivial entry for the corresponding homotopy group.
The line for $\nu=2$ is then identified with the next smallest $D$
with $\pi^{\,}_{D}(R^{\,}_{2})\ne0$, and so on.

\subsubsection{The symmetry class CI when $d=3$}

Let $X^{\,}_{\mu\nu\lambda}\equiv
\tau^{\,}_{\mu}\otimes\rho^{\,}_{\nu}\otimes\sigma^{\,}_{\lambda}$ 
with $\mu,\nu,\lambda=0,1,2,3$.
Consider the three-dimensional bulk single-particle Dirac Hamiltonian
(with Dirac matrices of dimension $r=8\equiv r^{\,}_{\mathrm{min}}$),
\begin{subequations}
\label{eq: def 3D H(0) CI} 
\begin{equation}
\mathcal{H}^{(0)}(\bm{x}):=
-\mathrm{i}\partial^{\,}_{1}
X^{\,}_{310}
-\mathrm{i}\partial^{\,}_{2}\,
X^{\,}_{020} 
-\mathrm{i}\partial^{\,}_{3}\,
X^{\,}_{110}
+
m(\bm{x})\,
X^{\,}_{030}.
\label{eq: def 3D H(0) CI a} 
\end{equation}
This single-particle Hamiltonian belongs to the 
three-dimensional symmetry class CI, for
\begin{align}
&
\mathcal{T}\,
\mathcal{H}^{(0)}(\bm{x})\,
\mathcal{T}^{-1}
=
+
\mathcal{H}^{(0)}(\bm{x}),
\label{eq: def 3D H(0) CI b} 
\\
&
\mathcal{C}\,
\mathcal{H}^{(0)}(\bm{x})\,
\mathcal{C}^{-1}
=
-
\mathcal{H}^{(0)}(\bm{x}),
\label{eq: def 3D H(0) CI c} 
\end{align}
where
\begin{align}
\mathcal{T}&:=
X^{\,}_{202}\,\mathsf{K},
&
\mathcal{C}&:=
\mathrm{i}
X^{\,}_{012}\,\mathsf{K}.
\label{eq: def 3D H(0) CI d} 
\end{align}
\end{subequations}
The multiplicative factor $\mathrm{i}$ in the definition of
$\mathcal{C}$ is needed for $\mathcal{T}$ 
to commute with $\mathcal{C}$.

The single-particle Hamiltonian
(\ref{eq: def 3D H(0) CI a}) 
is the direct product of the single-particle Hamiltonian
(\ref{eq: def 3D H(0) DIII a}) 
with the unit $2\times2$ matrix $\sigma^{\,}_{0}$.
If we interpret the degrees of freedom encoded by $\sigma^{\,}_{0}$
and the Pauli matrices $\bm{\sigma}$ as carrying spin-1/2 degrees of freedom,
we may then interpret Eqs.\ (\ref{eq: def 3D H(0) CI}) as defining
a spin-singlet superconductor that preserves TRS.
 
The Dirac mass matrix 
$X^{\,}_{030}$
is here the only one allowed for dimension eight Dirac matrices
under the constraints (\ref{eq: def 3D H(0) CI b})
and (\ref{eq: def 3D H(0) CI c}).
Consequently,
the domain wall
\begin{subequations}
\label{eq: def domain wall d=3 CI}
\begin{equation}
m(\bm{x})\equiv m(y):=
m^{\,}_{\infty}\,
\mathrm{sgn}(y),
\qquad
m^{\,}_{\infty}\in\mathbb{R},
\label{eq: def domain wall d=3 CI a}
\end{equation}
at $y=0$, binds the zero mode
\begin{equation}
e^{
-\mathrm{i}
X^{\,}_{020}\,X^{\,}_{030}\,
\int\limits_{0}^{y}\mathrm{d}y'\,m(y')
  }\,\chi
=
e^{
-|m^{\,}_{\infty}\,y|
  }
\chi,
\label{eq: def domain wall d=3 CI b}
\end{equation}
where
\begin{equation}
X^{\,}_{010}\,
\chi=
-\mathrm{sgn}(m^{\,}_{\infty})\,
\chi
\label{eq: def domain wall d=3 CI c}
\end{equation}
\end{subequations}
with $\chi$ independent of $x$ and $z$.
The kinetics of the gapless boundary states is
governed by the Dirac Hamiltonian
\begin{equation}
\mathcal{H}^{(0)}_\mathrm{bd}(x,z)=
-\mathrm{i}\partial^{\,}_{x}\tau_3^{\,}\otimes\sigma_0^{\,}
-\mathrm{i}\partial^{\,}_{z}\tau_1^{\,}\otimes\sigma_0^{\,},
\end{equation}
where we have chosen $m^{\,}_{\infty}<0$.

On the boundary, the operations for reversal of time and charge conjugation
are now represented by
\begin{subequations}
\label{eq: boundary dyn H d=3 class CI}
\begin{align}
\mathcal{T}^{\,}_{\mathrm{bd}\,\nu}&:=
\tau^{\,}_{2}\otimes\sigma^{\,}_{2}
\otimes
\openone
\mathsf{K},
&
\mathcal{C}^{\,}_{\mathrm{bd}\,\nu}&:=
\mathrm{i}\,
\tau^{\,}_{0}\otimes\sigma^{\,}_{2}
\otimes\openone\mathsf{K},
\end{align}
where $\openone$ is the $\nu\times\nu$ unit matrix.
We seek the single-particle Hamiltonian on the boundary
that encodes the fluctuations arising from the Hubbard-Stratonovich
decoupling of quartic interactions through a generic dynamical mass
that respects the PHS on the boundary.
It is given by
\begin{align}
\mathcal{H}^{(\mathrm{dyn})}_{\mathrm{bd}\,\nu}(\tau,x,z):=&\,
-\mathrm{i}\partial^{\,}_{x}
\tau^{\,}_{3}\otimes\sigma^{\,}_{0}\otimes\openone
-\mathrm{i}\partial^{\,}_{z}
\tau^{\,}_{1}\otimes\sigma^{\,}_{0}\otimes\openone
\nonumber\\
&\,
+\tau^{\,}_{2}\otimes M(\tau,x,z),
\label{eq: boundary dyn H d=3 class CI a}
\end{align}
with the $2\nu\times2\nu$ Hermitian matrix 
\begin{equation}
M(\tau,x,z)=
+
(\sigma^{\,}_{2}\otimes\openone)\,
M^{*}(\tau,x,z)\,
(\sigma^{\,}_{2}\otimes\openone).
\label{eq: boundary dyn H d=3 class CI b}
\end{equation}
\end{subequations}

The space of normalized Dirac mass matrices satisfying the condition
(\ref{eq: boundary dyn H d=3 class CI b})
is topologically equivalent to the space 
\begin{equation}
V^{\,}_{\nu}:=
\bigcup_{k=1}^{\nu} Sp(\nu)/\left[Sp(k)\times Sp(\nu-k)\right]
\end{equation}
for the symmetry class C in two-dimensional space.
The limit $\nu\to\infty$ of these spaces is the classifying space
$R^{\,}_{4}$.
In order to gap out dynamically the boundary zero modes 
without breaking the defining symmetries of the symmetry class CI,
we need to construct a (2+1)-dimensional
QNLSM for the (boundary) dynamical Dirac masses
from the two-dimensional symmetry class C without topological 
obstructions. We construct explicitly the spaces 
for the relevant normalized boundary dynamical Dirac mass matrices
of dimension $\nu=2^{n}$ with $n=0,1,2$ in the following.
The relevant homotopy groups are given in Table%
~\ref{tab: 3D CI}.

\begin{table}[tb]
\caption{
Reduction from $\mathbb{Z}$ 
to
$\mathbb{Z}^{\,}_{4}$ 
for the topologically equivalent classes of the
three-dimensional SPT phases in the symmetry class CI
that arises from interactions. 
We denote by $V^{\,}_{\nu}$ the space of
$\nu\times\nu$ normalized Dirac mass matrices
in boundary ($d=2$) Dirac Hamiltonians
belonging to the symmetry class C. 
The limit $\nu\to\infty$ of these spaces is the classifying space
$R^{\,}_{4}$. 
The second column shows the stable $D$-th homotopy groups 
of the classifying space $R^{\,}_{4}$. 
The third column gives the number $\nu$ of copies of 
boundary (Dirac) fermions for which a topological obstruction is permissible.
The fourth column gives the type of topological obstruction
that prevents the gapping of the boundary (Dirac) fermions.
\label{tab: 3D CI}
        }
\begin{tabular}{ccccccc}
\hline \hline
$D$ 
&\qquad\qquad& 
$\pi^{\,}_{D}(R^{\,}_{4})$ 
& \qquad\qquad& 
$\nu$ 
&\qquad\qquad& 
Topological obstruction 
\\
\hline
0
&\qquad\qquad&   
$\mathbb{Z}$ 
&\qquad\qquad&   
$1$ 
&\qquad\qquad&   
Domain wall  
\\
1
&\qquad\qquad&   
0  
&\qquad\qquad&   
\qquad\qquad&   
\\
2
&\qquad\qquad&   
0  
&\qquad\qquad&
\qquad\qquad&   
\\
3
&\qquad\qquad&   
0  
&\qquad\qquad&   
\qquad\qquad&     
\\
4
&\qquad\qquad&   
$\mathbb{Z}$ 
&\qquad\qquad&   
$2$ 
&\qquad\qquad&   
WZ term  
\\
5
&\qquad\qquad& 
$\mathbb{Z}^{\,}_{2}$  
&\qquad\qquad& 
4
&\qquad\qquad&
None   
\\
6
&\qquad\qquad& 
$\mathbb{Z}^{\,}_{2}$ 
&\qquad\qquad& 
\qquad\qquad&   
\\
7
&\qquad\qquad& 
0  
&\qquad\qquad& 
\qquad\qquad&   
\\
\hline \hline
\end{tabular}
\end{table}

\textit{Case $\nu=1$:} 
There is one $2\times2$ Hermitian matrix $M(\tau,x,z)$
on the boundary that
is proportional to $\sigma^{\,}_{0}$.
A domain wall in imaginary time such as
$m^{\,}_{2\,\infty}\,\mathrm{sign}(\tau)\,\tau^{\,}_{2}\otimes\sigma^{\,}_{0}$ 
prevents the dynamical generation of a spectral gap on the boundary.

\textit{Case $\nu=2$:}
We use the representation $\openone=\sigma^{\prime}_{0}$.
The Hermitian $4\times4$ matrix $M(\tau,x,z)$ is a linear combination with 
real-valued coefficients of the matrices
$X^{\,}_{\mu\mu'}\equiv\sigma^{\,}_{\mu}\otimes\sigma^{\prime}_{\mu'}$
with $\mu,\mu'=0,1,2,3$ such that
$X^{\,}_{20}\,X^{*}_{\mu\mu'}\,X^{\,}_{20}=+X^{\,}_{\mu\mu'}$.
Of these, one finds the five matrices
$X^{\,}_{12}$,
$X^{\,}_{22}$,
$X^{\,}_{32}$,
$X^{\,}_{01}$,
and
$X^{\,}_{03}$
that anticommute pairwise.
If $M(\tau,x,z)$ is a linear combinations with real-valued coefficients
of these five matrices and if $M(\tau,x,z)$ is normalized by demanding that
it squares to $X^{\,}_{00}$, then the set spanned by $M(\tau,x,z)$ is
homeomorphic to $S^{4}$.
As $\pi^{\,}_{2+1+1}(S^{4})=\mathbb{Z}$, it is then possible to add
a topological term to the (2+1)-dimensional QNLSM on the boundary 
that is of the WZ type.
This term is conjectured to prevent the gapping of the boundary states.

\textit{Case $\nu=4$:}
We use the representation 
$\openone=\sigma^{\prime}_{0}\otimes\sigma^{\prime\prime}_{0}$.
The Hermitian $8\times8$ matrix $M(\tau,x,z)$ is a linear combination with 
real-valued coefficients of the matrices
$X^{\,}_{\mu\mu'\mu''}\equiv
\sigma^{\,}_{\mu}\otimes\sigma^{\prime}_{\mu'}\otimes\sigma^{\prime\prime}_{\mu''}$
with $\mu,\mu',\mu''=0,1,2,3$ such that
$X^{\,}_{200}\,X^{*}_{\mu\mu'\mu''}\,X^{\,}_{200}=+X^{\,}_{\mu\mu'\mu''}$.
Of these, one finds the six matrices
$X^{\,}_{120}$,
$X^{\,}_{220}$,
$X^{\,}_{320}$,
$X^{\,}_{010}$,
$X^{\,}_{031}$
and
$X^{\,}_{033}$
that anticommute pairwise.
If $M(\tau,x,z)$ is a linear combinations with real-valued coefficients
of these six matrices and if $M(\tau,x,z)$ is normalized by demanding that
it squares to $X^{\,}_{000}$, then the set spanned by $M(\tau,x,z)$ is
homeomorphic to $S^{5}$. It is then impossible to add
a topological term to the (2+1)-dimensional QNLSM on the boundary.
The boundary zero modes can be gapped out.

We conclude that the effects of interactions on the three-dimensional
SPT phases in the symmetry class CI are to reduce
the topological classification $\mathbb{Z}$
in the noninteracting limit down to $\mathbb{Z}^{\,}_{4}$.
The logic used to reach this conclusion
is summarized by Table \ref{tab: 3D CI} once the line corresponding to
$\nu=1$ has been identified. It is given by the smallest $D$ that
accommodates a non-trivial entry for the corresponding homotopy group.

\subsubsection{The symmetry class AIII when $d=3$
\label{subsubsec: The symmetry class AIII when d=3}
}

By omitting the contributions arising from the gauge potentials,
the single-particle Hamiltonian 
(\ref{eq: def 3D H(0) DIII a})
does not specify uniquely the symmetry class.
For example, the single-particle Hamiltonian 
(\ref{eq: def 3D H(0) DIII a})
can also be interpreted as an insulator belonging to the
symmetry class AIII, for it anticommutes with the composition 
\begin{equation}
\Gamma^{\,}_{5}:=
-\mathrm{i}\mathcal{T}\,\mathcal{C}
=
X^{\,}_{21}
\end{equation}
of the operations 
$\mathcal{T}$ and $\mathcal{C}$
for time reversal and charge conjugation, respectively,
defined in Eq.\
(\ref{eq: def 3D H(0) DIII d}).

The direct product of the single-particle Hamiltonian 
(\ref{eq: def 3D H(0) DIII a})
with the $\nu\times\nu$ unit matrix $\openone$
supports $\nu$ zero modes bound to the boundary $y=0$,
for they are annihilated
by the boundary Hamiltonian
\begin{subequations}
\begin{equation}
\mathcal{H}^{(0)}_{\mathrm{bd}\,\nu}(x,z):=
-\mathrm{i}\partial^{\,}_{x}\,
\tau^{\,}_{3}\otimes\openone
-
\mathrm{i}\partial^{\,}_{z}\,
\tau^{\,}_{1}\otimes\openone  
\end{equation}
that anticommutes with
\begin{equation}
\Gamma^{(\mathrm{bd})}_{5}:=
\tau^{\,}_{2}\otimes\openone.
\end{equation}
\end{subequations}

The fate of these zero modes in the presence of fermion-fermion interactions
is investigated in two steps.
First, we include the effects of interactions by perturbing
the boundary Hamiltonian with all boundary dynamical Dirac masses from the
symmetry class A. Second, we introduce a BdG (Nambu)
grading to account for the interactions-driven superconducting fluctuations
by perturbing the boundary Hamiltonian 
$\mathcal{H}^{(0)}_{\mathrm{bd}\,\nu}(x,z)\otimes\rho^{\,}_{0}$
with all boundary dynamical Dirac masses that anticommute with
$\tau^{\,}_{0}\otimes\openone\otimes\rho^{\,}_{1}\,\mathsf{K}$,
i.e., with all boundary dynamical Dirac masses from the symmetry class D.
In the first step, the boundary dynamical single-particle Hamiltonian is
\begin{subequations}
\begin{align}
\mathcal{H}^{(\mathrm{dyn})}_{\mathrm{bd}\,\nu}(\tau,x,z):=&\,
\left(
-\mathrm{i}\partial^{\,}_{x}\,
\tau^{\,}_{3}
-
\mathrm{i}\partial^{\,}_{z}\,
\tau^{\,}_{1}
\right)
\otimes\openone  
\nonumber\\
&\,
+
\tau^{\,}_{2}\otimes M(\tau,x,z),
\label{eq: boundary dyn H d=3 classAIII a}
\end{align}
with the $\nu\times\nu$ Hermitian matrix 
\begin{equation}
M(\tau,x,z)=M^{\dag}(\tau,x,z).
\label{eq: boundary dyn H d=3 class AIII b}
\end{equation}
\end{subequations}
In the second step, the boundary dynamical single-particle Hamiltonian is
\begin{subequations}
\begin{align}
\mathcal{H}^{(\mathrm{dyn})}_{\mathrm{bd}\,\nu}(\tau,x,z):=&\,
\left(
-\mathrm{i}\partial^{\,}_{x}\,
\tau^{\,}_{3}
-
\mathrm{i}\partial^{\,}_{z}\,
\tau^{\,}_{1}
\right)
\otimes\left(\rho^{\,}_{0}\otimes\openone\right) 
\nonumber\\
&\,
+
\tau^{\,}_{2}\otimes M(\tau,x,z),
\label{eq: boundary dyn H d=3 class AIII bis a}
\end{align}
with the $2\nu\times2\nu$ Hermitian matrix 
\begin{equation}
M(\tau,x,z)=
+
(\rho^{\,}_{1}\otimes\openone)\,
M^{*}(\tau,x,z)\,
(\rho^{\,}_{1}\otimes\openone).
\label{eq: boundary dyn H d=3 class AIII bis b}
\end{equation}
\end{subequations}

The space of boundary dynamical Dirac mass matrices of the form
(\ref{eq: boundary dyn H d=3 class AIII b})
that square to the unit matrix
is homeomorphic to the classifying space $C^{\,}_{0}$
for the symmetry class A in two-dimensional space and
in the limit $\nu\to\infty$.
In order to gap out dynamically the boundary zero modes 
without breaking the defining symmetries of the symmetry class AIII,
we need to construct a (2+1)-dimensional
QNLSM for the (boundary) dynamical Dirac masses
from the two-dimensional symmetry class A without topological 
obstructions. When $\nu=1$, a domain wall such as
$M(\tau,x,z)=M^{\,}_{\infty}\,\mathrm{sgn}(\tau)$
prevents the gapping of the boundary zero modes. 
When $\nu=2$, we choose the representation $\openone=\sigma^{\,}_{0}$.
The set spanned by 
$M(\tau,x,z)= 
\sum^{3}_{j=1}m^{\,}_{j}(\tau,x,z)\,\sigma^{\,}_{j}$
with the real-valued functions
$m^{\,}_{j}(\tau,x,z)$
obeying the normalization condition
$\sum^{3}_{j=1}m^{2}_{j}(\tau,x,z)=1$ supports a monopole
that binds zero modes in (2+1)-dimensional space and time, 
as $\pi^{\,}_{2}(S^{2})=\mathbb{Z}$.
When $\nu=4$, we choose the representation $\openone=X^{\,}_{00}$
where $X^{\,}_{\mu\mu'}:=\sigma^{\,}_{\mu}\otimes\sigma^{\prime}_{\mu'}$
for $\mu,\mu'=0,1,2,3$. We may then write
$M(\tau,x,z)= 
\sum^{3}_{\mu,\mu'=0}m^{\,}_{\mu\mu'}(\tau,x,z)\,X^{\,}_{\mu\mu'}
$. Any $X^{\,}_{\mu\mu'}$ other than the unit matrix $X^{\,}_{00}$
belongs to a multiplet of five pairwise anticommuting matrices of
the form $X^{\,}_{\nu\nu'}\neq X^{\,}_{00}$. Hence, we may always construct
a set of normalized $M(\tau,x,z)$ homeomorphic to $S^{4}$. Since
$\pi^{\,}_{2+1+1}(S^{4})=\mathbb{Z}$, 
it is possible to augment the 
corresponding boundary QNLSM in (2+1)-dimensional space and time 
by a WZ term that modifies the equations of motion in a local way.
This term is conjectured to prevent the gapping of the boundary states.
When $\nu=2^{n}$ with $n\geq3$, 
we choose the representation $\openone=X^{\,}_{00\cdots}$
where
$X^{\,}_{\mu\mu'\cdots}:=\sigma^{\,}_{\mu}\otimes\sigma^{\prime}_{\mu'}\otimes\cdots$
for $\mu,\mu',\cdots=0,1,2,3$. Any $X^{\,}_{\mu\mu'\cdots}$ 
other than the unit matrix $X^{\,}_{00\cdots}$
belongs to a multiplet of no less than 
seven pairwise anticommuting matrices.
It is for this reason that the boundary states are then necessarily gapped, 
for it is not permissible to add a topological term to the action of 
the boundary QNLSM for a sphere of dimension larger than four. 
We conclude that the effects of interactions in the three-dimensional
SPT phases in the
symmetry class AIII is to reduce the topological classification $\mathbb{Z}$
in the noninteracting limit down to $\mathbb{Z}^{\,}_{8}$ under the
assumption that only fermion-number-conserving 
interacting channels are included in the 
stability analysis. The logic used to reach this conclusion
is summarized by Table \ref{tab: 3D AIII} once the line corresponding to
$\nu=1$ has been identified. It is given by the smallest $D$ that
accommodates a non-trivial entry for the corresponding homotopy group.

\begin{table}[tb]
\caption{
Reduction from $\mathbb{Z}$ 
to
$\mathbb{Z}^{\,}_{8}$ 
for the topologically equivalent classes of the
three-dimensional SPT phases in the symmetry classes AIII
that arises from 
the fermion-number-conserving interacting channels. 
We denote by $V^{\,}_{\nu}$ the space of $\nu\times\nu$
normalized Dirac mass matrices in boundary ($d=2$) Dirac Hamiltonians
belonging to the symmetry class A. 
The limit $\nu\to\infty$ of these spaces is the classifying spaces
$C^{\,}_{0}$. 
The second column shows the stable $D$-th homotopy groups 
of the classifying space $C^{\,}_{0}$. 
The third column gives the number $\nu$ of copies of 
boundary (Dirac) fermions for which a topological obstruction is permissible.
The fourth column gives the type of topological obstruction
that prevents the gapping of the boundary (Dirac) fermions.
\label{tab: 3D AIII}
        }
\begin{tabular}{ccccccc}
\hline \hline
$D$ 
&\qquad\qquad& 
$
\pi^{\,}_{D}(C^{\,}_{0})$
&\qquad\qquad& 
$\nu$ 
&\qquad\qquad&
Topological obstruction 
\\
\hline
0&&$\mathbb{Z}$ && $1$  && Domain wall \\
1&&0            &&      &&             \\
2&&$\mathbb{Z}$ && $2$  && Monopole    \\
3&&0            &&      &&             \\
4&&$\mathbb{Z}$ && $4$  &&  WZ term    \\
5&&0            &&      &&             \\
6&&$\mathbb{Z}$ && $8$  &&  None       \\
7&&0            &&      &&             \\
\hline \hline
\end{tabular}
\end{table}

The space of boundary dynamical matrices that satisfy the condition
(\ref{eq: boundary dyn H d=3 class AIII bis b})
and square to the unit matrix
is homeomorphic to the classifying space $R^{\,}_{2}$
for the symmetry class D in two-dimensional space and
in the limit $\nu\to\infty$. Because the dimension of the
boundary dynamical matrix
(\ref{eq: boundary dyn H d=3 class AIII bis b})
is twice that of the boundary dynamical matrix 
(\ref{eq: boundary dyn H d=3 class AIII b}),
one might have guessed that the gapping of the boundary
zero modes takes place for a value of $\nu$
smaller than eight. This is not so, however, because of two 
constraints. The first constraint is that of PHS.
The second constraint restricts the target space for the boundary QNLSM
that is built out of the boundary dynamical Dirac masses. 
The target space of the QNLSM must be invariant as a set under the action 
of a global gauge $U(1)$ transformation that is generated by 
$\tau^{\,}_{0}\otimes\rho^{\,}_{3}\otimes\openone$.
This global $U(1)$ symmetry implements conservation of the fermion number.
Indeed, one verifies
the following facts. When $\nu=1$, 
the boundary dynamical matrix $M(\tau,x,z)$
is a linear combination of $\rho^{\,}_{1}$ and $\rho^{\,}_{2}$
with real-valued functions as coefficients.
Hence, the space of normalized boundary dynamical Dirac mass matrices
is homeomorphic to $S^{1}$ 
and invariant as a set under any global gauge $U(1)$ transformation
when $\nu=1$. 
Because of $\pi^{\,}_{1}(S^{1})=\mathbb{Z}$,
vortex lines bind zero modes in (2+1)-dimensional space and time.
When $\nu=2$, we represent the unit $4\times4$ matrix by 
$\rho^{\,}_{0}\otimes\sigma^{\,}_{0}$ and 
we expand any $4\times4$ Hermitian matrix
as a linear combination with real-valued functions as coefficients of
the sixteen matrices 
$X^{\,}_{\mu\mu'}=\rho^{\,}_{\mu}\otimes\sigma^{\,}_{\mu'}$
with $\mu,\mu'=0,1,2,3$.
The boundary dynamical matrix
$M(\tau,x,z)$ 
is a linear combination with real-valued functions as coefficients
of the ten matrices 
$X^{\,}_{00}$,
$X^{\,}_{01}$,
$X^{\,}_{03}$,
$X^{\,}_{10}$,
$X^{\,}_{11}$,
$X^{\,}_{13}$,
$X^{\,}_{20}$,
$X^{\,}_{21}$,
$X^{\,}_{23}$,
$X^{\,}_{32}$
that satisfy the constraint
$X^{\,}_{\mu\mu'}=+X^{\,}_{10}\,X^{*}_{\mu\mu'}\,X^{\,}_{10}$.
Other than the unit matrix $X^{\,}_{00}$,
any one of these nine matrices belongs to a triplet
of pairwise anticommuting matrices. However, not all such triplets
are closed under the global $U(1)$ transformation
defined by multiplication from the left and right (conjugation) 
with $X^{\,}_{30}$. However, there exists a triplet
that is closed under conjugation by $X^{\,}_{30}$.
For example, each element from
the triplet of pairwise anticommuting matrices
$X^{\,}_{01}$,
$X^{\,}_{03}$,
$X^{\,}_{32}$
is invariant under conjugation with $X^{\,}_{30}$.
Moreover, no other matrix, satisfying the condition
$X^{\,}_{\mu\mu'}=+X^{\,}_{10}\,X^{*}_{\mu\mu'}\,X^{\,}_{10}$,
anticommutes with this triplet.
Hence, this triplet spans a set of normalized boundary dynamical Dirac masses
that is homeomorphic to $S^{2}$, each point of which is invariant under
the global $U(1)$ transformation associated with the conservation 
of the fermion number.
Because of $\pi^{\,}_{2}(S^{2})=\mathbb{Z}$,
monopoles bind zero modes in (2+1)-dimensional space and time
that prevent the gapping of the boundary states when $\nu=2$.
When $\nu=4$, we represent the unit $8\times8$ matrix by
$\rho^{\,}_{0}\otimes\sigma^{\,}_{0}\otimes\sigma^{\prime}_{0}$
and we expand any $8\times8$ matrix as a linear combination with
real-valued coefficients of the sixty-four matrices
$X^{\,}_{\mu\mu'\mu''}=
\rho^{\,}_{\mu}\otimes\sigma^{\,}_{\mu'}\otimes\sigma^{\prime}_{\mu''}$
with $\mu,\mu',\mu''=0,1,2,3$. The boundary dynamical matrix
$M(\tau,x,z)$ is a linear combination 
with real-valued functions as coefficients of those matrices 
$X^{\,}_{\mu\mu'\mu''}=+X^{\,}_{100}\,X^{*}_{\mu\mu'\mu''}\,X^{\,}_{100}$.
Other than the unit matrix $X^{\,}_{000}$,
any one of those matrices belong to a quintuplet
of pairwise anticommuting matrices. 
Among these, each element from the quintuplet 
$X^{\,}_{001}$,
$X^{\,}_{003}$,
$X^{\,}_{312}$,
$X^{\,}_{022}$,
$X^{\,}_{332}$
is invariant under conjugation by  $X^{\,}_{300}$.
Moreover, no other matrix, satisfying the condition
$X^{\,}_{\mu\mu'\mu''}=+X^{\,}_{100}\,X^{*}_{\mu\mu'\mu''}\,X^{\,}_{100}$,
anticommutes with this quintuplet. Hence, this quintuplet 
spans a set of normalized boundary dynamical Dirac masses
that is homeomorphic to $S^{4}$, each point of which is invariant under
the global $U(1)$ transformation associated with the conservation 
of the fermion number.
Because of $\pi^{\,}_{2+1+1}(S^{4})=\mathbb{Z}$,
it is possible to augment the corresponding
boundary QNLSM in (2+1)-dimensional space and time 
by a WZ term that modifies the equations of motion in a local way.
This term is conjectured to prevent the gapping of the boundary states.
When $\nu=2^{n-1}$ with $n\geq4$, any permissible matrix
$X^{\,}_{\mu\mu'\mu'''\cdots}=
+X^{\,}_{100\cdots}\,X^{*}_{\mu\mu'\mu'''\cdots}\,X^{\,}_{100\cdots}$
belongs to a $N(\nu)$-tuplet
of pairwise anticommuting permissible matrices with $N(\nu)>5$.~%
\footnote{
In general, $N(\nu)-1$ is determined from 
Table \ref{tab: 1-11D DIII}(a) 
by shifting the entries of $\nu$ downward by one non-trivial homotopy
group entry. 
For example, $N(4)-1=4$, $N(8)-1=8$, $N(16)-1=9$, $N(32)-1=10$, and so on.
         }
However, not all such $N(\nu)$-tuplets
are closed under the global $U(1)$ transformation
defined by conjugation with $X^{\,}_{300\cdots}$. 
The $N(\nu)$-tuplet that contains 
the pair of anticommuting matrices
$X^{\,}_{00\cdots01}=+X^{\,}_{10\cdots00}\,X^{*}_{00\cdots01}\,X^{\,}_{10\cdots00}$
and
$X^{\,}_{00\cdots03}=+X^{\,}_{10\cdots00}\,X^{*}_{00\cdots03}\,X^{\,}_{10\cdots00}$
has the particularity that each of its elements is invariant under
conjugation with $X^{\,}_{300\cdots}$ and cannot be augmented by one
more anticommuting $X^{\,}_{\mu\mu'\mu'''\cdots}=
+X^{\,}_{100\cdots}\,X^{*}_{\mu\mu'\mu'''\cdots}\,X^{\,}_{100\cdots}$.
Hence, this $N(\nu)$-tuplet
spans a set of normalized boundary dynamical Dirac masses
that is homeomorphic to $S^{N(\nu)-1}$, each point of which is invariant under
the global $U(1)$ transformation associated with the conservation 
of the fermion number. 
Since $N(\nu)>5$ for $\nu=2^{n-1}$ with $n\geq4$,
it follows that all homotopy groups 
of order less than four for the space of the normalized boundary dynamical 
Dirac masses  that are invariant under the global $U(1)$ transformation
are vanishing. The boundary states are then necessarily gapped. 

We conclude that the effects of interactions 
on the three-dimensional SPT phases in the symmetry class AIII are
to reduce the topological classification $\mathbb{Z}$
in the noninteracting limit down to $\mathbb{Z}^{\,}_{8}$.

\subsubsection{The symmetry class AII when $d=3$}
\label{subsubsec: The symmetry class AII when d=3}

We close the discussion of the stability to fermion-fermion interactions
of strong noninteracting topological insulators or superconductors 
in three-dimensional space by illustrating why the
$\mathbb{Z}^{\,}_{2}$ noninteracting classification is stable.

\begin{table}[tb]
\caption{
Stability to fermion-fermion interactions
of the noninteracting topological classification
$\mathbb{Z}^{\,}_{2}$ for three-dimensional strong topological
insulators belonging to the symmetry classes AII.
We denote by $V^{\,}_{\nu}$ the space of $\nu\times\nu$
normalized Dirac mass matrices in boundary ($d=2$) Dirac Hamiltonians
belonging to the symmetry class A. 
The limit $\nu\to\infty$ of these spaces is the classifying space
$C^{\,}_{0}$.  
The second column shows the stable $D$-th homotopy groups 
of the classifying space $C^{\,}_{0}$. 
The third column gives the number $\nu$ of copies of 
boundary (Dirac) fermions for which a topological obstruction is permissible.
The fourth column gives the type of topological obstruction
that prevents the gapping of the boundary (Dirac) fermions.     
\label{tab: 3D AII}
        }
\begin{tabular}{ccccccc}
\hline \hline
$D$ 
&\qquad\qquad& 
$\pi^{\,}_{D}(C^{\,}_{0})$ 
&\qquad\qquad& 
$\nu$ 
&\qquad\qquad& 
Topological obstruction 
\\
\hline
0
&\qquad\qquad& 
$\mathbb{Z}$ 
&\qquad\qquad& 
$1$ 
&\qquad\qquad&  
Domain wall  
\\
1
&\qquad\qquad& 
0   
&\qquad\qquad&
&\qquad\qquad&
\\
\hline \hline
\end{tabular}
\end{table}

To this end, consider the single-particle bulk Dirac Hamiltonian
\begin{subequations}
\label{eq: def mathcal H 0 AII d=3}
\begin{equation}
\mathcal{H}^{(0)}(\bm{x}):=
-\mathrm{i}\partial^{\,}_{x}\,
X^{\,}_{21}
-\mathrm{i}\partial^{\,}_{y}\,
X^{\,}_{11}
-\mathrm{i}\partial^{\,}_{z}\,
X^{\,}_{02}
+
m(\bm{x})\,
X^{\,}_{03},
\label{eq: def mathcal H 0 AII d=3 a}
\end{equation}
where $X^{\,}_{\mu\nu}:=\sigma^{\,}_{\mu}\otimes\tau^{\,}_{\mu'}$
for $\mu,\mu'=0,1,2,3$. Because
\begin{equation}
\mathcal{H}^{(0)}(\bm{x})=
+
\mathcal{T}\,
\mathcal{H}^{(0)}(\bm{x})\,
\mathcal{T}^{-1},
\qquad
\mathcal{T}:=\mathrm{i}X^{\,}_{20}\,\mathsf{K},
\label{eq: def mathcal H 0 AII d=3 b}
\end{equation}
\end{subequations}
we interpret this Hamiltonian as realizing a noninteracting
topological insulator in the three-dimensional symmetry class AII. 
The domain wall in the mass
\begin{subequations}
\label{eq: def mathcal H 0 AII d=3 domain wall}
\begin{equation}
m(x,y,z)=
m^{\,}_{\infty}\,\mathrm{sgn}(z)
\label{eq: def mathcal H 0 AII d=3 domain wall a}
\end{equation}
binds a zero mode to the boundary $z=0$ that is annihilated by
the boundary single-particle Hamiltonian
\begin{align}
\mathcal{H}^{(0)}_{\mathrm{bd}}(x,y)=&\,
-\mathrm{i}\partial^{\,}_{x}\,
\sigma^{\,}_{2}
-\mathrm{i}\partial^{\,}_{y}\,
\sigma^{\,}_{1}
\nonumber\\
=&\,
\mathcal{T}^{\,}_{\mathrm{bd}}\,
\mathcal{H}^{(0)}_{\mathrm{bd}}(x,y)\,
\mathcal{T}^{-1}_{\mathrm{bd}},
\label{eq: def mathcal H 0 AII d=3 domain wall b}
\end{align}
where
\begin{equation}
\mathcal{T}^{\,}_{\mathrm{bd}}:=
\mathrm{i}\sigma^{\,}_{2}\,\mathsf{K}.
\label{eq: def mathcal H 0 AII d=3 domain wall c}
\end{equation}
\end{subequations}
The boundary dynamical Dirac Hamiltonian
\begin{equation}
\mathcal{H}^{(\mathrm{dyn})}_{\mathrm{bd}}(\tau,x,y)=
-\mathrm{i}\partial^{\,}_{x}\,
\sigma^{\,}_{2}
-\mathrm{i}\partial^{\,}_{y}\,
\sigma^{\,}_{1}
+
M(\tau,x,y)\,
\sigma^{\,}_{3}
\label{eq: def mathcal H 0 AII d=3 dyn bd}
\end{equation}
belongs to the symmetry class A,
as the Dirac mass $M\sigma^{\,}_{3}$ breaks TRS unless
$M(-\tau,x,y)=-M(\tau,x,y)$.
The space of normalized boundary dynamical Dirac mass matrices
$\{\pm\sigma^{\,}_{3}\}$ is homeomorphic to the
space of normalized Dirac mass matrices for the two-dimensional system 
in the symmetry class A
\begin{equation}
V^{\,}_{\nu=1}=\bigcup_{k=0}^{1} U(\nu)/\left[U(k)\times U(\nu-k)\right].
\label{eq: def mathcal H 0 AII d=3 dyn bd V nu=1}
\end{equation}
The domain wall in imaginary time 
$M(\tau,x,y)=M^{\,}_{\infty}\,\mathrm{sgn}(\tau)$ 
prevents the gapping of the boundary zero modes.

We conclude that the 
noninteracting topological classification $\mathbb{Z}^{\,}_{2}$
of three-dimensional insulators in the symmetry class AII
is robust to the effects of interactions under the
assumption that only fermion-number-conserving 
interacting channels are included in the 
stability analysis. The logic used to reach this conclusion
is summarized by Table \ref{tab: 3D AII} once the line corresponding to
$\nu=1$ has been identified. It is given by the smallest $D$ that
accommodates a non-trivial entry for the corresponding homotopy group.
Moreover, one verifies 
by introducing a BdG (Nambu) grading that this robustness 
extends to interaction-driven dynamical superconducting fluctuations.

\subsection{Higher dimensions} 
\label{sec: higher dims}

By working out explicitly the effects of fermion-fermion interactions
on the boundary states supported by single-particle
Dirac Hamiltonians representing strong topological insulators and 
superconductors when the dimensionality of space ranges
from $d=1$ to $d=8$, the following rules can be deduced.~%
\footnote{
With the usual caveat that the interactions are strong
on the boundary but not too strong in the bulk.
         } 

\textit{Rule 1:} The $\mathbb{Z}^{\,}_{2}$ topological classification of 
strong topological insulators and superconductors 
is robust to interactions in all dimensions.

\textit{Rule 2:} The $\mathbb{Z}$ topological classification of 
strong topological insulators and superconductors 
is robust to interactions in all even dimensions.

\textit{Rule 3:} The $\mathbb{Z}$ topological classification of 
strong topological insulators and superconductors 
is unstable to interactions in all odd dimensions.
    
We prove Rules 1 and 2 in Secs.\ 
\ref{subsubsec: The case of Z2 classification}
and \ref{subsubsec: The case of even dimensions}, respectively. 
Finally, we work out explicitly the reduction pattern of the noninteracting
$\mathbb{Z}$
topological classification for any odd dimension in Sec.\
\ref{subsubsec: The case of odd dimensions}.

\subsubsection{The case of $\mathbb{Z}_2$ classification}
\label{subsubsec: The case of Z2 classification}

The proof of Rule 1 follows the same logic as 
in the example of the three-dimensional
symmetry class AII in Sec.\ 
\ref{subsubsec: The symmetry class AII when d=3}.
When $d=1$, there are two symmetry classes
with $\pi^{\,}_{0}(V)=\mathbb{Z}^{\,}_{2}$, the symmetry classes D and DIII
(see Table \ref{table: topo classification short-rnaged entangles AZ classes}).
No dynamical Dirac mass is allowed in class D, since there is no 
protecting symmetry to break. For the symmetry class DIII,
the normalized boundary dynamical Dirac masses
are taken from Dirac masses in the symmetry class D and
belong to the classifying space $R^{\,}_{2}$,
according to Table 
\ref{table: topo classification short-rnaged entangles AZ classes}.
According to Table \ref{table: all homotopies classifying spaces}
in Appendix~\ref{app: tenfold way}, 
$\pi^{\,}_{0}(R^{\,}_{2})=\mathbb{Z}^{\,}_{2}$.
Hence, the two noninteracting $\mathbb{Z}^{\,}_{2}$ topological
classification are stable in one-dimension.
To treat the case of $d\ge2$,
let $V$ denote any one of the eight real classifying spaces $V$ and
observe that, according to
Table~\ref{table: all homotopies classifying spaces},
at least one of the homotopy groups $\pi^{\,}_{D}(V)$ 
with $D=0,1,2,3$ is non-trivial. 
We specialize to any one of the two 
symmetry classes in $d$ dimensions
with the classifying space (the space of normalized bulk Dirac masses)
$V$ such that $\pi^{\,}_{0}(V)=\mathbb{Z}^{\,}_{2}$.
By assumption $d+1\geq3$.
Let $V^{\,}_{\mathrm{bd}}$ denote the space 
of the boundary dynamical Dirac masses.
If this space is empty, the $\mathbb{Z}^{\,}_{2}$ classification is stable.
If this space is not empty, then we know that at least one of
$\pi^{\,}_{D}(V^{\,}_{\mathrm{bd}})$ with $D=0,1,\cdots,d+1$
is nonvanishing.
In turn, this implies that at least one
of the homotopy groups from Eq.%
~(\ref{eq: family of homotopy groups from 0 to d+1 if any nu})
is non-trivial.
As the sphere $S^{N(1)-1}$ entering Eq.%
~(\ref{eq: family of homotopy groups from 0 to d+1 if any nu})
is the target space for the QNLSM in $(d-1)+1$ space and time dimensions
obtained from integrating the $\nu=1$ boundary Dirac fermions 
subjected to dynamical masses, the QNLSM accommodates a topological term that
prevents the gapping of the $\nu=1$ boundary zero mode. 
Hence, the two noninteracting $\mathbb{Z}^{\,}_{2}$ topological
classification are stable in any spatial dimension.

\subsubsection{The case of even dimensions}
\label{subsubsec: The case of even dimensions}

Because of Rule 1, we only need to 
consider the symmetry classes in even dimensions
which have $\mathbb{Z}$ topological classification
for gapped noninteracting fermions.
According to Table
\ref{table: topo classification short-rnaged entangles AZ classes}
and the Bott periodicity of two (eight)
for the complex (real) symmetry classes,
these are the symmetry classes 
(i) A for $d=0$ mod $2$,
(ii) AI and AII for $d=4,8$ mod $8$,
(iii) D and C for $d=2,6$ mod $8$.

\textit{Proof for case (i):} 
We start with the complex symmetry class A in even dimensions.
We proceed in two steps. First we rule out dynamical superconducting
fluctuations. We then show that the inclusion of dynamical superconducting
fluctuations is harmless.

Without dynamical superconducting fluctuations, 
the classifying space for the normalized dynamical Dirac masses 
is that for the complex symmetry class A.
Because there is no symmetry that is violated by such dynamical Dirac masses, 
dynamical Dirac masses are forbidden altogether.

With dynamical superconducting fluctuations, 
normalized dynamical Dirac masses
form the space of Dirac masses in the symmetry class D.
The original single-particle Hamiltonian 
$\mathcal{H}^{(0)}_{\nu}$ 
that annihilates $\nu$ zero modes 
is extended to a  BdG (Nambu) single-particle Hamiltonian 
$\mathcal{H}^{(0)}_{\mathrm{BdG}\,\nu}$ 
that commutes with $\rho^{\,}_{3}$ and anticommutes with
$\rho^{\,}_{1}\mathsf{K}$. 
Here, $\rho^{\,}_{0}$ is the unit $2\times2$ matrix 
and $\bm{\rho}$ are the Pauli matrices 
acting on the auxiliary particle-hole degrees of freedom.
Boundary dynamical Dirac masses may then exist. However, they must 
anticommute with $\rho^{\,}_{3}$, 
since no boundary Dirac mass that commutes with $\rho^{\,}_{3}$ 
is allowed after restricting 
$\mathcal{H}^{(0)}_{\mathrm{BdG}\,\nu}$ 
to the boundary.
Upon integrating the boundary Dirac fermions, 
a QNLSM in $(d-1)+1$ space and time dimensions ensues.
The target space of this QNLSM has to be closed under the action of 
a global $U(1)$ gauge transformation. This is to say that
a generic boundary dynamical Dirac mass must be of the form
\begin{subequations}
\begin{equation}
\gamma'=
\left[
\cos(\theta)\rho^{\,}_{1}  
+ 
\sin(\theta)\rho^{\,}_{2}
\right]
\otimes M,
\end{equation}
with the $(r^{\,}_{\mathrm{min}}\,\nu/2)\times(r^{\,}_{\mathrm{min}}\,\nu/2)$ 
matrix $M$ satisfying
\begin{equation}
M=M^{\dag}=-M^{*}.
\end{equation}
\label{eq: dynamical mass in class A with SC fluctuations}
\end{subequations}
We recall that
$r^{\,}_{\mathrm{min}}$ is the minimal rank of the BdG Hamiltonian.
The action on $\gamma'$ of a global $U(1)$ gauge transformation
parametrized by the global phase $\alpha$
is simply the shift $\theta\mapsto\theta+\alpha$.
If so, the target space of the QNLSM is homeomorphic to 
$S^{1}\times V^{\,}_{\mathrm{BdG}\,\nu}$ 
whereby $S^{1}$ is parametrized by $\theta$ and 
$V^{\,}_{\mathrm{BdG}\,\nu}$
is parametrized by $M$ squaring to the unit matrix.
For such a target space, we can always assign a topological term
to account for the vortices supported by the parameter $\theta$
for the $S^{1}$ factor, as $\pi^{\,}_{1}(S^{1})=\mathbb{Z}$. 
These vortices bind $\nu$ zero modes.

We conclude that the noninteracting topological classification 
$\mathbb{Z}$
for the symmetry class A in even dimensions
survives strong interactions on the boundary provided the
fermion-number conservation is neither explicitly nor spontaneously broken.

\textit{Proof {for case} (ii):} First, we show the statement for: 
(a) cases with dynamical Dirac masses that preserve 
the fermion-number $U(1)$ symmetry.
We then proceed to:
(b) cases with $U(1)$-breaking dynamical Dirac masses.

(a)
We consider the massive Dirac Hamiltonian 
\begin{equation}
\mathcal{H}^{(0)}(\bm{x})=
\sum_{j=1}^{d} 
(-\mathrm{i}\partial^{\,}_{j})\,
\gamma^{\,}_{j}
+
m(\bm{x})\,\gamma^{\,}_{0},
\quad
m(\bm{x})\in\mathbb{R},
\end{equation}
obeying the TRS represented by $\mathcal{T}$
for classes AI and AII in $d=4n$ for $n=1,2,\cdots$.
Here, the Dirac matrices are of dimension 
$r\geq r^{\,}_{\mathrm{min}}$.
They obey the Clifford algebra
$\{\gamma^{\,}_{\mu},\gamma^{\,}_{\mu'}\}=2\,\delta^{\,}_{\mu\mu'}$
with $\mu,\mu'=0,\cdots,d$. 
The Dirac matrices entering $\mathcal{H}^{(0)}(\bm{x})$ 
and the operator $\mathcal{T}$ that represents reversal of time
can be used to define the following pair of Clifford algebras.%
~\cite{Morimoto13,Morimoto15}

For the symmetry class AI, 
reversal of time is represented by an element of the Clifford
algebra $\mathcal{T}$
that squares to the unit matrix.
It and $\mathrm{i}\mathcal{T}$,
together with the gamma matrices
$\gamma^{\,}_1$, $\cdots$, $\gamma^{\,}_d$ satisfying the conditions
$\gamma^{\,}_{1}=-\mathcal{T}\,\gamma^{\,}_{1}\,\mathcal{T}^{-1}$,
$\cdots$,
$\gamma^{\,}_{d}=-\mathcal{T}\,\gamma^{\,}_{d}\,\mathcal{T}^{-1}$,
are generators of the Clifford algebra.
On the other hand, the Dirac mass matrix
$\mathrm{i}\gamma^{\,}_{0}$ 
is chosen as the generator that squares to minus the unit matrix, 
for $\gamma^{\,}_{0}=+\mathcal{T}\,\gamma^{\,}_{0}\,\mathcal{T}^{-1}$.
We arrive at the Clifford algebra
\begin{equation}
Cl^{\,}_{1,2+d}:=
\{
J\gamma^{\,}_{0};
\mathcal{T},J\mathcal{T},
\gamma^{\,}_{1},\cdots,\gamma^{\,}_{d}
\}
\label{eq: def Clifford algebra AI}
\end{equation}
for the symmetry class AI, where $J$ is the generator that
represents the imaginary unit ``$\mathrm{i}$'' and satisfies
the relations $J^{2}=-1$ and $\{\mathcal{T},J\}=0$.%
~\footnote{
We have used a simplified notation for real Clifford algebras
as defined below.
A real Clifford algebra 
$Cl^{\,}_{p,q}=\{e^{\,}_{1},\ldots,e^{\,}_{p};e^{\,}_{p+1},\ldots,e^{\,}_{p+q}\}$
is a real algebra that is generated by $p+q$ pairwise anticommuting
generators $(e^{\,}_{1},\ldots,e^{\,}_{p+q})$ satisfying the conditions
$e^{2}_{j}=-1$ for $j=1,\ldots,p$ and
$e^{2}_{j}=+1$ for $j=p+1,\ldots,p+q$.
          }

For the symmetry class AII, 
reversal of time is represented by an element of the Clifford
algebra $\mathcal{T}$
that squares to minus the unit matrix.
It and $\mathrm{i}\mathcal{T}$
enter on equal footing with 
$\mathrm{i}\gamma^{\,}_{0}$ 
as the generators that square to minus the unit matrix, 
for $\gamma^{\,}_{0}=+\mathcal{T}\,\gamma^{\,}_{0}\,\mathcal{T}^{-1}$.
On the other hand,
gamma matrices $\gamma^{\,}_{1}$, $\cdots$, $\gamma^{\,}_{d}$, satisfying
the conditions
$\gamma^{\,}_{1}=-\mathcal{T}\,\gamma^{\,}_{1}\,\mathcal{T}^{-1}$,
$\cdots$,
$\gamma^{\,}_{d}=-\mathcal{T}\,\gamma^{\,}_{d}\,\mathcal{T}^{-1}$,
are the generators that square to the unit matrix
in the Clifford algebra.
We arrive at the Clifford algebra
\begin{equation}
Cl^{\,}_{3,d}:=
\{
J\gamma^{\,}_{0},\mathcal{T},J\mathcal{T};
\gamma^{\,}_{1},\cdots,\gamma^{\,}_{d}
\label{eq: def Clifford algebra AII}
\}
\end{equation}
for the symmetry class AII.

In both symmetry classes 
the choice of $\gamma^{\,}_{0}$ is unique, up to a sign, 
as a consequence of the fact that the zeroth homotopy groups
of the classifying spaces for the symmetry classes
AI and AII is $\mathbb{Z}$ in $4n$ dimensions.
In other words,
no other Dirac mass matrix that is invariant
under reversal of time anticommutes
with $\gamma^{\,}_{0}$.%
~\cite{Morimoto15}
This leaves open the possibility that the Clifford algebras
(\ref{eq: def Clifford algebra AI})
and
(\ref{eq: def Clifford algebra AII})
for $d=4n$ could admit the addition of a generator
$\gamma^{\prime}_{0}$ that anticommutes with $\mathcal{H}^{(0)}$
and is odd under reversal of time,
$\gamma^{\prime}_{0}=-\mathcal{T}\,\gamma^{\prime}_{0}\,\mathcal{T}^{-1}$.
If so, the choice of $\gamma^{\,}_{1}$ to $\gamma^{\,}_{d}$
in the Clifford algebras
(\ref{eq: def Clifford algebra AI})
and
(\ref{eq: def Clifford algebra AII})
would not be unique in an uncountable (in a continuous) way. 
The existence of $\gamma^{\prime}_{0}$
is thus tied to the task of parametrizing in a continuous way
the representation of the generator (e.g., $\gamma^{\,}_{d}$) present
in $Cl^{\,}_{p,q+1}$ but absent in $Cl^{\,}_{p,q}$ 
applied to the cases
$(p,q)=(1,4n+1)$ and $(p,q)=(3,4n-1)$ 
for the $4n$-dimensional symmetry classes AI and AII, respectively.%
~\footnote{
These tasks correspond to the following classification problem.
How does one parametrize the generators
of $Cl^{\,}_{p,q+1}$
that enter the kinetic contribution to the Dirac Hamiltonian?
This classification problem is thus distinct from the one
in which one seeks to parametrize the generators 
that enter the Dirac Hamiltonian as a Dirac mass.
          }
Both tasks are denoted by the extension problem of Clifford algebras
\begin{align}
Cl^{\,}_{p,q} &\to Cl^{\,}_{p,q+1},
\end{align}
with the classifying spaces
\begin{align}
R^{\,}_{q-p}=
\begin{cases}
R^{\,}_{4n},
&
(p,q)=(1,4n+1),
\\
R^{\,}_{4n-4},
&
(p,q)=(3,4n-1),
\end{cases}
\label{eq: ext for AI and AII id d=4n}
\end{align}
as solutions for the set of representations of
possible $\gamma^{\prime}_{0}$ in the 
symmetry classes AI and AII, respectively.
Hereto, it is the zeroth homotopy group of the
classifying space $R^{\,}_{p-q}$ that seals the fate of the existence
of $\gamma^{\prime}_{0}$.
As $\pi^{\,}_{0}(R^{\,}_{4n})=\pi^{\,}_{0}(R^{\,}_{4n-4})=\mathbb{Z}$,
it follows that $\gamma^{\prime}_{0}$ does not exist, i.e.,
no dynamical Dirac mass that breaks the TRS symmetry
but preserves the global $U(1)$ gauge symmetry 
is permissible for the symmetry classes AI and AII
when $d=4n$.

(b)
After having established the absence of $U(1)$-preserving 
dynamical Dirac masses,
the stability analysis in the presence of 
dynamical superconducting fluctuations 
for the symmetry classes AI and AII 
is the same as that for the symmetry class A.
The boundary dynamical Dirac mass takes the form
(\ref{eq: dynamical mass in class A with SC fluctuations}).
The target space of the QNLSM is homeomorphic to 
$S^{1}\times V^{\,}_{\t{BdG}\, \nu}$
since it has to be closed under the action of 
a global $U(1)$ gauge transformation.
Vortices that bind boundary zero modes
originate from the $S^{1}$ manifold.

\textit{Proof for case (iii):}
The symmetry classes D and C for $d=2,6$ (mod 8) do not support
dynamical Dirac masses along the boundary, 
because the PHS is kept as a fundamental symmetry.
Their noninteracting topological classification $\mathbb{Z}$
survives strong interactions on the boundary provided the
PHS is neither explicitly nor spontaneously broken.

\begin{table*}[t]
\caption{
Application of the Bott periodicity obeyed by the homotopy groups
$\pi^{\,}_{D}(V)$ for $D=0,1,\cdots$ 
of a given classifying space $V^{\prime}_{d-1}$ 
of dynamical boundary Dirac masses
to deduce the reduction pattern
$\mathbb{Z}\to\mathbb{Z}^{\,}_{\nu^{\,}_{\mathrm{max}}}$
for the symmetry class BDI in dimensions
(a) $d=1$ for which $V^{\prime}_{d-1}=R^{\,}_{2}$ and $\nu^{\,}_{\mathrm{max}}=8$,
(b) $d=5$ for which $V^{\prime}_{d-1}=R^{\,}_{6}$ and $\nu^{\,}_{\mathrm{max}}=16$,
and (c) $d=9$ for which $V^{\prime}_{d-1}=R^{\,}_{2}$ and $\nu^{\,}_{\mathrm{max}}=128$.
The column $\nu$ fixes the rank $r:=r^{\,}_{\mathrm{min}}\,\nu$ 
of the Dirac Hamiltonian in the symmetry class BDI.
The fourth column gives the
target manifold of the QNLSM with the action $S^{\,}_{\mathrm{QNLSM}}$ 
that encodes the fermion-fermion interactions
on the $(d-1)$-dimensional boundary.
The fifth column indicates if a topological action
$S^{\,}_{\mathrm{top}}$ can be added to $S^{\,}_{\mathrm{QNLSM}}$. 
\label{tab: 1-9D BDI}
        }
\begin{tabular}{ccccc}
(a) Symmetry class BDI in $d=1$
&\qquad\qquad &
(b) Symmetry class BDI in $d=5$
&\qquad\qquad &
(c) Symmetry class BDI in $d=9$
\\
%1D class BDI
\begin{tabular}[t]{ccccc}
\hline \hline
$D$ 
&
$
\pi^{\,}_{D}(R^{\,}_{2})$
&
$\nu$ 
&
$S^{\,}_{\mathrm{QNLSM}}$
&
$S^{\,}_{\mathrm{top}}$
\\
\hline
0&$\mathbb{Z}^{\,}_{2}$ & 2         & $S^{0 }$ & \checkmark  \\
1&0                   &            &         &             \\
2&$\mathbb{Z}$        & 4          & $S^{2 }$ & \checkmark  \\
3&0                   &            &          &            \\
4&0                   &            &          &            \\
5&0                   &            &          &            \\
6&$\mathbb{Z}$        & 8          & $S^{6}$   & --         \\
7&$\mathbb{Z}^{\,}_{2}$& 16         & $S^{7}$   & --          \\
\hline \hline
\end{tabular}
&\qquad\qquad &
%5D class BDI
\begin{tabular}[t]{ccccc}
\hline \hline
$D$ 
& 
$\pi^{\,}_{D}(R^{\,}_{6})$ 
& 
$\nu$ 
& 
$S^{\,}_{\mathrm{QNLSM}}$
&
$S^{\,}_{\mathrm{top}}$
\\
\hline
0  & 0                    &           &         &           \\
1  & 0                    &           &         &           \\
2  & $\mathbb{Z}$         & 1         & $S^{2 }$ & \checkmark\\
3  & $\mathbb{Z}^{\,}_{2}$ & 2         & $S^{3 }$ & \checkmark\\
4  & $\mathbb{Z}^{\,}_{2}$ & 4         & $S^{4 }$ & \checkmark\\
5  & 0                    &           &         &           \\
6  & $\mathbb{Z}$         & 8         & $S^{6 }$ & \checkmark\\
7  & 0                    &           &         &           \\
\hline
8  & 0                    &           &         &           \\
9  & 0                    &           &         &           \\
10 & $\mathbb{Z}$         & 16        & $S^{10}$ & --        \\
\hline \hline
\end{tabular}
&\qquad\qquad &
%9D class BDI
\begin{tabular}[t]{ccccc}
\hline \hline
$D$
& 
$\pi^{\,}_{D}(R^{\,}_{2})$
&
$\nu$
&
$S^{\,}_{\mathrm{QNLSM}}$
&
$S^{\,}_{\mathrm{top}}$
\\
\hline
0  & $\mathbb{Z}^{\,}_{2}$ &  2        & $S^{0 }$ & \checkmark \\
1  & 0                   &            &         &            \\
2  & $\mathbb{Z}$        &  4         & $S^{2 }$ & \checkmark \\
3  & 0                   &            &         &            \\
4  & 0                   &            &         &            \\
5  & 0                   &            &         &            \\
6  & $\mathbb{Z}$        &  8         & $S^{6 }$ & \checkmark \\
7  & $\mathbb{Z}^{\,}_{2}$ & 16         & $S^{7 }$ & \checkmark \\  
\hline
8  & $\mathbb{Z}^{\,}_{2}$ & 32         & $S^{8 }$ & \checkmark \\             
9  & 0                   &            &         &            \\
10 &$\mathbb{Z}$         & 64         & $S^{10}$ & \checkmark \\
11 & 0                   &            &         &            \\
12 & 0                   &            &         &            \\
13 & 0                   &            &         &            \\
14 & $\mathbb{Z}$        & 128        & $S^{14}$ & --         \\
\hline \hline
\end{tabular}
\\
\end{tabular}
\end{table*}

\begin{table*}[tb]
\caption{
Application of the Bott periodicity obeyed by the homotopy groups
$\pi^{\,}_{D}(V)$ for $D=0,1,\cdots$ 
of a given classifying space $V^{\prime}_{d-1}$ 
of dynamical boundary Dirac masses
to deduce the reduction pattern
$\mathbb{Z}\to\mathbb{Z}^{\,}_{\nu^{\,}_{\mathrm{max}}}$
for the symmetry class DIII in dimensions
(a) $d=3$ for which $V^{\prime}_{d-1}=R^{\,}_{0}$ and $\nu^{\,}_{\mathrm{max}}=16$,
(b) $d=7$ for which $V^{\prime}_{d-1}=R^{\,}_{4}$ and $\nu^{\,}_{\mathrm{max}}=32$,
and (c) $d=11$ for which $V^{\prime}_{d-1}=R^{\,}_{0}$ and 
$\nu^{\,}_{\mathrm{max}}=256$.  
The column $\nu$ fixes the rank $r:=r^{\,}_{\mathrm{min}}\,\nu$ 
of the Dirac Hamiltonian in the symmetry class DIII.
The fourth column gives the
target manifold of the QNLSM with the action $S^{\,}_{\mathrm{QNLSM}}$ 
that encodes the fermion-fermion interactions
on the $(d-1)$-dimensional boundary.
The fifth column indicates if a topological action
$S^{\,}_{\mathrm{top}}$ can be added to $S^{\,}_{\mathrm{QNLSM}}$.
\label{tab: 1-11D DIII}
        }
\begin{tabular}{ccccc}
(a) Symmetry class DIII in $d=3$
&\qquad\qquad &
(b) Symmetry class DIII in $d=7$
&\qquad\qquad &
(c) Symmetry class DIII in $d=11$
\\
%3D DIII
\begin{tabular}[t]{ccccc}
\hline \hline
$D$ 
&
$\pi^{\,}_{D}(R^{\,}_{0})$ 
&
$\nu$ 
&
$S^{\,}_{\mathrm{QNLSM}}$
&
$S^{\,}_{\mathrm{top}}$
\\
\hline
0 & $\mathbb{Z}$        &  1       & $S^{0}$ & \checkmark \\
1 & $\mathbb{Z}^{\,}_{2}$ &  2       & $S^{1}$ & \checkmark \\
2 & $\mathbb{Z}^{\,}_{2}$ &  4       & $S^{2}$ & \checkmark \\
3 & 0                   &           &        &            \\
4 & $\mathbb{Z}$        &  8        & $S^{4}$ & \checkmark \\
5 & 0                   &           &        &            \\
6 & 0                   &           &        &            \\
7 & 0                   &           &        &            \\
\hline
8 & $\mathbb{Z}$        &  16       & $S^{8}$ & --         \\
\hline \hline
\end{tabular}
&\qquad\qquad&
%7D DIII
\begin{tabular}[t]{ccccc}
\hline \hline
$D$ 
& 
$\pi^{\,}_{D}(R^{\,}_{4})$ 
& 
$\nu$ 
& 
$S^{\,}_{\mathrm{QNLSM}}$ 
&
$S^{\,}_{\mathrm{top}}$ 
\\
\hline
0&$\mathbb{Z}$         & 1  & $S^{0}$ &\checkmark \\
1&0  &  \\
2&0  &  \\
3&0  &  \\
4&$\mathbb{Z}$         & 2  & $S^{4}$  &\checkmark \\
5&$\mathbb{Z}^{\,}_{2}$  & 4  & $S^{5}$  &\checkmark \\
6&$\mathbb{Z}^{\,}_{2}$  & 8  & $S^{6}$  &\checkmark \\
7&0  &  \\
\hline
8&$\mathbb{Z}$         & 16 & $S^{8}$  &\checkmark \\
9&$0$   &  \\
10&$0$  &  \\
11&$0$  &  \\
12&$\mathbb{Z}$        & 32 & $S^{12}$  &--         \\
\hline \hline
\end{tabular}
&\qquad\qquad&
%11D DIII
\begin{tabular}[t]{ccccc}
\hline \hline
$D$ 
&
$\pi^{\,}_{D}(R^{\,}_{0})$ 
&
$\nu$ 
&
$S^{\,}_{\mathrm{QNLSM}}$
&
$S^{\,}_{\mathrm{top}}$
\\
\hline
0  & $\mathbb{Z}$        & 1   & $S^{0}$  & \checkmark \\
1  & $\mathbb{Z}^{\,}_{2}$ & 2   & $S^{1}$  & \checkmark \\
2  & $\mathbb{Z}^{\,}_{2}$ & 4   & $S^{2}$  & \checkmark \\
3  & 0                   &     &          &            \\
4  & $\mathbb{Z}$        & 8   & $S^{4}$   & \checkmark \\
5  & 0                   &     &          &            \\
6  & 0                   &     &          &            \\
7  & 0                   &     &          &            \\
\hline
8  & $\mathbb{Z}$        & 16  & $S^{8}$  & \checkmark \\
9  & $\mathbb{Z}^{\,}_{2}$ & 32  & $S^{9}$  & \checkmark \\
10 & $\mathbb{Z}^{\,}_{2}$ & 64  & $S^{10}$ & \checkmark \\
11 & 0                   &      &         &            \\
12 & $\mathbb{Z}$        & 128 & $S^{12}$ & \checkmark \\
13 & 0                   &     &         &            \\
14 & 0                   &     &         &            \\
15 & 0                   &     &         &            \\
\hline
16 & $\mathbb{Z}$        & 256 & $S^{16}$ & --         \\
\hline \hline
\end{tabular}
\\
\end{tabular}
\end{table*}

\begin{table*}[tb]
\caption{
Application of the Bott periodicity obeyed by the homotopy groups
$\pi^{\,}_{D}(V)$ for $D=0,1,\cdots$ 
of a given classifying space 
$V^{\prime}_{d-1}$ of dynamical boundary Dirac masses
to deduce the reduction pattern
$\mathbb{Z}\to\mathbb{Z}^{\,}_{\nu^{\,}_{\mathrm{max}}}$
for the symmetry class CII in dimensions
(a) $d=1$ for which $V^{\prime}_{d-1}=R^{\,}_{6}$ and $\nu^{\,}_{\mathrm{max}}=2$,
(b) $d=5$ for which $V^{\prime}_{d-1}=R^{\,}_{2}$ and $\nu^{\,}_{\mathrm{max}}=16$,
and (c) $d=9$ for which $V^{\prime}_{d-1}=R^{\,}_{6}$ and $\nu^{\,}_{\mathrm{max}}=32$.
The column $\nu$ fixes the rank $r:=r^{\,}_{\mathrm{min}}\,\nu$ 
of the Dirac Hamiltonian in the symmetry class CII.
The fourth column gives the
target manifold of the QNLSM with the action $S^{\,}_{\mathrm{QNLSM}}$ 
that encodes the fermion-fermion interactions
on the $(d-1)$-dimensional boundary.
The fifth column indicates if a topological action
$S^{\,}_{\mathrm{top}}$ can be added to $S^{\,}_{\mathrm{QNLSM}}$. 
\label{tab: 1-11D CII}
        }
\begin{tabular}{ccccc}
(a) Symmetry class CII in $d=1$
&\qquad\qquad &
(b) Symmetry class CII in $d=5$
&\qquad\qquad &
(c) Symmetry class CII in $d=9$
\\
%1D class CII
\begin{tabular}[t]{ccccc}
\hline \hline
$D$ 
& 
$\pi^{\,}_{D}(R^{\,}_{6})$ 
& 
$\nu$ 
& 
$S^{\,}_{\mathrm{QNLSM}}$
&
$S^{\,}_{\mathrm{top}}$
\\
\hline
0  & 0                    &           &         &           \\
1  & 0                    &           &         &           \\
2  & $\mathbb{Z}$         & 1         & $S^{2 }$ & \checkmark\\
3  & $\mathbb{Z}^{\,}_{2}$ & 2         & $S^{3 }$ & --        \\
4  & $\mathbb{Z}^{\,}_{2}$ & 4         & $S^{4 }$ & --        \\
5  & 0                    &           &         &           \\
6  & $\mathbb{Z}$         & 8         & $S^{6 }$ & --        \\
7  & 0                    &           &         &           \\
\hline \hline
\end{tabular}
&\qquad\qquad &
%5D class CII
\begin{tabular}[t]{ccccc}
\hline \hline
$D$ 
&
$
\pi^{\,}_{D}(R^{\,}_{2})$
&
$\nu$ 
&
$S^{\,}_{\mathrm{QNLSM}}$
&
$S^{\,}_{\mathrm{top}}$
\\
\hline
0&$\mathbb{Z}^{\,}_{2}$ & 2         & $S^{0 }$ & \checkmark  \\
1&0                   &            &         &             \\
2&$\mathbb{Z}$        & 4          & $S^{2 }$ & \checkmark  \\
3&0                   &            &          &            \\
4&0                   &            &          &            \\
5&0                   &            &          &            \\
6&$\mathbb{Z}$        & 8          & $S^{6}$   & \checkmark \\
7&$\mathbb{Z}^{\,}_{2}$& 16          & $S^{7}$  & --          \\
\hline \hline
\end{tabular}
&\qquad\qquad &
%9D class CII
\begin{tabular}[t]{ccccc}
\hline \hline
$D$ 
& 
$\pi^{\,}_{D}(R^{\,}_{6})$ 
& 
$\nu$ 
& 
$S^{\,}_{\mathrm{QNLSM}}$
&
$S^{\,}_{\mathrm{top}}$
\\
\hline
0  & 0                    &           &         &           \\
1  & 0                    &           &         &           \\
2  & $\mathbb{Z}$         & 1         & $S^{2 }$ & \checkmark\\
3  & $\mathbb{Z}^{\,}_{2}$ & 2         & $S^{3 }$ & \checkmark\\
4  & $\mathbb{Z}^{\,}_{2}$ & 4         & $S^{4 }$ & \checkmark\\
5  & 0                    &           &         &           \\
6  & $\mathbb{Z}$         & 8         & $S^{6 }$ & \checkmark\\
7  & 0                    &           &         &           \\
\hline
8  & 0                    &           &         &           \\
9  & 0                    &           &         &           \\
10 & $\mathbb{Z}$         & 16        & $S^{10}$ & \checkmark\\
11 & $\mathbb{Z}^{\,}_{2}$ & 32         & $S^{11}$ & --        \\
\hline \hline
\end{tabular}
\\
\end{tabular}
\end{table*}

\begin{table*}[tb]
\caption{
Application of the Bott periodicity obeyed by the homotopy groups
$\pi^{\,}_{D}(V)$ for $D=0,1,\cdots$ 
of a given classifying space 
$V^{\prime}_{d-1}$ of dynamical boundary Dirac masses
to deduce the reduction pattern
$\mathbb{Z}\to\mathbb{Z}^{\,}_{\nu^{\,}_{\mathrm{max}}}$
for the symmetry class CI in dimensions
(a) $d=3$ for which $V^{\prime}_{d-1}=R^{\,}_{4}$ and $\nu^{\,}_{\mathrm{max}}=4$,
(b) $d=7$ for which $V^{\prime}_{d-1}=R^{\,}_{0}$ and $\nu^{\,}_{\mathrm{max}}=32$,
and (c) $d=11$ for which $V^{\prime}_{d-1}=R^{\,}_{4}$ and $\nu^{\,}_{\mathrm{max}}=64$.
The column $\nu$ fixes the rank $r:=r^{\,}_{\mathrm{min}}\,\nu$ 
of the Dirac Hamiltonian in the symmetry class CI.
The fourth column gives the
target manifold of the QNLSM with the action $S^{\,}_{\mathrm{QNLSM}}$ 
that encodes the fermion-fermion interactions
on the $(d-1)$-dimensional boundary.
The fifth column indicates if a topological action
$S^{\,}_{\mathrm{top}}$ can be added to $S^{\,}_{\mathrm{QNLSM}}$.
\label{tab: 1-11D CI}
        }
\begin{tabular}{ccccc}
(a) Symmetry class CI in $d=3$
&\qquad\qquad &
(b) Symmetry class CI in $d=7$
&\qquad\qquad &
(c) Symmetry class CI in $d=11$
\\
%3D CI
\begin{tabular}[t]{ccccc}
\hline \hline
$D$ 
& 
$\pi^{\,}_{D}(R^{\,}_{4})$ 
& 
$\nu$ 
& 
$S^{\,}_{\mathrm{QNLSM}}$ 
&
$S^{\,}_{\mathrm{top}}$ 
\\
\hline
0&$\mathbb{Z}$         & 1       & $S^{0}$ &\checkmark \\
1&0  &  \\
2&0  &  \\
3&0  &  \\
4&$\mathbb{Z}$         & 2       & $S^{4}$  &\checkmark \\
5&$\mathbb{Z}^{\,}_{2}$  & 4       & $S^{5}$  &--        \\
6&$\mathbb{Z}^{\,}_{2}$  & 8       & $S^{6}$  &--        \\
7&0  &  \\
\hline \hline
\end{tabular}
&\qquad\qquad&
%7D CI
\begin{tabular}[t]{ccccc}
\hline \hline
$D$ 
&
$\pi^{\,}_{D}(R^{\,}_{0})$ 
&
$\nu$ 
&
$S^{\,}_{\mathrm{QNLSM}}$
&
$S^{\,}_{\mathrm{top}}$
\\
\hline
0 & $\mathbb{Z}$        &  1       & $S^{0}$ & \checkmark \\
1 & $\mathbb{Z}^{\,}_{2}$ &  2       & $S^{1}$ & \checkmark \\
2 & $\mathbb{Z}^{\,}_{2}$ &  4       & $S^{2}$ & \checkmark \\
3 & 0                   &           &        &            \\
4 & $\mathbb{Z}$        &  8        & $S^{4}$ & \checkmark \\
5 & 0                   &           &        &            \\
6 & 0                   &           &        &            \\
7 & 0                   &           &        &            \\
\hline
8 & $\mathbb{Z}$        & 16        & $S^{8}$ & \checkmark \\
9 & $\mathbb{Z}^{\,}_{2}$ & 32        & $S^{9}$ &--         \\
\hline \hline
\end{tabular}
&\qquad\qquad&
%1D CI
\begin{tabular}[t]{ccccc}
\hline \hline
$D$ 
& 
$\pi^{\,}_{D}(R^{\,}_{4})$ 
& 
$\nu$ 
& 
$S^{\,}_{\mathrm{QNLSM}}$ 
&
$S^{\,}_{\mathrm{top}}$ 
\\
\hline
0&$\mathbb{Z}$         & 1  & $S^{0}$ &\checkmark \\
1&0  &  \\
2&0  &  \\
3&0  &  \\
4&$\mathbb{Z}$         & 2  & $S^{4}$  &\checkmark \\
5&$\mathbb{Z}^{\,}_{2}$  & 4  & $S^{5}$  &\checkmark \\
6&$\mathbb{Z}^{\,}_{2}$  & 8  & $S^{6}$  &\checkmark \\
7&0  &  \\
\hline
8&$\mathbb{Z}$         & 16 & $S^{8}$  &\checkmark \\
9&$0$   &  \\
10&$0$  &  \\
11&$0$  &  \\
12&$\mathbb{Z}$         & 32 & $S^{12}$  &\checkmark \\
13&$\mathbb{Z}^{\,}_{2}$ & 64  & $S^{13}$  &--        \\
\hline \hline
\end{tabular}
\\
\end{tabular}
\end{table*}

\subsubsection{The case of odd dimensions}
\label{subsubsec: The case of odd dimensions}

The topological classification $\mathbb{Z}$ of noninteracting
strong topological insulators and superconductors in odd dimensions of space
is reduced to the coarser classification 
$\mathbb{Z}^{\,}_{\nu^{\,}_{\mathrm{max}}}$
with $\nu^{\,}_{\mathrm{max}}$ an integer:
\begin{subequations}
\begin{equation}
\mathbb{Z}\to\mathbb{Z}^{\,}_{\nu^{\,}_{\mathrm{max}}}.
\end{equation}
The label ``max'' stands here for maximum. The task at hand is thus to
compute the integer $\nu^{\,}_{\mathrm{max}}$.
Computing $\nu^{\,}_{\mathrm{max}}$ proceeds with the following algorithm
(see Tables \ref{tab: 1-9D BDI}-\ref{tab: 1-11D CI}).
\\
\textit{Step 1:} 
Choose any one of the ten AZ symmetry classes from Table
\ref{table: topo classification short-rnaged entangles AZ classes}.
\\
\textit{Step 2:} 
Choose any odd dimension $d$ for which the zero-th homotopy group of
the classifying space of the chosen symmetry class
$V^{\,}_{d}$ is the set of integers 
[$\pi^{\,}_{0}(V^{\,}_{d})=\mathbb{Z}$].
This step restricts the symmetry classes to the complex
symmetry class AIII and the real symmetry classes BDI, DIII, CII, and CI.
\\
\textit{Step 3:}
Identify the parent symmetry class and its classifying space 
$V^{\prime}_{d}$
that follows if CHS is broken for the complex 
symmetry class AIII or if TRS is broken for the real symmetry classes.
This step restricts the parent symmetry classes to the complex
symmetry class A if the symmetry class AIII is interpreted as realizing
an insulator, the real symmetry class D if the symmetry classes
BDI and DIII are interpreted as superconductors, and the
real symmetry class C if the symmetry classes
CII and CI are interpreted as superconductors.
\\
\textit{Step 4:}
Assign the minimal value 
\begin{equation}
\nu^{\,}_{\mathrm{min}}:=
\begin{cases}
1,&\pi^{\,}_{0}(V^{\prime}_{d})=0,\\
&\\
2,&\pi^{\,}_{0}(V^{\prime}_{d})\neq0,
\end{cases}
\end{equation}
if the zero-th homotopy group of $V^{\prime}_{d}$
is trivial or non-trivial, respectively.
\\
\textit{Step 5:}
Identify the classifying space $V^{\prime}_{d-1}$
that determines the dynamical Dirac mass matrices
induced by the fermion-fermion interactions on the boundary.
\\
\textit{Step 6:}
Construct a table with lines labeled by the 
integer $D=0,1,2,\cdots$. The first column gives the order
$D$ of the homotopy group $\pi^{\,}_{D}(V^{\prime}_{d-1})$
given in the second column.
The third column is the number $\nu$ of
boundary zero modes in the selected symmetry class.
Enter the value $\nu^{\,}_{\mathrm{min}}$ in the third column
for the smallest value of $D$
for which $\pi^{\,}_{D}(V^{\prime}_{d-1})$ is non-trivial.
The value of $\nu$ is then doubled for
each successive line with $\pi^{\,}_{D}(V^{\prime}_{d-1})$ non-trivial.
The fourth column denotes the target space of the QNLSM
with the action $S^{\,}_{\mathrm{QNLSM}}$
defined by integrating out all the boundary Dirac fermions 
when coupled to $(D+1)$ real-valued bosonic fields,
each one of which couples to a Dirac mass matrix from
a $(D+1)$-tuplet of pairwise anticommuting
Dirac mass matrices allowed on the boundary by the parent symmetry class.
The fifth column indicates when a topological term 
$S^{\,}_{\mathrm{top}}$ can be added to the 
action $S^{\,}_{\mathrm{QNLSM}}$.
\footnote{
When identifying non-trivial homotopy groups and topological terms,
we assume that the homotopy groups 
$\pi^{\,}_{D}(V^{\,}_{\nu})$ for the space $V^{\,}_{\nu}$ of 
$\nu\times\nu$ normalized Dirac mass matrices 
with the relevant finite $\nu$ 
are nontrivial whenever $\pi^{\,}_{D}(R^{\,}_{q})$ is nontrivial.
This is valid when $\nu$ is larger than a certain value determined by $D$.
Here, $R^{\,}_{q}$ is the space of 
normalized Dirac mass matrices in the limit $\nu\to\infinity$,
and $\pi^{\,}_{D}(R^{\,}_{q})$ obeys the Bott periodicity 
and are known from the mathematic literature.
We are not able to prove that this assumption is true for all dimensions, 
but we have observed that it always holds in  one, two, and three
dimensions. 
          }
\\
\textit{Step 7:}
Let $n^{\,}_{\mathrm{WZ}}$ be the number of lines with 
$\pi^{\,}_{D}(V^{\prime}_{d-1})$ non-trivial when $D$ takes the values
$D=0,1,\cdots,d+1$. It then follows that
\begin{equation}
\nu^{\,}_{\mathrm{max}}=
\nu^{\,}_{\mathrm{min}}\times
2^{n^{\,}_{\mathrm{WZ}}}.
\end{equation}
\end{subequations}

For the complex symmetry class AIII in dimension
$d=2n+1$ with $n=0,1,2,\cdots$, 
the reduction pattern induced by the fermion-fermion interactions is
\begin{equation}
\mathbb{Z}\to\mathbb{Z}^{\,}_{2^{n+2}}.
\end{equation}
By making use of the eight-fold Bott periodicity, 
one verifies that the reduction patterns are
\begin{equation}
\begin{split}
&
\begin{tabular}{lc|c|c|c|c}
& &
$d=8n+1$
&
$d=8n+3$
&
$d=8n+5$
&
$d=8n+7$
\\\hline
BDI &
&
$\mathbb{Z}\to\mathbb{Z}^{\,}_{2^{4n+3}}$
&
--
&
$\mathbb{Z}\to\mathbb{Z}^{\,}_{2^{4n+4}}$
&
--
\\
DIII &
&
--
&
$\mathbb{Z}\to\mathbb{Z}^{\,}_{2^{4n+4}}$
&
--
&
$\mathbb{Z}\to\mathbb{Z}^{\,}_{2^{4n+5}}$
\\
CII &
&
$\mathbb{Z}\to\mathbb{Z}^{\,}_{2^{4n+1}}$
&
--
&
$\mathbb{Z}\to\mathbb{Z}^{\,}_{2^{4n+4}}$
&
--
\\
CI &
&
--
&
$\mathbb{Z}\to\mathbb{Z}^{\,}_{2^{4n+2}}$
&
--
&
$\mathbb{Z}\to\mathbb{Z}^{\,}_{2^{4n+5}}$
\end{tabular}
\\
&
\end{split}
\end{equation}
if we interpret the symmetry classes BDI, DIII, CII, and CI as
superconductors, or
\begin{equation}
\begin{split}
&
\begin{tabular}{lc|c|c|c|c}
& &
$d=8n+1$
&
$d=8n+3$
&
$d=8n+5$
&
$d=8n+7$
\\\hline
BDI &
&
$\mathbb{Z}\to\mathbb{Z}^{\,}_{2^{4n+2}}$
&
--
&
$\mathbb{Z}\to\mathbb{Z}^{\,}_{2^{4n+3}}$
&
--
\\
CII &
&
$\mathbb{Z}\to\mathbb{Z}^{\,}_{2^{4n+1}}$
&
--
&
$\mathbb{Z}\to\mathbb{Z}^{\,}_{2^{4n+4}}$
&
--
\end{tabular}
\\
&
\end{split}
\end{equation}
if we interpret the symmetry classes BDI and CII as insulators.
We have thus found two different patterns for the reduction of the 
topological classification of
the symmetry class BDI depending on these two interpretations,
as we have observed for $d=1$
in Sec.~\ref{subsec: The chiral symmetry classes when d=1}.

\section{Reduction for TCS and TCI}
\label{eq: Reduction for TCS and TCI}

The addition of discrete symmetries (such as spatial ones) 
to the PHS, TRS, and CHS
enriches the classification of noninteracting fermions.
A general method to account for additional symmetries
that square to the unity
has been proposed in Ref.\ \onlinecite{Morimoto13}.
Hereto, local fermion-fermion interactions can reduce the noninteracting 
topological classification over $\mathbb{Z}$, 
as was first observed in Refs.\ 
\onlinecite{Ryu12,Qi13,Yao13,Gu13}
by way of explicit examples.
We are interested in the robustness to local fermion-fermion interactions
of noninteracting topological phases with reflection symmetry.
We treat two-dimensional topological superconductors
in the symmetry class DIII
with an additional reflection symmetry
(the Yao-Ryu model from Ref.~\onlinecite{Yao13})
and three-dimensional topological insulators
in the symmetry class AII
with an additional reflection symmetry
(a topological crystalline insulator is realized in SnTe,
as was shown in Ref.~\onlinecite{hsieh12}).
We show that the reduction $\mathbb{Z}\to\mathbb{Z}^{\,}_{8}$
holds in both cases by applying the approach detailed in 
Sec.~\ref{sec: Definition and strategy}.

The notation
$X^{\,}_{\mu\mu'\mu''\cdots}:=
\tau^{\,}_{\mu}\otimes\tau^{\prime}_{\mu'}
\otimes\tau^{\prime\prime}_{\mu''}\otimes\cdots$
for $\mu,\mu',\mu'',\cdots=0,1,2,3$ where 
$\tau^{\,}_{0}$, $\tau^{\prime}_{0}$, $\tau^{\prime\prime}_{0}$, $\cdots$ 
are unit $2\times2$ matrices and
the other $\tau^{\,}_\mu$, $\tau^{\prime}_{\mu'}$, $\tau^{\prime\prime}_{\mu''}$,
$\cdots$
are the corresponding Pauli matrices.

\subsection{Two-dimensional superconductors with time-reversal
and reflection symmetries (DIII$+R$)}
\label{sec: 2D DIII+R--}

Let $\bm{x}\equiv (x,y)\equiv(x^{\,}_{1},x^{\,}_{2})$ 
denote a point in two-dimensional space.
Let $X^{\,}_{\mu\mu'}:=\sigma^{\,}_{\mu}\otimes\tau^{\,}_{\mu'}$
with $\mu,\mu'=0,1,2,3$  denote a basis for the vector space
of all Hermitian $4\times4$ matrices.
Following Yao and Ryu,\cite{Yao13}
consider the two-dimensional bulk single-particle Dirac Hamiltonian%
%\footnote{
%This model was proposed by Yao and Ryu to study the reduction of 
%integer-labeled topological phases with interactions in 
%two-dimensional space.
%         }
\begin{subequations}
\label{eq: def single particle mathcal H 0 DIII R--x d=2}
\begin{equation}
\mathcal{H}^{(0)}(x,y):=
-\mathrm{i}\partial^{\,}_{x}
X^{\,}_{31}
-\mathrm{i}\partial^{\,}_{y}
X^{\,}_{02} 
+ 
m(x,y)
X^{\,}_{03}. 
\label{eq: def single particle mathcal H 0 DIII R--x d=2 a}
\end{equation}
This single-particle Dirac Hamiltonian
belongs to the symmetry class DIII, for
\begin{align}
&
\mathcal{T}\,\mathcal{H}^{(0)}(x,y)\,\mathcal{T}^{-1}=
+
\mathcal{H}^{(0)}(x,y),
\label{eq: def single particle mathcal H 0 DIII R--x d=2 b}
\\
&
\mathcal{C}\,\mathcal{H}^{(0)}(x,y)\,\mathcal{C}^{-1}=
-
\mathcal{H}^{(0)}(x,y),
\label{eq: def single particle mathcal H 0 DIII R--x d=2 c}
\end{align}
where
\begin{align}
\mathcal{T}:=
\mathrm{i}X^{\,}_{20}\,\mathsf{K},
\qquad 
\mathcal{C}:=
X^{\,}_{01}\,\mathsf{K}.
\label{eq: def single particle mathcal H 0 DIII R--x d=2 d}
\end{align}
In addition, the Dirac Hamiltonian is invariant under reflection
in the $x$ direction,
\begin{equation}
\mathcal{R}^{\,}_{x}\,\mathcal{H}^{(0)}(-x,y)\,(\,\mathcal{R}^{\,}_{x})^{-1}=
+
\mathcal{H}^{(0)}(x,y),
\label{eq: def single particle mathcal H 0 DIII R--x d=2 e}
\end{equation}
where
\begin{equation}
\mathcal{R}^{\,}_{x}=
\mathrm{i}X^{\,}_{20}.
\end{equation}
\end{subequations}
The operators $\mathcal{T}$, $\mathcal{C}$, and $\mathcal{R}^{\,}_x$
commute pairwise and square to
$\mathcal{T}^{2}=-1$, $\mathcal{C}^{2}=+1$, and $\mathcal{R}^{2}_{x}=-1$.

The Dirac mass matrix $X^{\,}_{03}$ is here the only one allowed for dimension 
$r=r^{\,}_{\mathrm{min}}=4$ Dirac matrices
under the symmetry constraints
(\ref{eq: def single particle mathcal H 0 DIII R--x d=2 b}),
(\ref{eq: def single particle mathcal H 0 DIII R--x d=2 c}), and
(\ref{eq: def single particle mathcal H 0 DIII R--x d=2 e}).
The domain wall
\begin{subequations}
\label{eq: y=0 DIII R--2 d=2}
\begin{equation}
m(x,y)=
m^{\,}_{\infty}\,\mathrm{sgn}(y)
\label{eq: y=0 DIII R--2 d=2 a}
\end{equation}
at $y=0$ binds the zero mode
\begin{equation}
e^{-\mathrm{i}X^{\,}_{02}\,X^{\,}_{03}
\int\limits_{0}^{y}\mathrm{d}y\,m(x,y')}\,\chi=
e^{-|m^{\,}_{\infty}y|}\,
\chi,
\label{eq: y=0 DIII R--2 d=2 b}
\end{equation}
where
\begin{equation}
X^{\,}_{01}\,\chi=-\mathrm{sgn}(m_\infty)\,\chi
\label{eq: y=0 DIII R--2 d=2 c}
\end{equation}
with $\chi$ independent of $x$ and $z$,
that is annihilated by the boundary Hamiltonian
\begin{equation}
\mathcal{H}^{(0)}_{\mathrm{bd}}(x):=
-\mathrm{i}\partial^{\,}_{x}
\sigma^{\,}_{3},
\label{eq: y=0 DIII R--2 d=2 d}
\end{equation}
where we have chosen $m_\infty<0$.
On the boundary $y=0$, the symmetries 
(\ref{eq: def single particle mathcal H 0 DIII R--x d=2 e}) 
are realized by
\begin{equation}
\mathcal{T}^{\,}_{\mathrm{bd}}:=
\mathrm{i}\sigma^{\,}_{2}\,\mathsf{K},
\qquad
\mathcal{C}^{\,}_{\mathrm{bd}}:=
\mathsf{K},
\qquad
\mathcal{R}^{\,}_{x\,\mathrm{bd}}:=
\mathrm{i}\sigma^{\,}_{2}.
\label{eq: y=0 DIII R--2 d=2 f}
\end{equation}
\end{subequations}

The boundary single-particle Hamiltonian
(\ref{eq: y=0 DIII R--2 d=2 d})
and the operators
(\ref{eq: y=0 DIII R--2 d=2 f})
are denoted by
$\mathcal{H}^{(0)}_{\mathrm{bd}\,\nu}(x)$,
$\mathcal{T}^{\,}_{\mathrm{bd}\,\nu}$,
$\mathcal{C}^{\,}_{\mathrm{bd}\,\nu}$,
and
$\mathcal{R}^{\,}_{x\,\mathrm{bd}\,\nu}$,
respectively,
when tensored with the $\nu\times\nu$ unit matrix $\openone$.
The single-particle Hamiltonian
$\mathcal{H}^{(0)}_{\mathrm{bd}\,\nu}(x)$
supports $\nu$ linearly independent zero modes.
Their stability to interactions that preserve the symmetries 
is probed by studying the dynamical single-particle boundary Hamiltonian
\begin{subequations}
\label{eq: gamma' DIII R--x}
\begin{equation}
\mathcal{H}^{(\mathrm{dyn})}_{\mathrm{bd}\,\nu}(\tau,x):=
-\mathrm{i}\partial^{\,}_{x}
\sigma^{\,}_{3}\otimes\openone
+
\gamma'(\tau,x),
\label{eq: gamma' DIII R--x a}
\end{equation}
where the boundary dynamical Dirac mass matrix $\gamma'(x)$
satisfies the particle-hole symmetry
\begin{equation}
\mathcal{C}^{\,}_{\mathrm{bd}\,\nu}\gamma'
\mathcal{C}_{\mathrm{bd}\,\nu}^{-1}
=-\gamma'
\end{equation}
and takes the form
\begin{equation}
\gamma'(x)=
\sigma^{\,}_{2}\otimes M^{\,}_{1}(\tau,x)
+
\sigma^{\,}_{1}\otimes M^{\,}_{2}(\tau,x)
\label{eq: gamma' DIII R--x b}
\end{equation}
with the $\nu\times\nu$ Hermitian matrices
\begin{equation}
M^{\,}_{1}(\tau,x)=+M^{*}_{1}(\tau,x),
\qquad
M^{\,}_{2}(\tau,x)=-M^{*}_{2}(\tau,x),
\label{eq: gamma' DIII R--x c}
\end{equation}
i.e., $M^{\,}_{1}(\tau,x)$ is a real-valued symmetric matrix
while $M^{\,}_{2}(\tau,x)$ is an imaginary-valued antisymmetric matrix.
This is to say that the normalized 
boundary dynamical Dirac mass matrix $\gamma'(x)$ 
belongs to the space
\begin{align}
V^{\,}_{\nu}:=O(\nu),
\qquad
R^{\,}_{1}=\lim_{\nu\to\infty} O(\nu),
\label{eq: gamma' DIII R--x d}
\end{align}
\end{subequations}
for the Dirac matrices in the boundary ($d=1$) Dirac
Hamiltonians belonging to the symmetry class D.%
~\footnote{ 
Let $M:=M^{\,}_{1}+\mathrm{i}M^{\,}_{2}$
with $M^{\,}_{1}=+M^{*}_{1}=+M^{\mathsf{T}}_{1}$
and $M^{\,}_{2}=-M^{*}_{2}=-M^{\mathsf{T}}_{2}$
defined by Eqs.\ (\ref{eq: gamma' DIII R--x b})
and (\ref{eq: gamma' DIII R--x c}).
It follows that
$$
\mathrm{i}\gamma'=
\begin{pmatrix}
0       & +M \\
-M^{\mathsf{T}} & 0
\end{pmatrix}.
$$
Now, demand that $\gamma'$ squares to the unit matrix $\openone$.
This implies that $M\,M^{\mathsf{T}}=M^{\mathsf{T}}\,M=\openone$,
i.e., $M\in O(\nu)$. Hence,
the classifying space is homeomorphic to $O(\nu)$. 
         }
Integrating the boundary Dirac fermions delivers a QNLSM in (1+1)-dimensional 
space and time. In order to gap out dynamically the boundary zero modes 
without breaking the time-reversal, particle-hole,
and reflection symmetries,
this QNLSM must be free of topological obstructions.
We construct explicitly the spaces 
for the relevant normalized boundary dynamical Dirac mass matrices
of dimension $\nu=2^{n}$ with $n=0,1,2,3$ in the following.
The relevant homotopy groups are given in 
Table~\ref{tab: 2D DIII+R--}.

\begin{subequations}
\textit{Case $\nu=1$:}
There is a topological obstruction of the domain wall type 
as the target space is
\begin{equation}
S^{0}=\{ \pm \sigma^{\,}_{y}\}
\end{equation}
and $\pi^{\,}_{0}(S^{0})\neq 0$.

\textit{Case $\nu=2$:}
There is a topological obstruction of the vortex type as
the target space is
\begin{align}
S^1=\{
c^{\,}_{1}X^{\,}_{21}+c^{\,}_{2}X^{\,}_{23}
|~c^{2}_{1}+c^{2}_{2}=1,~c^{\,}_{i}\in\mathbb{R}
\}
\end{align}
and $\pi^{\,}_{1}(S^{1})=\mathbb{Z}$.

\textit{Case $\nu=4$:}
There is a topological obstruction of the WZ type as
the target space is
\begin{align}
S^3=&
\biggl\{
c^{\,}_{1}
X^{\,}_{210}
+c^{\,}_{2}
X^{\,}_{230}
+c^{\,}_{3}
X^{\,}_{102}
+c^{\,}_{4}
X^{\,}_{222}
\nonumber\\
&\quad
\bigg|~
\sum_{i=1}^{4}c^{2}_{i}=1,
~c^{\,}_{i}\in\mathbb{R}
\biggr\}
\end{align}
and $\pi^{\,}_{3}(S^{3})=\mathbb{Z}$.

\textit{Case $\nu=8$:}
There is no topological obstruction
as one can find more than four
pairwise anticommuting matrices such as the set
\begin{align}
\{
X^{\,}_{2100},
X^{\,}_{2310},
X^{\,}_{2331},
X^{\,}_{2333},
X^{\,}_{1120}
\}.
\end{align}
\end{subequations}

We conclude that the effects of interactions
on the two-dimensional topological superconductors
in the symmetry class DIII
with additional reflection symmetry are
to reduce the topological classification $\mathbb{Z}$
in the noninteracting limit down to $\mathbb{Z}^{\,}_{8}$.

\begin{table}[tb]
\caption{
Reduction from $\mathbb{Z}$ 
to
$\mathbb{Z}^{\,}_{8}$ due to interactions
for the topologically equivalent classes of the
two-dimensional topological superconductors protected by
time-reversal and reflection symmetries (DIII$+R$).
We denote by $V^{\,}_{\nu}$ the space of
$\nu\times\nu$ normalized Dirac mass matrices in boundary ($d=1$)
Dirac Hamiltonians
belonging to the symmetry class D. 
The limit $\nu\to\infty$ of these spaces is the classifying space
$R^{\,}_{1}$. 
The second column shows the stable $D$-th homotopy groups 
of the classifying space $R^{\,}_{1}$. 
The third column gives the number $\nu$ of copies of 
boundary (Dirac) fermions for which a topological obstruction is permissible.
The fourth column gives the type of topological obstruction
that prevents the gapping of the boundary (Dirac) fermions.
\label{tab: 2D DIII+R--}
        }
\begin{tabular}{ccccccc}
\hline \hline
$D$ 
&\qquad\qquad&
$\pi^{\,}_{D}(R^{\,}_{1})$ 
&\qquad\qquad&
$\nu$  
&\qquad\qquad& 
Topological obstruction 
\\
\hline
0&&$\mathbb{Z}^{\,}_{2}$ && 1  && Domain wall  \\
1&&$\mathbb{Z}^{\,}_{2}$ && 2  && Vortex       \\
2&&0                   &&    &&              \\
3&&$\mathbb{Z}$        && 4  &&  WZ term     \\
4&&0                   &&    &&              \\
5&&0                   &&    &&              \\
6&&0                   &&    &&              \\
7&&$\mathbb{Z}$        && 8  &&  None        \\
\hline \hline
\end{tabular}
\end{table}

\subsection{Three-dimensional insulators with time-reversal
and reflection symmetries (AII$+R$)}

We consider again the bulk, boundary, and dynamical boundary Hamiltonian
defined in Sec.\ \ref{subsubsec: The symmetry class AII when d=3},
i.e., Eqs.\ 
(\ref{eq: def mathcal H 0 AII d=3})--%
(\ref{eq: def mathcal H 0 AII d=3 dyn bd V nu=1}).
We observe that the single-particle Hamiltonian
(\ref{eq: def mathcal H 0 AII d=3 a})
has the symmetry
\begin{subequations}
\begin{equation}
\mathcal{R}^{\,}_{x}\,\mathcal{H}^{(0)}(-x,y,z)\,(\,\mathcal{R}^{\,}_{x})^{-1}
=
+
\mathcal{H}^{(0)}(x,y,z),
\end{equation}
where
\begin{align}
\mathcal{R}^{\,}_{x}:=
\mathrm{i}X^{\,}_{10},
\quad
\mathcal{R}^{2}_{x}=-1,
\quad
[\mathcal{T},\mathcal{R}^{\,}_{x}]=0,
\end{align}
\label{eq: def R in AII+R d=3}
\end{subequations}
in addition to the TRS (\ref{eq: def mathcal H 0 AII d=3 b}).
The presence of the additional reflection symmetry
allows one to define a mirror Chern number ($n^{\,}_{+}\in\mathbb{Z}$)
for the sector with the eigenvalue $\mathcal{R}^{\,}_{x}=+\mathrm{i}$
on the two-dimensional mirror plane ($k^{\,}_{x}=0$) in the three-dimensional
Brillouin zone.%
~\cite{hsieh12}
(The mirror Chern number for the eigensector 
$\mathcal{R}^{\,}_{x}=-\mathrm{i}$
is $-n^{\,}_{+}$.)
Thus, the $\nu$ linearly independent zero modes that follow from tensoring
the single-particle Hamiltonian
(\ref{eq: def mathcal H 0 AII d=3 a})
with the $\nu\times\nu$ unit matrix $\openone$ 
along the domain wall (\ref{eq: def mathcal H 0 AII d=3 domain wall a})
are stable to strong one-body perturbations on the boundary
that preserve the reflection symmetry (\ref{eq: def R in AII+R d=3}).
(If we forget the reflection symmetry and keep only the TRS,
as we did in Sec.\ \ref{subsubsec: The symmetry class AII when d=3},
it is only the parity of $\nu$ that is stable to
strong one-body perturbations on the boundary.)
If we only consider dynamical masses that preserve the 
fermion-number $U(1)$ symmetry,
the space of normalized boundary dynamical Dirac mass matrices 
after tensoring the boundary dynamical Dirac Hamiltonian%
~(\ref{eq: def mathcal H 0 AII d=3 dyn bd V nu=1})
with $\openone$ is homeomorphic to the space of normalized Dirac masses
for the two-dimensional system in the symmetry class A
\begin{align}
V^{\,}_{\nu}=\bigcup_{k=0}^{\nu} U(\nu)/[U(k)\times U(\nu-k)].
\end{align}
The limit $\nu\to\infty$ of these spaces is the classifying space
$C^{\,}_{0}$.

Integrating the boundary Dirac fermions delivers a 
QNLSM in (2+1)-dimensional space and time.
In order to gap out dynamically the boundary zero modes 
without breaking the symmetries,
this QNLSM must be free of topological obstructions.
We construct explicitly the spaces 
for the relevant normalized boundary dynamical Dirac mass matrices
[$M(\tau,x,y)$ in Eq.~(\ref{eq: def mathcal H 0 AII d=3 dyn bd})]
of dimension $\nu=2^{n}$ with $n=0,1,2,3$ in the following.
The relevant homotopy groups are given in 
Table~\ref{tab: 2D AII+R-}.

\begin{subequations}
\textit{Case $\nu=1$:}
There is a topological obstruction of the domain wall type as
the target space is
\begin{equation}
S^{0}=\{\pm1\}
\end{equation}
and $\pi^{\,}_{0}(S^{0}) \neq 0$.

\textit{Case $\nu=2$:}
There is a topological obstruction of the monopole type as
the target space is
\begin{align}
S^{2}=
\{
c^{\,}_{1}X^{\,}_{1}+c^{\,}_{2}X^{\,}_{2}+c^{\,}_{3}X^{\,}_{3}
|~c^{2}_{1}+c^{2}_{2}+c^{2}_{3}=1
\}
\end{align}   
and $\pi^{\,}_{2}(S^{2})=\mathbb{Z}$.

\textit{Case $\nu=4$:}
There is a topological obstruction of the WZ type as
the target space is
\begin{align}
S^{4}=&
\biggl\{
c^{\,}_{1}
X^{\,}_{13}
+c^{\,}_{2}
X^{\,}_{23}
+c^{\,}_{3}
X^{\,}_{33}
+c^{\,}_{4}
X^{\,}_{01}
+c^{\,}_{5}
X^{\,}_{02}
\nonumber\\
&\quad
\bigg|~
\sum_{i=1}^{5}c^{2}_{i}=1,
~c^{\,}_{i}\in\mathbb{R}
\biggr\}
\end{align}   
and $\pi^{\,}_{4}(S^{4})=\mathbb{Z}$.

\textit{Case $\nu=8$:}
There is no topological obstruction
as one can find more than five
pairwise anticommuting matrices such as the set
\begin{align}
\{
X^{\,}_{133},
X^{\,}_{233},
X^{\,}_{333},
X^{\,}_{013},
X^{\,}_{023},
X^{\,}_{001},
X^{\,}_{002}
\}.
\end{align}
\end{subequations}

We conclude that the effects of interactions
on the three-dimensional topological insulators
in the symmetry class AII with an additional symmetry are
to reduce the topological classification $\mathbb{Z}$
in the noninteracting limit down to $\mathbb{Z}^{\,}_{8}$.

\begin{table}[tb]
\caption{
Reduction from $\mathbb{Z}$ 
to
$\mathbb{Z}^{\,}_{8}$ due to interactions
for the topologically equivalent classes of the
three-dimensional topological insulators 
with time-reversal and reflection symmetries (AII$+R$).
We denote by $V^{\,}_{\nu}$ the space of
$\nu\times\nu$ normalized Dirac mass matrices in boundary ($d=2$)
Dirac Hamiltonians
belonging to the symmetry class A.
The limit $\nu\to\infty$ of these spaces is the classifying space
$C^{\,}_{0}$. 
The second column shows the stable $D$-th homotopy groups 
of the classifying space $C^{\,}_{0}$. 
The third column gives the number $\nu$ of copies of 
boundary (Dirac) fermions for which a topological obstruction is permissible.
The fourth column gives the type of topological obstruction
that prevents the gapping of the boundary (Dirac) fermions.
\label{tab: 2D AII+R-}
        }
\begin{tabular}{ccccccc}
\hline \hline
$D$ 
&\qquad\qquad&
$\pi^{\,}_{D}(C^{\,}_{0})$ 
&\qquad\qquad&
$\nu$ 
&\qquad\qquad&
Topological obstruction 
\\
\hline
0&&$\mathbb{Z}$ && 1 && Domain wall  \\
1&&0            &&   &&              \\
2&&$\mathbb{Z}$ && 2 && Monopole     \\
3&&0            &&   &&              \\
4&&$\mathbb{Z}$ && 4 && WZ term      \\
5&&0            &&   &&              \\
6&&$\mathbb{Z}$ && 8 && None         \\
7&&0            &&   &&              \\
\hline \hline
\end{tabular}
\end{table}

This $\mathbb{Z}^{\,}_{8}$ classification is unchanged if
all boundary dynamical masses 
that break the fermion-number $U(1)$ symmetry
are accounted for.
The corresponding target spaces for 
the boundary dynamical masses and their topological obstructions 
are derived as was done in the stability analysis made for the
symmetry class AIII in $d=3$
in Sec.~\ref{subsubsec: The symmetry class AIII when d=3}.
Namely, we extend the single-particle Hamiltonian 
[Eq.~(\ref{eq: def mathcal H 0 AII d=3 dyn bd})]
to a BdG Hamiltonian
\begin{align}
\mathcal{H}^{(\mathrm{dyn})}_{\mathrm{bd}}
&=
\left(
-
\mathrm{i}\partial^{\,}_{x}\,
\sigma^{\,}_{2}\tensor\rho^{\,}_{3}
-
\mathrm{i}\partial^{\,}_{y}\,
\sigma^{\,}_{1}\tensor\rho^{\,}_{0}
\right)
\tensor\openone
+
\gamma'(\tau,x,y),
\end{align}
where $\rho^{\,}_{0}$ and $\rho^{\,}_{\mu}$ are unit $2\times 2$ and
Pauli matrices, respectively, acting on the particle-hole (Nambu) space
and the particle-hole symmetry is given by 
$\mathcal{C}=\rho^{\,}_{1}\,\mathsf{K}$.
In this case, the target spaces of the QNLSM made of 
normalized boundary dynamical Dirac mass matrices $\gamma'$
of dimension $\nu=2^{n}$ with $n=0,1,2,3$ 
are modified as listed in the following with the notation
$X^{\,}_{\mu \mu' \mu'' \mu''' \ldots}
=\sigma^{\,}_{\mu} \tensor \rho^{\,}_{\mu'} \tensor
 \tau^{\,}_{\mu''} \tensor \tau^{\,}_{\mu'''} \ldots$.
The relevant homotopy groups are given in 
Table~\ref{tab: 2D AII+R- BdG}.
We note that these target spaces are closed 
under the global $U(1)$ transformation generated by $\rho^{\,}_{3}$.

\begin{subequations}
\textit{Case $\nu=1$:}
There is a topological obstruction of the vortex type as
the target space is
\begin{equation}
S^1=\{
c^{\,}_{1}X_{21} 
+c^{\,}_{2}X_{22}
|~c^{2}_{1}+c^{2}_{2}=1
\}
\end{equation}
and $\pi^{\,}_{1}(S^{1}) = \mathbb{Z}$.

\textit{Case $\nu=2$:}
There is a topological obstruction of the monopole type as
the target space is
\begin{align}
S^{2}=
\{
c^{\,}_{1}X^{\,}_{210} 
+c^{\,}_{2}X^{\,}_{220} 
+c^{\,}_{3}X^{\,}_{302}
|~c^{2}_{1}+c^{2}_{2}+c^{2}_{3}=1
\}
\end{align}   
and $\pi^{\,}_{2}(S^{2})=\mathbb{Z}$.

\textit{Case $\nu=4$:}
There is a topological obstruction of the WZ type as
the target space is
\begin{align}
S^{4}=&
\biggl\{
c^{\,}_{1}
X^{\,}_{2100}
+c^{\,}_{2}
X^{\,}_{2200}
+c^{\,}_{3}
X^{\,}_{3020}
+c^{\,}_{4}
X^{\,}_{3012}
+c^{\,}_{5}
X^{\,}_{3032}
\nonumber\\
&\quad
\bigg|~
\sum_{i=1}^{5}c^{2}_{i}=1,
~c^{\,}_{i}\in\mathbb{R}
\biggr\}
\end{align}   
and $\pi^{\,}_{4}(S^{4})=\mathbb{Z}$.

\textit{Case $\nu=8$:}
There is no topological obstruction
as one can find more than five
pairwise anticommuting matrices such as the set
\begin{align}
\{
X^{\,}_{21000},
X^{\,}_{22000},
X^{\,}_{30200},
X^{\,}_{30120},
X^{\,}_{30312},
X^{\,}_{30332}
\}.
\end{align}
\end{subequations}
Therefore, the topological classification $\mathbb{Z}^{\,}_{8}$ for
three-dimensional TCIs
in the symmetry class AII with an additional reflection symmetry
is unchanged when the superconducting fluctuations are accounted for.
This $\mathbb{Z}^{\,}_{8}$ classification is consistent with the results
obtained recently in Refs.~\onlinecite{Isobe15,Yoshida15}.
We note that the classifying space $R^{\,}_{0}$ for the  
dynamical masses is the same as that in the case of three-dimensional TSs
in the symmetry class DIII. There is an important difference, however. 
Namely,
the line corresponding to $\nu=1$ is moved to $D=1$
in Table~\ref{tab: 2D AII+R- BdG} from $D=0$
in Table~\ref{tab: 3D DIII}.
This change originates from the fact that
the minimum matrix dimension of
the BdG Hamiltonian $\mathcal{H}^{(\mathrm{dyn})}_{\mathrm{bd}}$ for
three-dimensional TCIs in the symmetry class AII+$R$ is four,
while that for three-dimensional TSs in the symmetry class DIII is two.
Hence, the breakdown of the topological classification $\mathbb{Z}$
for three-dimensional TCIs in the symmetry class AII+$R$ takes place 
at $\nu=8$, which is the half of $\nu=16$ for three-dimensional
TSs in the symmetry class DIII.

\begin{table}[tb]
\caption{
Reduction from $\mathbb{Z}$ 
to
$\mathbb{Z}^{\,}_{8}$ due to interactions
for the topologically equivalent classes of the
three-dimensional topological insulators 
with time-reversal and reflection symmetries (AII$+R$)
when the superconducting fluctuations are accounted for.
We denote by $V^{\,}_{\nu}$ the space of
$\nu\times\nu$ normalized Dirac mass matrices in boundary ($d=2$)
Dirac Hamiltonians
belonging to the symmetry class D.
The limit $\nu\to\infty$ of these spaces is the classifying space
$R^{\,}_{0}$. 
The second column shows the stable $D$-th homotopy groups 
of the classifying space $R^{\,}_{0}$. 
The third column gives the number $\nu$ of copies of 
boundary (Dirac) fermions for which a topological obstruction is permissible.
The fourth column gives the type of topological obstruction
that prevents the gapping of the boundary (Dirac) fermions.
\label{tab: 2D AII+R- BdG}
        }
\begin{tabular}{ccccccc}
\hline \hline
$D$ 
&\qquad\qquad&
$\pi^{\,}_{D}(R^{\,}_{0})$ 
&\qquad\qquad&
$\nu$ 
&\qquad\qquad&
Topological obstruction 
\\
\hline
0&&$\mathbb{Z}$ &&  &&   \\
1&&$\mathbb{Z}_2$ && 1 && Vortex       \\
2&&$\mathbb{Z}_2$ && 2 && Monopole     \\
3&&0            &&   &&              \\
4&&$\mathbb{Z}$ && 4 && WZ term      \\
5&&0            &&   &&              \\
6&&0            &&   &&              \\
7&&0            &&   &&              \\
8&&$\mathbb{Z}$ && 8 && None  \\
\hline \hline
\end{tabular}
\end{table}

\subsection{Massless Dirac fermions on the surfaces of SnTe}

The crystal SnTe is a three-dimensional topological crystalline
insulator protected by time-reversal and reflection symmetries
(AII$+R$).
SnTe supports four Dirac cones on the [001] surface and
six Dirac cones on the [111] surface.
If strong interaction effects are present,
we expect that the $\nu=4$ phase described by a QNLSM with a WZ term 
should be realized on the [001] surface.
At the [111] surface, we expect that the $\nu=6=4+2$ phase be realized,
whereby the effective field theory is that of a
QNLSM with a WZ term for 4 out of the six surface Dirac cones
and that of a QNLSM with a topological term arising 
from a gas of monopoles of the remaining two surface Dirac cones.

\acknowledgments
This work was supported in part by Grants-in-Aid from the Japan Society for
Promotion of Science (Grant No.~15K05141)
and by the RIKEN iTHES Project.

\appendix

\section{Defining symmetries of strong topological insulators 
(superconductors)}
\label{appsec: Defining symmetries of strong topological insulators}

Define the many-body quadratic form
\begin{subequations}
\begin{equation}
\widehat{H}=
\int\mathrm{d}^{d}\bm{x}\,
\int\mathrm{d}^{d}\bm{y}\,
\sum_{ij}
\hat{\psi}^{\dag}_{i}(t,\bm{x})\,
\mathcal{H}^{\,}_{ij}(\bm{x},\bm{y})\,
\hat{\psi}^{\,}_{j}(t,\bm{y}),
\end{equation}
where
\begin{equation}
\mathcal{H}^{\,}_{ij}(\bm{x},\bm{y})=
\mathcal{H}^{*}_{ji}(\bm{y},\bm{x}),
\end{equation}
and
\begin{equation}
\{\hat{\psi}^{\,}_{i}(t,\bm{x}),\hat{\psi}^{\dag}_{j}(t,\bm{y})\}=
\delta^{\,}_{ij}\,
\delta(\bm{x}-\bm{y})
\end{equation}
\end{subequations}
are the only non-vanishing equal-time anticommutators.

\textit{Time-reversal symmetry.}
Let $\mathsf{K}$ denote complex conjugation.
Define the time-reversal transformation by the antiunitary transformation
\begin{subequations} 
\begin{equation}
\hat{\mathrm{T}}:=
\hat{\mathcal{T}}\,
\mathsf{K}
\end{equation}
that reverses time but leaves space unchanged
by demanding that
\begin{equation}
\hat{\mathcal{T}}^{-1}=\hat{\mathcal{T}}^{\dag}
\end{equation}
and
\begin{equation}
\hat{\mathrm{T}}\,
\hat{\psi}^{\,}_{j}(t,\bm{y})\,
\hat{\mathrm{T}}^{-1}=
\sum_{j'}
\mathcal{T}^{*}_{j'j}\,
\hat{\psi}^{\,}_{j'}(-t,\bm{y}).
\end{equation}
\end{subequations}
One verifies that
\begin{subequations}
\begin{equation}
\hat{\mathrm{T}}\,
\widehat{H}\,
\hat{\mathrm{T}}^{-1}=
\widehat{H}
\end{equation}
if and only if
\begin{equation}
\sum_{ij}
\mathcal{T}^{\,}_{i'i}\,
\mathcal{H}^{*}_{ij}(\bm{x},\bm{y})\,
\mathcal{T}^{-1}_{jj'}=
\mathcal{H}^{\,}_{i'j'}(\bm{x},\bm{y}).
\end{equation}
\end{subequations}

\textit{Particle-hole (charge-conjugation) symmetry.}
Assume that
\begin{equation}
\sum_{i}\mathcal{H}^{\,}_{ii}(\bm{x},\bm{y})=0.
\end{equation}
Define the particle-hole transformation by the unitary transformation
\begin{subequations} 
\begin{equation}
\hat{\mathrm{C}}:=
\hat{\mathcal{C}}
\end{equation}
that reverses the sign of the fermion number
\begin{equation}
\hat{n}^{\,}_{i}(x)
-
\frac{1}{2}\,\delta(\bm{x}=0):=
\hat{\psi}^{\dag}_{i}(\bm{x})\,
\hat{\psi}^{\,  }_{i}(\bm{x})
-
\frac{1}{2}\,\delta(\bm{x}=0)
\end{equation}
measured relative to the background of the fermion density 1/2
but leaves space unchanged by demanding that
\begin{equation}
\hat{\mathcal{C}}^{-1}=\hat{\mathcal{C}}^{\dag}
\end{equation}
and
\begin{equation}
\hat{\mathrm{C}}\,
\hat{\psi}^{\,}_{j}(t,\bm{y})\,
\hat{\mathrm{C}}^{-1}=
\sum_{j'}
\mathcal{C}^{\,}_{j'j}\,
\hat{\psi}^{\dag}_{j'}(t,\bm{y}).
\end{equation}
\end{subequations}
One verifies that
\begin{subequations}
\begin{equation}
\hat{\mathrm{C}}\,
\widehat{H}\,
\hat{\mathrm{C}}^{-1}=
\widehat{H}
\end{equation}
if and only if
\begin{equation}
\sum_{ij}
\mathcal{C}^{\,}_{i'i}\,
\mathcal{H}^{*}_{ij}(\bm{y},\bm{x})\,
\mathcal{C}^{-1}_{jj'}=
-
\mathcal{H}^{\,}_{i'j'}(\bm{y},\bm{x}).
\end{equation}
\end{subequations}

\textit{Chiral symmetry.}
Assume that
\begin{equation}
\sum_{i}\mathcal{H}^{\,}_{ii}(\bm{x},\bm{y})=0.
\end{equation}
Define the chiral transformation by the antiunitary transformation
\begin{subequations} 
\begin{equation}
\hat{\mathrm{S}}:=
\hat{\mathcal{S}}\,
\mathsf{K}
\end{equation}
that reverses the sign of the fermion number
\begin{equation}
\hat{n}^{\,}_{i}(x)
-
\frac{1}{2}\,\delta(\bm{x}=0):=
\hat{\psi}^{\dag}_{i}(\bm{x})\,
\hat{\psi}^{\,  }_{i}(\bm{x})
-
\frac{1}{2}\,\delta(\bm{x}=0)
\end{equation}
measured relative to the background of the fermion density 1/2
but leaves space unchanged by demanding that
\begin{equation}
\hat{\mathcal{S}}^{-1}=\hat{\mathcal{S}}^{\dag}
\end{equation}
and
\begin{equation}
\hat{\mathrm{S}}\,
\hat{\psi}^{\,}_{j}(t,\bm{y})\,
\hat{\mathrm{S}}^{-1}=
\sum_{j'}
\mathcal{S}^{\,}_{j'j}\,
\hat{\psi}^{\dag}_{j'}(t,\bm{y}).
\end{equation}
\end{subequations}
One verifies that
\begin{subequations}
\begin{equation}
\hat{\mathrm{S}}\,
\widehat{H}\,
\hat{\mathrm{S}}^{-1}=
\widehat{H}
\end{equation}
if and only if
\begin{equation}
\sum_{ij}
\mathcal{S}^{\,}_{i'i}\,
\mathcal{H}^{\,}_{ij}(\bm{y},\bm{x})\,
\mathcal{S}^{-1}_{jj'}=
-
\mathcal{H}^{\,}_{i'j'}(\bm{y},\bm{x}).
\end{equation}
\end{subequations}

The unitary symmetry under $\hat{\mathrm{C}}$ 
is called charge conjugation symmetry
or particle-hole symmetry (PHS). The antiunitary symmetry under 
$\hat{\mathrm{S}}$ 
is called the chiral symmetry (CHS). The antiunitary symmetry under 
$\hat{\mathrm{T}}$
is called time-reversal symmetry (TRS).

\ \par

\section{Tenfold way and classifying spaces}
\label{app: tenfold way}

\begin{table*}[tb]
\begin{center}
\caption{
Complex and real classifying spaces and their stable homotopy groups.
Homotopy groups $\pi^{\,}_{D}(V)$ for complex and real classifying spaces
are periodic in $D$ with periods of 2 and 8, respectively.
\label{table: all homotopies classifying spaces}
        }
\begin{tabular}[t]{cccccccccc}
\hline \hline
Label 
& 
Classifying space $V$
& 
$\pi^{\,}_{0}(V)$ 
&
$\pi^{\,}_{1}(V)$ 
&
$\pi^{\,}_{2}(V)$
&
$\pi^{\,}_{3}(V)$ 
 &
$\pi^{\,}_{4}(V)$ 
&
$\pi^{\,}_{5}(V)$ 
&
$\pi^{\,}_{6}(V)$ 
&
$\pi^{\,}_{7}(V)$
\\
\hline
$C^{\,}_{0}$
&
$\cup_{n=0}^{N}\big\{U(N)/\big[U(n)\times U(N-n)\big]\big\}$
&
$\mathbb{Z}$
&
$0$
&
$\mathbb{Z}$
&
$0$
&
$\mathbb{Z}$
&
$0$
&
$\mathbb{Z}$
&
$0$
\\
~$C^{\,}_{1}$
&
~$U(N)$
&
$0$
&
$\mathbb{Z}$
&
$0$
&
$\mathbb{Z}$
&
$0$
&
$\mathbb{Z}$
&
$0$ 
&
$\mathbb{Z}$
\\
\hline 
~$R^{\,}_{0}$
&
~$\cup_{n=0}^{N}\big\{O(N)/\big[O(n)\times O(N-n)\big]\big\}$
& 
$\mathbb{Z}$
&
$\mathbb{Z}^{\,}_{2}$
&
$\mathbb{Z}^{\,}_{2}$
&
$0$
&
$\mathbb{Z}$
&
$0$
& 
$0$
&
$0$
\\
~$R^{\,}_{1}$
& 
~$O(N)$
& 
$\mathbb{Z}^{\,}_{2}$ 
&
$\mathbb{Z}^{\,}_{2}$
&
$0$
&
$\mathbb{Z}$
&
$0$
&
$0$
&
$0$
&
$\mathbb{Z}$
\\
~$R^{\,}_{2}$
& 
$O(2N)/U(N)$ 
& 
$\mathbb{Z}^{\,}_{2}$
&
$0$
&
$\mathbb{Z}$
&
$0$
&
$0$
&
$0$
&
$\mathbb{Z}$
&
$\mathbb{Z}^{\,}_{2}$
\\
~$R^{\,}_{3}$ 
& 
~$U(2N)/Sp(N)$
&
$0$
&
$\mathbb{Z}$
&
$0$
&
$0$
&
$0$   
&
$\mathbb{Z}$
&
$\mathbb{Z}^{\,}_{2}$
&
$\mathbb{Z}^{\,}_{2}$
\\
~$R^{\,}_{4}$
& 
~$\cup_{n=0}^{N}\big\{Sp(N)/\big[Sp(n)\times Sp(N-n)\big]\big\}$
& 
$\mathbb{Z}$
&
$0$
&
$0$
&
$0$ 
&
$\mathbb{Z}$
&
$\mathbb{Z}^{\,}_{2}$
&
$\mathbb{Z}^{\,}_{2}$
&
$0$
\\
~$R^{\,}_{5}$
&
~$Sp(N)$
& 
$0$
&
$0$         
&
$0$
&
$\mathbb{Z}$ 
&
$\mathbb{Z}^{\,}_{2}$
&
$\mathbb{Z}^{\,}_{2}$
&
$0$
&
$\mathbb{Z}$
\\
~$R^{\,}_{6}$
& 
~$Sp(N)/U(N)$
& 
$0$
&
$0$   
&
$\mathbb{Z}$         
&
$\mathbb{Z}^{\,}_{2}$
&
$\mathbb{Z}^{\,}_{2}$
&
$0$
&
$\mathbb{Z}$         
&
$0$
\\
~$R^{\,}_{7}$ 
& 
~$U(N)/O(N)$
& 
$0$
&
$\mathbb{Z}$  
&
$\mathbb{Z}^{\,}_{2}$
&
$\mathbb{Z}^{\,}_{2}$
&
$0$
&
$\mathbb{Z}$
&
$0$
&
$0$       
\\
\hline \hline
\end{tabular}
\end{center}
\end{table*} 

In this appendix, we summarize the classification of gapped phases
of noninteracting fermions in terms of the tenfold way.
We also define the classifying spaces of
normalized Dirac masses.
The ten Altland-Zirnbauer (AZ) symmetry classes 
for Hermitian matrices are shown in
Table~\ref{table: topo classification short-rnaged entangles AZ classes}.
There, two complex and eight real symmetry classes are characterized
by the presence or the absence of 
time-reversal symmetry ($T$), 
particle-hole symmetry ($C$), 
and chiral symmetry ($\Gamma$).
Their presence is indicated
by the sign entering the squared operators, 
$T^{2}=\pm1$ or $C^{2}=\pm1$, 
and by $1$ for $\Gamma$.
Their absence is indicated by 0. 
For each symmetry class and for any dimension $d=0,1,2,\ldots$
of space, the classifying space $V^{\,}_{d}$,
which is the space of normalized Dirac masses, 
is given in the last column by labels to symmetric spaces.
We list the ten relevant symmetric spaces and 
their homotopy groups in the stable homotopy regime
in Table~\ref{table: all homotopies classifying spaces}.
The number $N$ is related to the dimension $r=r^{\,}_{\min}N$
of the Dirac matrices, i.e.,
$N=1,2,\ldots$ is the number of copies of the minimal massive
Dirac Hamiltonian of rank $r^{\,}_{\min}$.
The stable homotopy regime refers to the limit $N\to \infinity$.
According to the Bott periodicity,
the complex classifying spaces obey the periodicity condition
\begin{align}
\pi^{\,}_{D}(C^{\,}_{q})&=\pi^{\,}_{D+2}(C^{\,}_{q}),
& 
(q&=0,1),
\end{align}
and the real classifying spaces obey the periodicity condition
\begin{align}
\pi^{\,}_{D}(R^{\,}_{q})&=\pi^{\,}_{D+8}(R^{\,}_{q}),
&
(q&=0,\dots,7).
\end{align}

\bibliography{Z16}

\end{document}